%% file: ks.tex
   \documentclass[pre,twocolumn,groupedaddress,showpacs,showkeys,letterpaper,floatfix]{revtex4}

\usepackage{amssymb}
\usepackage{amsmath}
\usepackage{mathrsfs}
\usepackage{dsfont} 
\usepackage[dvips]{graphicx}
\usepackage{natbib}       
\usepackage{tableKS}	
            \input defsKS  
\begin{document}
                \title{
Unstable recurrent patterns in Kuramoto-Sivashinsky dynamics
                        }
                        \author{
Yueheng Lan
                }
                \email{ylan2@engineering.ucsb.edu}
                        \affiliation{
    Department of Mechanical and Environmental Engineering,\\
    University of California,
    Santa Barbara, CA 93106
                }
                \author{
Predrag Cvitanovi\'c
                }
                  \email{Predrag.Cvitanovic@physics.gatech.edu}
                \affiliation{
    Center for Nonlinear Science, School of Physics,\\
    Georgia Institute of Technology, Atlanta, GA 30332-0430
                }

                 \date{\today}

\begin{abstract}

We undertake an exploration of recurrent patterns
in the antisymmetric subspace of 1-dimensional Kuramoto-Sivashinsky system.
For a small, but
already rather ``turbulent'' system, the long-time dynamics takes place on a
low-dimensional invariant manifold. A set of equilibria offers a coarse
geometrical partition of this manifold. The \descent\ method
enables us to determine
numerically a large number of
unstable spatiotemporally periodic solutions.
The attracting set appears surprisingly thin - its backbone are several
Smale horseshoe repellers, well approximated by intrinsic local 1-dimensional
return maps, each with
an approximate symbolic dynamics. The
dynamics appears decomposable into chaotic dynamics within such local
repellers, interspersed by rapid jumps between them.

\end{abstract}

\PC{your list:
02.50.Ey:stochastic processes;
03.20.+k:Theory of quantized fields;
03.65.Sq:semiclassical theories and applications;
05.45.-a:Nonlinear dynamics and nonlinear dynamical systems;
seems unrelated to PACS above?
    }

                \maketitle

%

Statistical approaches to the study of turbulence\rf{frisch},
rely on assumptions which break
down in presence of large-scale coherent structures typical
of fluid motions\rf{HLBcoh98}.
Description of such coherent structures requires detailed
understanding of the dynamics of underlying equations of
motion.
In E. Hopf's dynamical systems vision\rf{hopf48} turbulence explores a
repertoire of distinguishable patterns; as we watch
a turbulent system evolve,
every so often we catch a glimpse of a familiar whorl. 
At any instant and a given finite spatial resolution the system
approximately tracks for a finite time a pattern belonging to a finite
alphabet of admissible patterns, and the dynamics can be thought of as a
walk through the space of such patterns, just as chaotic dynamics with a
low dimensional attractor can be thought of as a succession of nearly
periodic (but unstable) motions.

Exploration of Hopf's program close to the onset of
spatiotemporal chaos was initiated in
\refref{ks} which was the first to
extend the periodic orbit theory to a PDE,
the
1-spatial dimension Kuramoto-Sivashinsky\rf{kuturb78,siv} system,
a flow embedded in
an infinite-dimensional \statesp.
Many recurrent patterns were determined
numerically, and the recurrent-patterns theory predictions tested for
several parameter values.

In this paper
(and, in a much greater detail, in
 \refref{lanthe})
we venture into a bigger {\KS} system, just large
enough to exhibit ``turbulent'' dynamics arising through
competition of several unstable coherent structures.
Basic properties of the {\KSe} are reviewed in \refsect{sect:ksprop}.
In \refsect{sect:ksrecur} we sketch the \descent\ method
that we deploy in our searches for recurrent patterns.
\Eqva, which play a key role
in organizing the \statesp\ dynamics, are investigated
in \refsect{sect:kseqlb}.
We then fix the
size of  {\KS} system  in order to illustrate our
methodology on a concrete example.
In \refsect{sect:kspatt} we show how
intrinsic curvilinear coordinates are built along unstable manifolds,
leading to low-dimensional
Poincar\'{e} return maps. These enable us
to search systematically
for periodic orbits that build up local Smale horseshoe repellers.
Our results are summarized in \refsect{sect:sum}.

\section{Kuramoto-Sivashinsky equation}
\label{sect:ksprop}

The Kuramoto-Sivashinsky system\rf{kuturb78,siv,HLBcoh98} \beq
 u_t=(u^2)_x-u_{xx}-\nu u_{xxxx},
 \ee{kseq}
arises as an amplitude equation for interfacial
instabilities in a variety of physical contexts\rf{siv,chang94,kschang86}.
We shall study $u(x,t)$ on a periodic domain  $x \in [0,L]$,
$u(x,t)=u(x+L,t)$.
In the Fourier space,
\beq
  u(x,t)=i \sum_{k=-\infty}^{+\infty} a_k (t) e^{ i  k q x}
  \,,\qquad q=2\pi/L
\,,
\ee{eq:ksexp}
\KS\ PDE is represented by the infinite ladder of
coupled ODEs for complex Fourier coefficients,
 \beq
 \dot{a}_k=\left(kq\right)^2
           \left(1-\nu\left(kq\right)^2 \right) a_k
              - kq \sum_{m=-\infty}^{+\infty}a_m a_{k-m}
\,.
 \ee{eq:ksexp2}
In this paper
we restrict our investigation to the subspace of odd solutions
$ u(x,t)=-u(-x,t)$ for which $a_k$'s are real.
$a_0$ mode is conserved since $\dot{a_0}=0$ and
$a_0= \int dx \, u = 0$ since $u$ has odd parity.
The linear term controls the stability of the
$u(x,t)=0$ \eqv, with each Fourier mode $a_k$ an eigenvector of the
linearized equation,
with eigenvalue $\omega (kq)=(kq)^2 (1-\nu (kq)^2)$.
If all eigenvalues are
non-positive, the equilibrium $u(x,t)=0$  is globally stable. 
In general, the
longest wavelengths are unstable, while the higher $k$
contract rapidly, restricting dynamics to a finite-dimensional
inertial manifold~\cite{FNSTks88,infdymnon}.
The peak of the $\omega (kq)$ stability curve
identifies the maximally
unstable mode at $kq=1/\sqrt{2\nu}$ and sets the typical
wavelength of the large system size
spatiotemporal patterns of the {\KSe}.

Rescaling
$t \to \nu t$,
$a_k \to \nu^{-1/2} a_k$,
$L = 2\pi \nu^{1/2} \tildeL$, results in
\beq
\dot{a}_k=(k/\tildeL)^2\left( 1 - (k/\tildeL)^2  \right)a_k
           - (k/\tildeL) \sum_{m=-\infty}^{+\infty} a_m a_{k-m}
\,,
\ee{expan}
where we trade in the ``hyper-viscosity'' $\nu$ and the system size $L$ for
a single
dimensionless length parameter
\beq
    \tildeL=L/\left(2 \pi \sqrt{\nu}\right)
    \,,
\ee{tildeL}
which plays the role of a ``Reynolds number''
for the Kuramoto-Sivashinsky system.
In the literature sometimes
$L$ is used as the control parameter, with $\nu$ fixed to $1$, and
at other times $\nu$ is varied with $L$ fixed. 
In what follows we find it most convenient to set $\nu=1$ and
compare different
calculations in terms of either $\tildeL$ or $L$.

\begin{figure} 
\centering
(a)%
\includegraphics[width=0.46\textwidth]{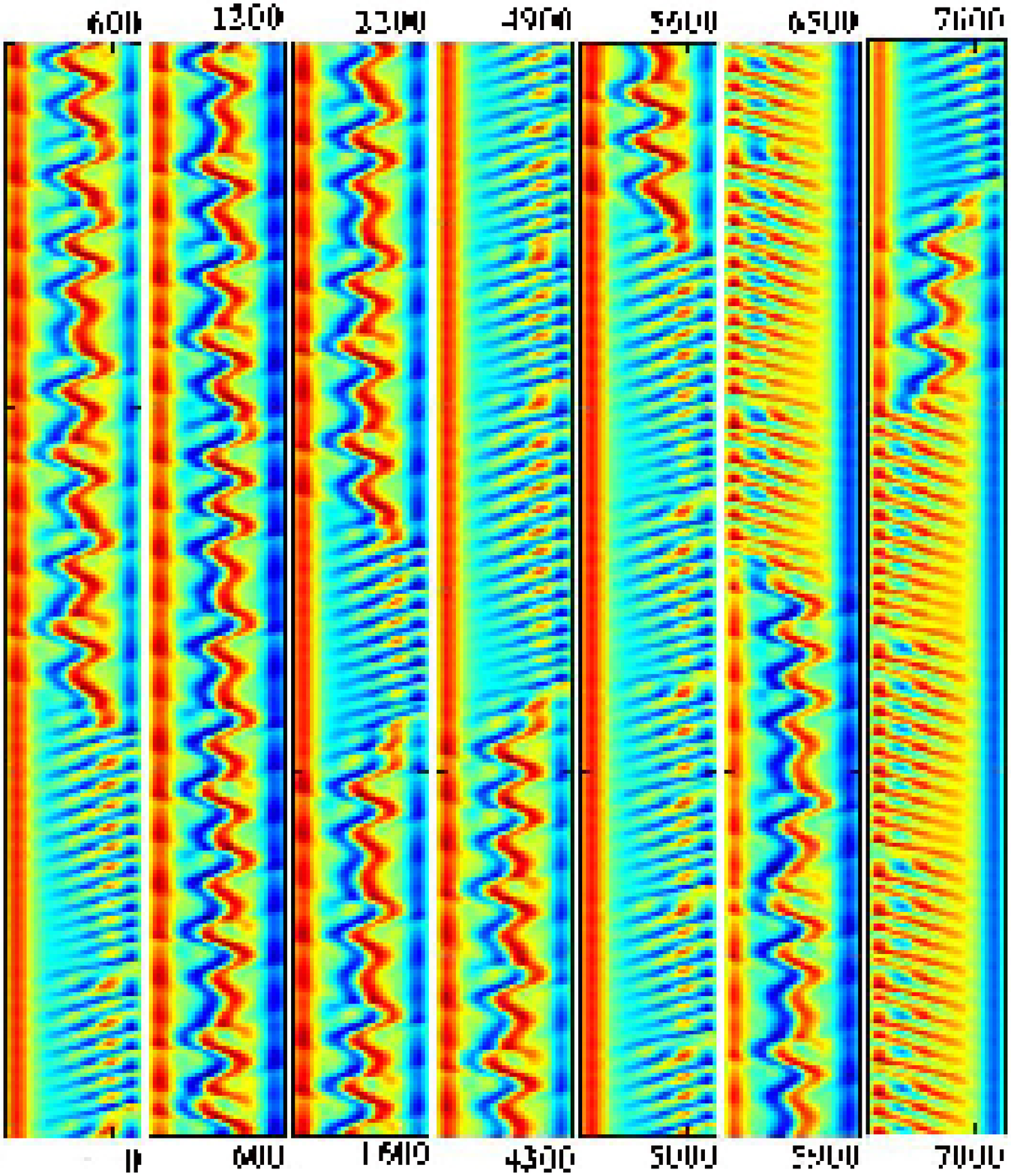}
\\
\hspace{-0.22\textwidth}
\includegraphics[width=0.23\textwidth]{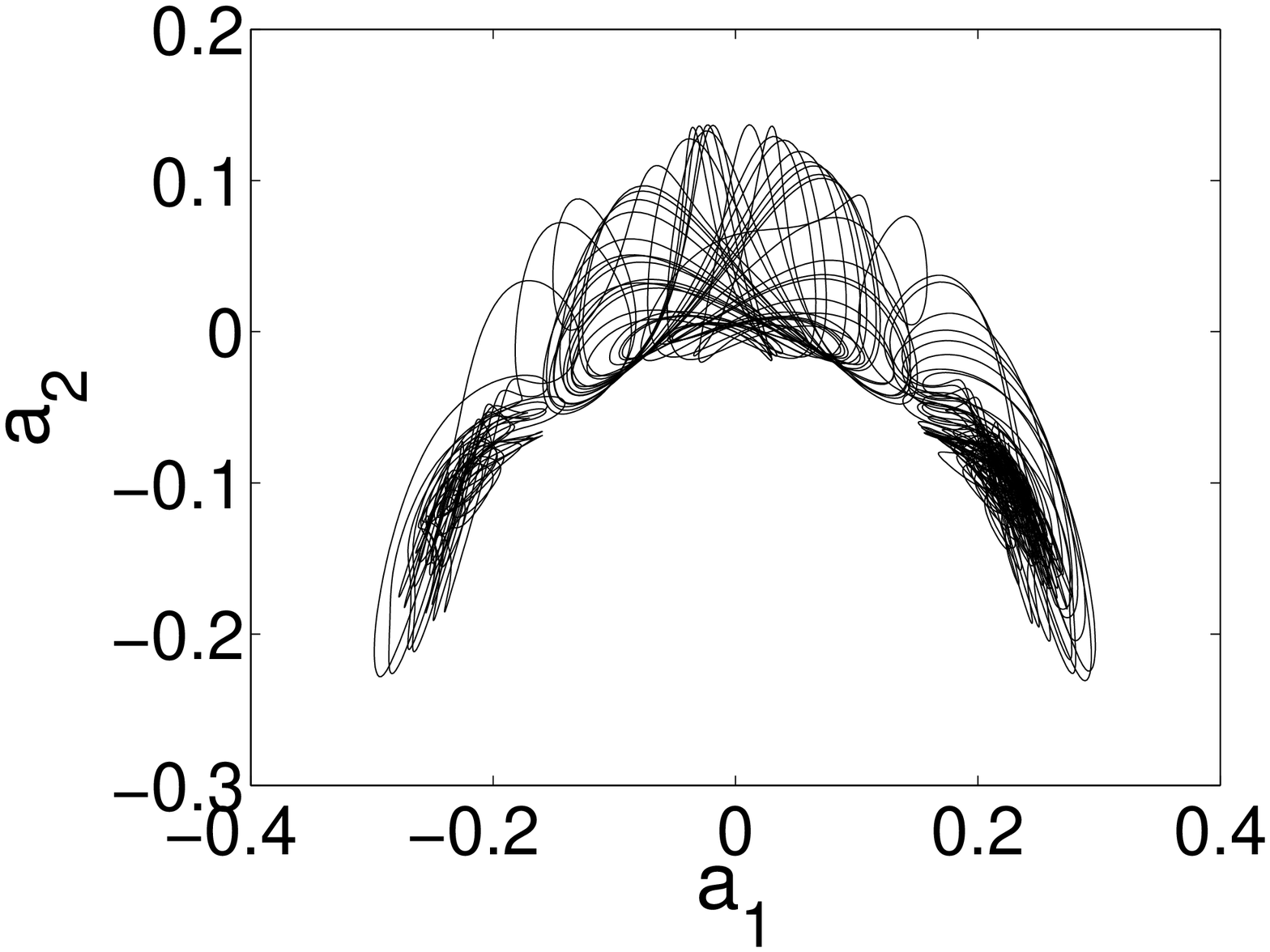}
\hspace{-0.24\textwidth} (b) \hspace{0.22\textwidth}
\includegraphics[width=0.23\textwidth]{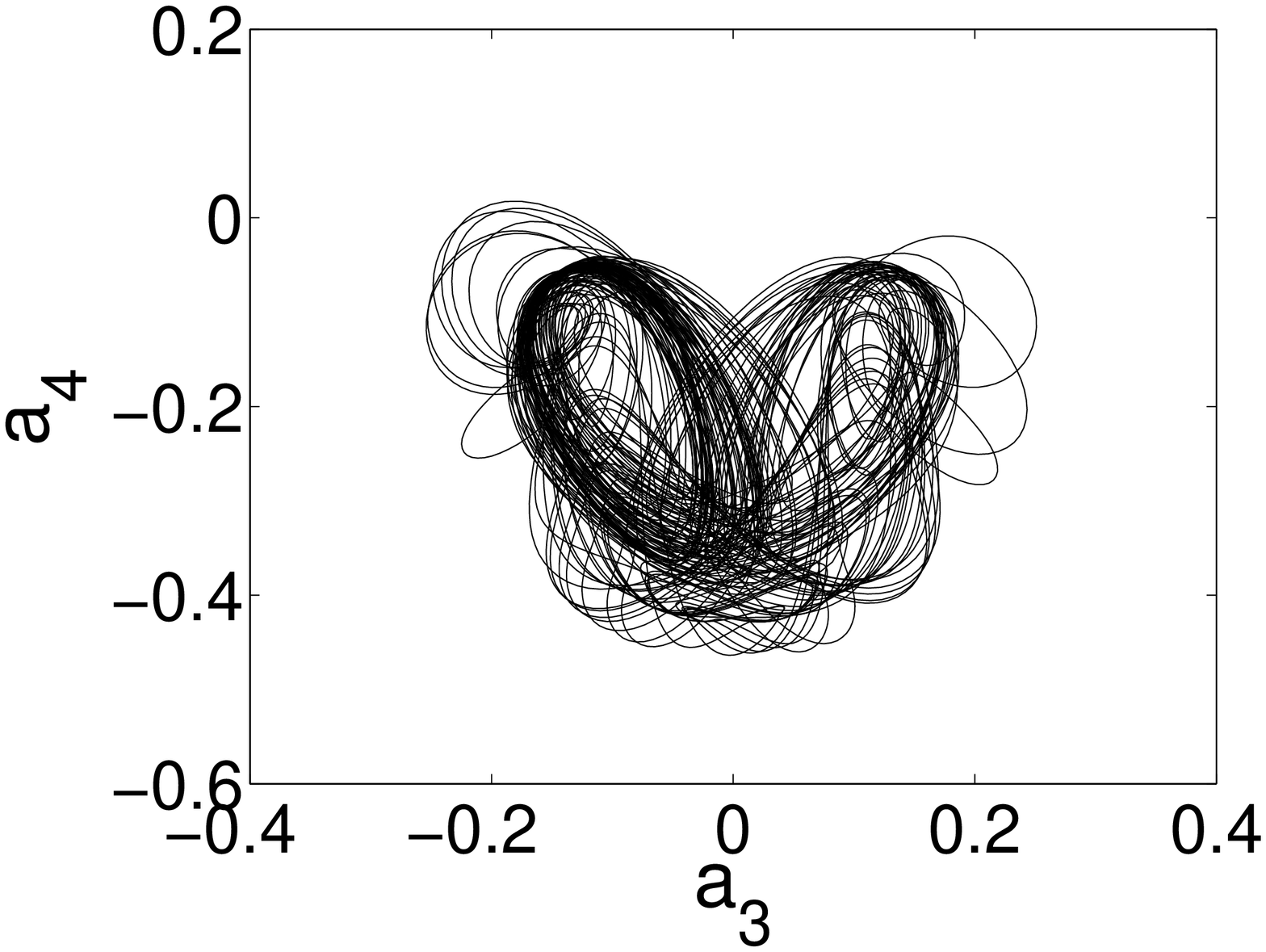}
\hspace{-0.24\textwidth} (c)
\caption{
Long-time evolution of a typical
``sustained turbulence'' trajectory for
$L=38.5$
in:
(a) The space-time representation
$(u(x,t),t)$ for $x \in [0,L/2]]$,
$t \in [0, 7600]$. The color (grayscale) represents
the magnitude of $u(x,t)$.
(b) $[a_1,a_2]$ Fourier modes projection,
(c) $[a_3,a_4]$ projection.
The typical time scale is set
by the shortest periods of the \UPO s embedded in the
central, ``wobbly'' and side, ``traveling wave''
patterns of order $\period{}=20 \sim 25$, see
\reffig{f:ant1p2}, so this is a very long
simulation, over 300 ``turnover'' times. The goal of this
paper is to describe the characteristic unstable ``wobble'' and
``traveling wave'' patterns in terms of a hierarchy of invariant
\po\ solutions.
    }
\label{f:antorbspt}
\end{figure}

For small $\tildeL$, the dynamics and bifurcation sequences
are investigated in
\refrefs{HNks86,KNSks90,AGHks89,HLBcoh98,kev01ks}.
For $\tildeL<1$
\eqv\ $u=0$ is the global attractor.
As the system size $\tildeL$  is increased,
the ``flame front'' becomes increasingly unstable and turbulent.
While
for $\tildeL$ sufficiently large
existence of many
coexisting attractors is an open possibility%
\rf{HNZks86},
in numerical studies most initial
conditions settle down in the same region of \statesp,
the attractor with the largest basin of attraction.
This is illustrated by \reffig{f:antorbspt}
for system size $L=38.5$, 
$\tildeL = 6.12\cdots$
that we shall focus on in this paper.

In the antisymmetric subspace the translational
invariance of the full system reduces
to the invariance under discrete translation by $x \to x+L/2$.
In the Fourier representation \refeq{expan},
the corresponding solution is obtained by reflection
\beq
a_{2m} \to a_{2m}\,, a_{2m+1} \to -a_{2m+1}
\,.
\ee{FModInvSymm}

\section{Method of {\descent}}
\label{sect:ksrecur}
We will
investigate the properties of the {\KSe}
in a weakly ``turbulent'' (or ``chaotic'')
regime from the perspective of
periodic orbit theory\rf{DasBuch}, and refer to
the application of this theory to PDEs
as the {\em recurrent pattern program} since here
the coordination of spatial degrees of freedom
plays a major role.

Christiansen {\em et al.}\rf{ks} proposed in 1996   
that the periodic orbit theory be applied to spatio-temporally chaotic
systems, using the Kuramoto-Sivashinsky system
as a laboratory for exploring viability of the program.
They
examined the dynamics of the periodic b.c., antisymmetric subspace,
for system sizes
$\tildeL \approx 5.8$,
close to the onset of chaos, where a 
truncation of expansion \refeq{expan} to 16
Fourier modes already yields accurate results.
The main result was that the high-dimensional (16-64 dimensions)
dynamics of this dissipative flow
could be reduced to an approximately 1\dmn\ Poincar\'e return map,
by constructing an invariant, unstable manifold-based curvilinear coordinate
passing close to all unstable periodic orbits embedded within the
strange attractor.
A binary symbolic dynamics arising from this surprisingly simple
return map made possible a
systematic determination of {\em all}
nearby unstable \po s up to a given
number of Poincar\'e section returns.

The essential limitation on the numerical studies
undertaken in  \refref{ks} were
computational constraints: in truncation of high modes in the
expansion (\ref{expan}), sufficiently many have to be retained to ensure
that
the dynamics is accurately represented; on the other hand, recurrent patterns
have to be located in this high-dimensional \statesp. High wave number modes
have large
negative coefficients in the linear term of \refeq{expan},
making  the
system stiff and the integration slow.
Basic difficulties also exist in the application of commonly used
cycle-searching techniques\rf{lanthe,DasBuch}, due to
the intricate orbit structure induced by strong nonlinearity. The
integration of
the associated Jacobian matrix can also be expensive
due to the high dimensionality.

\begin{figure}[h]
\centering

\hspace{-0.22\textwidth}
\includegraphics[width=0.22\textwidth]{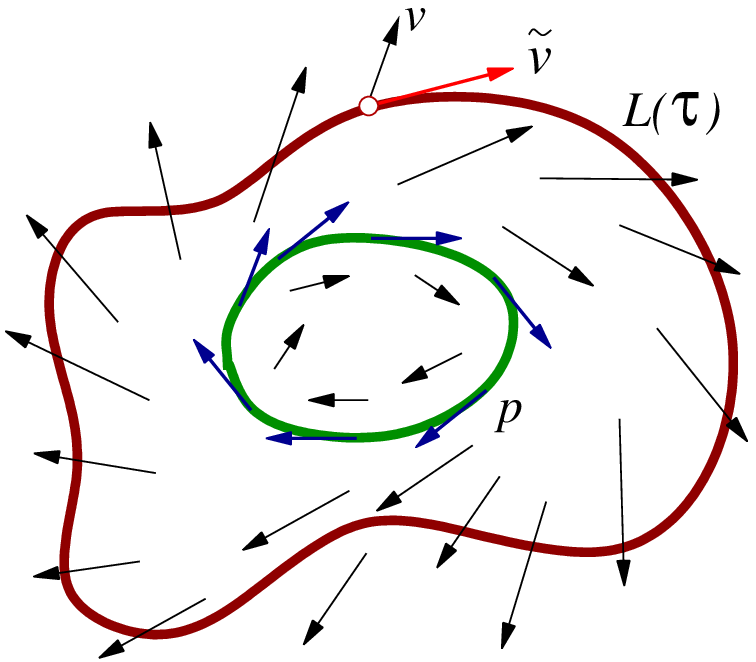}%
\hspace{-0.24\textwidth} (a) \hspace{0.22\textwidth}
\includegraphics[width=0.21\textwidth]{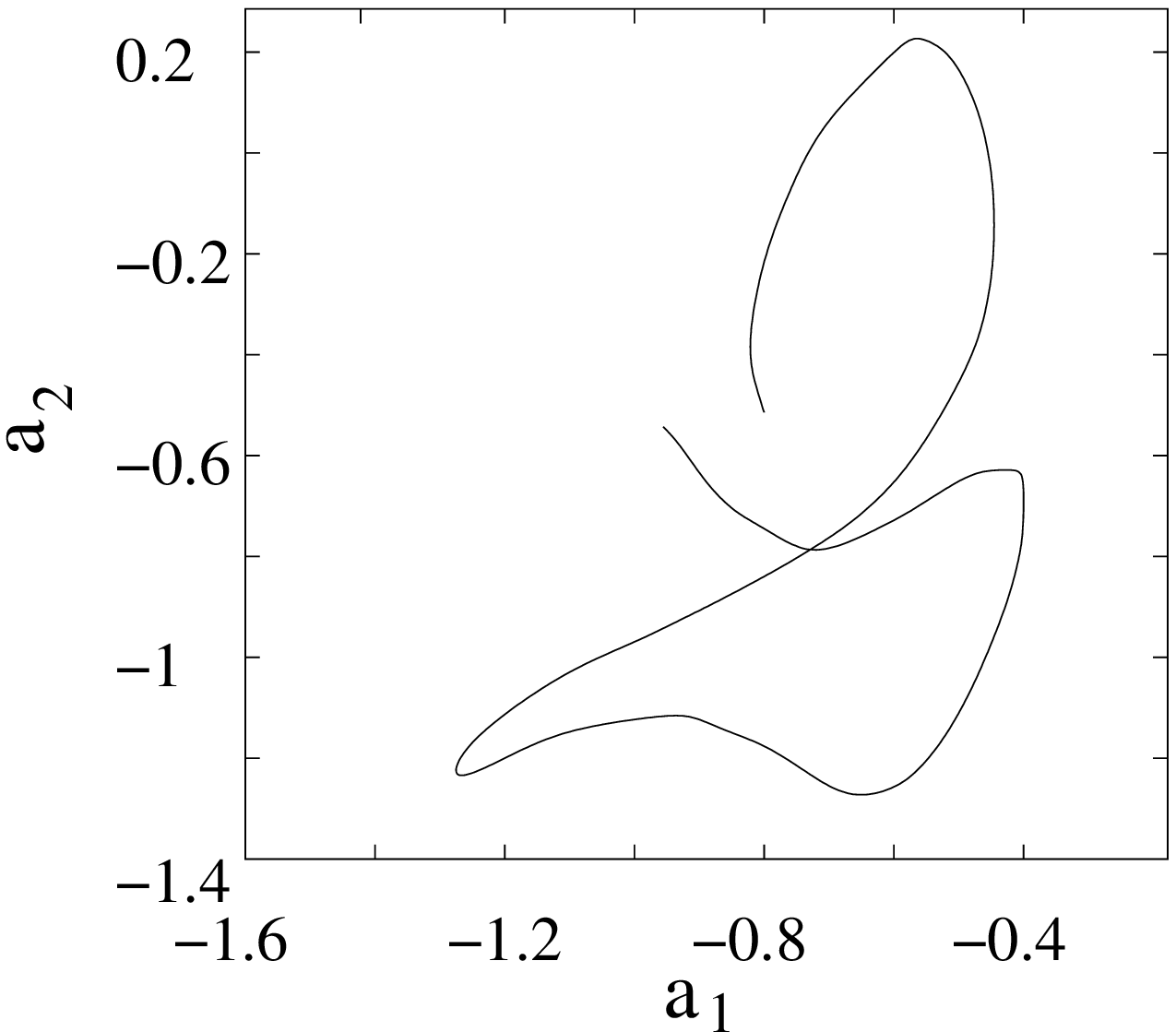}%
\hspace{-0.22\textwidth}(b)
\\
\hspace{-0.22\textwidth}
\includegraphics[width=0.22\textwidth]{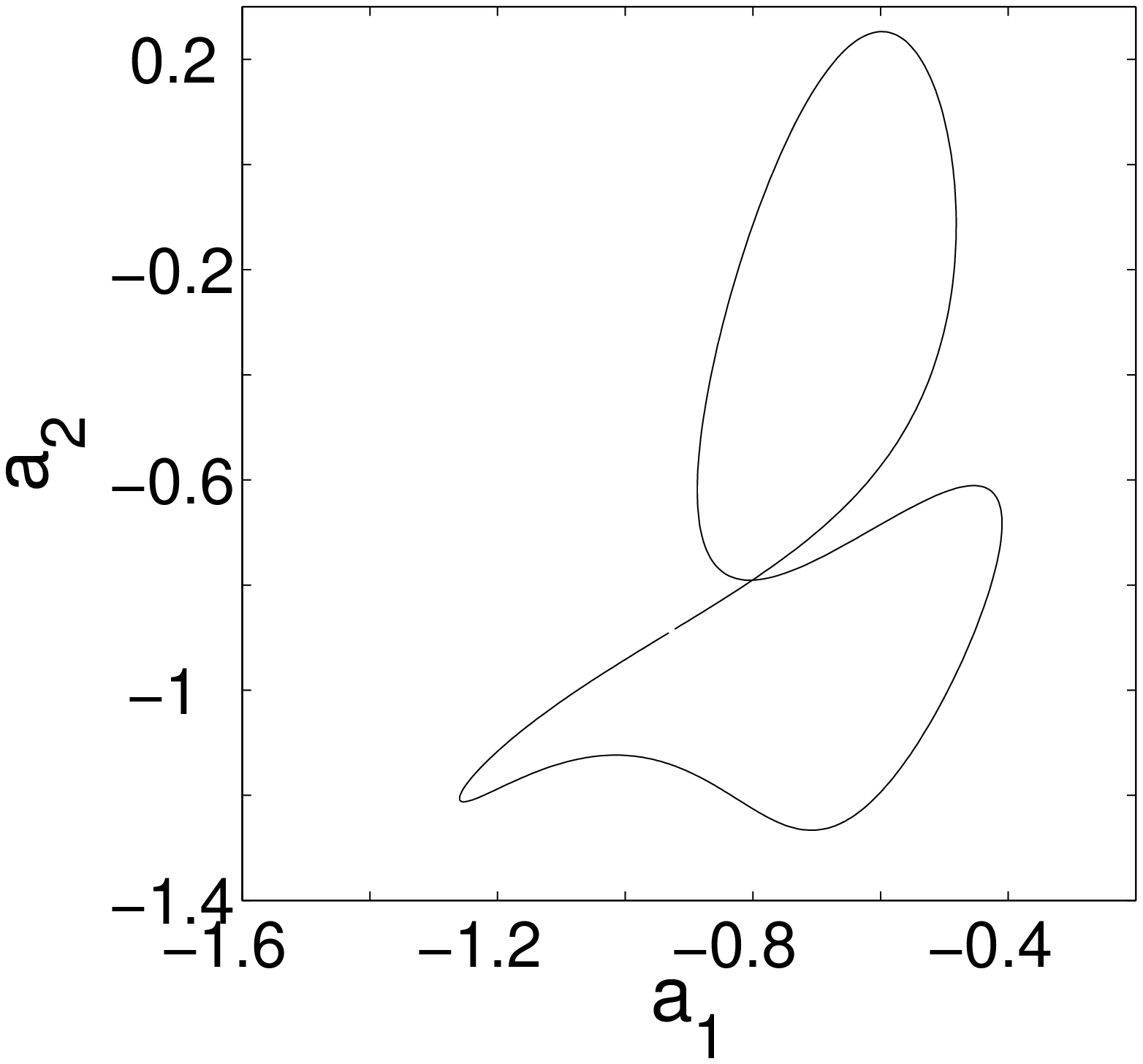}%
\hspace{-0.22\textwidth}(c) \hspace{0.22\textwidth}
\includegraphics[width=0.22\textwidth]{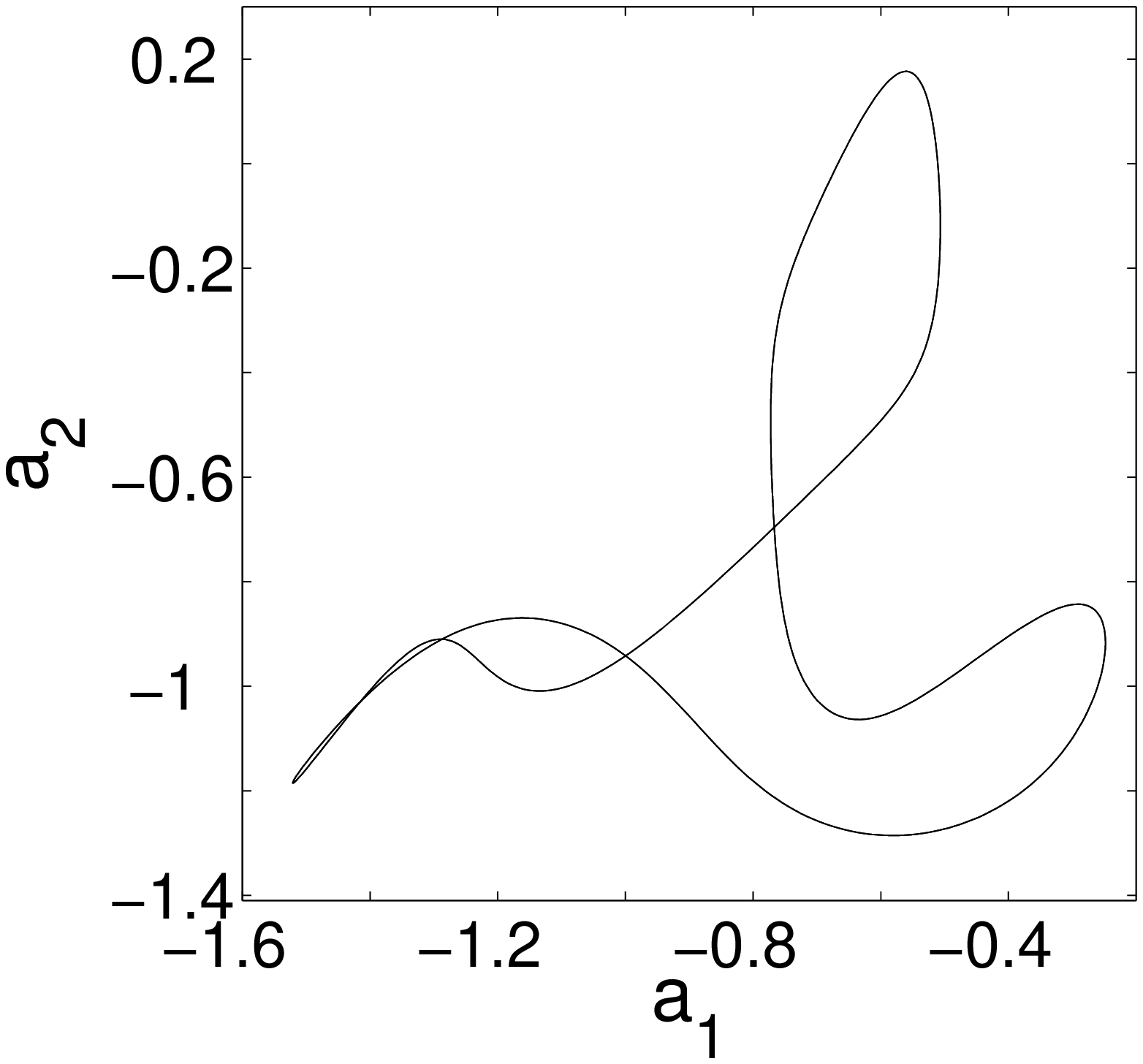}
\hspace{-0.22\textwidth}(d)
%
\caption{
(a)
 The orientation of a tangent $\lVeloc(\lSpace)$
of the guess loop $\Loop(\tau)$ does not coincide with
the orientation of the velocity field $\pVeloc(\lSpace)$;
for a periodic orbit $p$ it  does so at every $x \in p$.
The \descent\ method aligns the closed loop tangent to the given
vector field by driving the loop to a periodic
orbit.
\descent\ at work for a Kuramoto-Sivashinsky system:
 (b) a near return extracted from a long-time orbit,
 (c) initial guess loop crafted from it,
 (d) the periodic orbit $p$ reached by the \descent. 
$N=512$ points representation of the loop,
$[a_1,a_2]$ Fourier modes projection.
        }
\label{f:loops}
\end{figure}

The {``\descent''} method for determining unstable
spatio-temporally periodic solutions of extended systems
has been formulated and explored numerically
in \refrefs{CvitLanCrete02,lanVar1}. 
The idea of the method
is to make a rough but informed guess of what the desired pattern
looks like globally, and then use a variational method to drive the
initial guess toward the exact solution,
by minimizing a {\costFct} computed from the
deviation of the approximate flow from the true flow,
\reffig{f:loops}\,(a).

We initiate our searches by a long-time numerical run
of the dynamics, in order to identify the frequently
visited regions of the
\statesp\ (natural measure),
then search for close recurrences\rf{pchaot}.  
An initial loop guess $\Loop(0)$ is crafted by taking
a nearly recurring segment of the orbit,
smoothed and made periodic by
a FFT into the wave number representation,
dropping the high frequency components, and
an FFT back to the \statesp.
In a loop discretization
 each point has to be specified in all $d$~dimensions.
A typical initial loop guess is displayed in
\reffig{f:loops}\,(b,c),
along with the periodic orbit
found by
the \descent\ method in \reffig{f:loops}\,(d).

\section{\Eqva\ of the KS equation}
\label{sect:kseqlb}

\Eqva\ (or the steady solutions)
are the simplest invariant objects in the \statesp.
Some of them are dynamically important as they,
together with their unstable/stable manifolds,
partition the
\statesp\ into qualitatively different regions
and offer a first, coarse description of
typical \statesp\ \recurrStr s.
As we shall show here, each such region owns its own
local Smale horseshoe hierarchy of \UPO s,
and there are orbits communicating between different regions.


The {\eqva}  of the {\KSe} \refeq{kseq} satisfy
\[
(u^2)_x-u_{xx}-u_{xxxx}=0 \,.
\]
Integrating once we get
\beq
u^2-u_x-u_{xxx}=E
\,,
\label{eq:stdks}
\eeq
where $E$ is an integration constant\rf{ksgreene88}.
Written as a 3\dmn\ ODE,
with spatial coordinate $x$ playing the role of ``time,''
this is a dynamical system\rf{Mks86}
\beq
u_x = v \,,\qquad
v_x = w \,,\qquad
w_x = u^2-v-E \,,
  \label{eq:3dks}
\eeq
with the ``time reversal'' symmetry,
\[
x \to -x,\quad u \to -u, \quad v \to v, \quad w \to -w \,.
\]
Rewriting \refeq{eq:3dks} as
\[
(u+w)_x=u^2-E \,,
\]
we see that
for $E<0$, $u+w$ increases without bound as $x \to \infty$,
and every solution escapes to infinity.
If $E=0$, the origin $(0,0,0)$ is the
only bounded  solution, a marginally stable center with
eigenvalues $(0, i,-i)$.

For $E>0$ there is rich
$E$-dependent dynamics, with
fractal sets of bounded solutions.
The solutions of
\refeq{eq:3dks} are themselves in turn organized by its own
{\eqva} and
the connections between them\rf{Mks86}.
    For $E>0$ the {\eqv}  points of \refeq{eq:3dks} are
$c_{+}=(\sqrt{E},0,0)$ and $c_{-}=(-\sqrt{E},0,0)$.
Linearization of the flow around
$c_{+}$ shows that $c_{+}$ has a {1\dmn}
unstable manifold and a 2\dmn\ stable manifold
along which solutions spiral in.
By the $x \to -x$ ``time reversal'' symmetry, the
invariant manifolds of $c_{-}$
have reversed stability properties.
Most orbits escape quickly even if initiated close to the \SIS, and that
renders the numerical calculations
difficult\rf{ksham95,kshooper88,pimyk,pimsimp}.
The \descent\ method\rf{lanVar1,CvitLanCrete02}
that we employ appears more robust and effective than
the earlier approaches.

For a fixed spatial size
$L$ with periodic boundary condition, the only {\eqva}  are
those with spatial periodicity $L$.
The \eqva, represented as
$\sum_{k=-\infty}^{\infty}a_k e^{ikqx}$, with $q=2\pi/L$
and $a_k^*=-a_{-k}$, satisfy\rf{ksgreene88}
\beq
(kq)^2\left(1-(kq)^2\right) a_k
    +ikq\sum_{m=-\infty}^{\infty}a_{k-m}a_m=0
\,.
\label{eq:stfks}
\eeq
In analogy with Hamiltonian dynamics,
we can say that we are looking for solutions of \refeq{kseq}
of a given spatial period $L$, on
any ``$E$ shell,'' rather than looking for solutions of arbitrary
period for a fixed ``$E$ shell.''

In the antisymmetric subspace considered here,
the invariance \refeq{FModInvSymm}
under discrete translation by $x \to x+L/2$
implies that every \eqv\ solution
is either invariant under 1/2-cell shift, or has
a half-cell translated partner.

\subsection{Search for dynamically important {\eqva} }
\label{sect:sdyneq}

For small system sizes $L$ the number of {\eqva} is small and
concentrated on the low wavenumber end of the Fourier spectrum.
In a high-dimensional \statesp\ not all {\eqva}
influence dynamics significantly, so we need
to classify them according to their importance
 in shaping the long-time dynamics of \refeq{kseq}.
We gauge the relative importance of an \eqv\ by its
proximity to the most recurrent \statesp\ regions.
Empirically, an {\eqv} plays at least two roles.
The more unstable eigendirections it has (for example, the
$u=0$ solution), the more unlikely it is  that
an orbit will recur in its neighborhood:
thus a highly unstable \eqv\ can
help elucidate the topology of an asymptotic attracting set
by the ``hole'' that it cuts in the natural measure.
On the other hand, the asymptotic dynamics
can spend  a large fraction of time in
neighborhoods of a few ``least unstable''
{\eqva}, {\eqva} with only a few unstable eigendirections.
Unstable manifolds of a set of such {\eqva}  tile
\statesp\ with a set of regions
explored by the asymptotic dynamics.

\begin{figure}[tbp] 
    \centering
\hspace{-0.22\textwidth}
\includegraphics[width=0.21\textwidth]{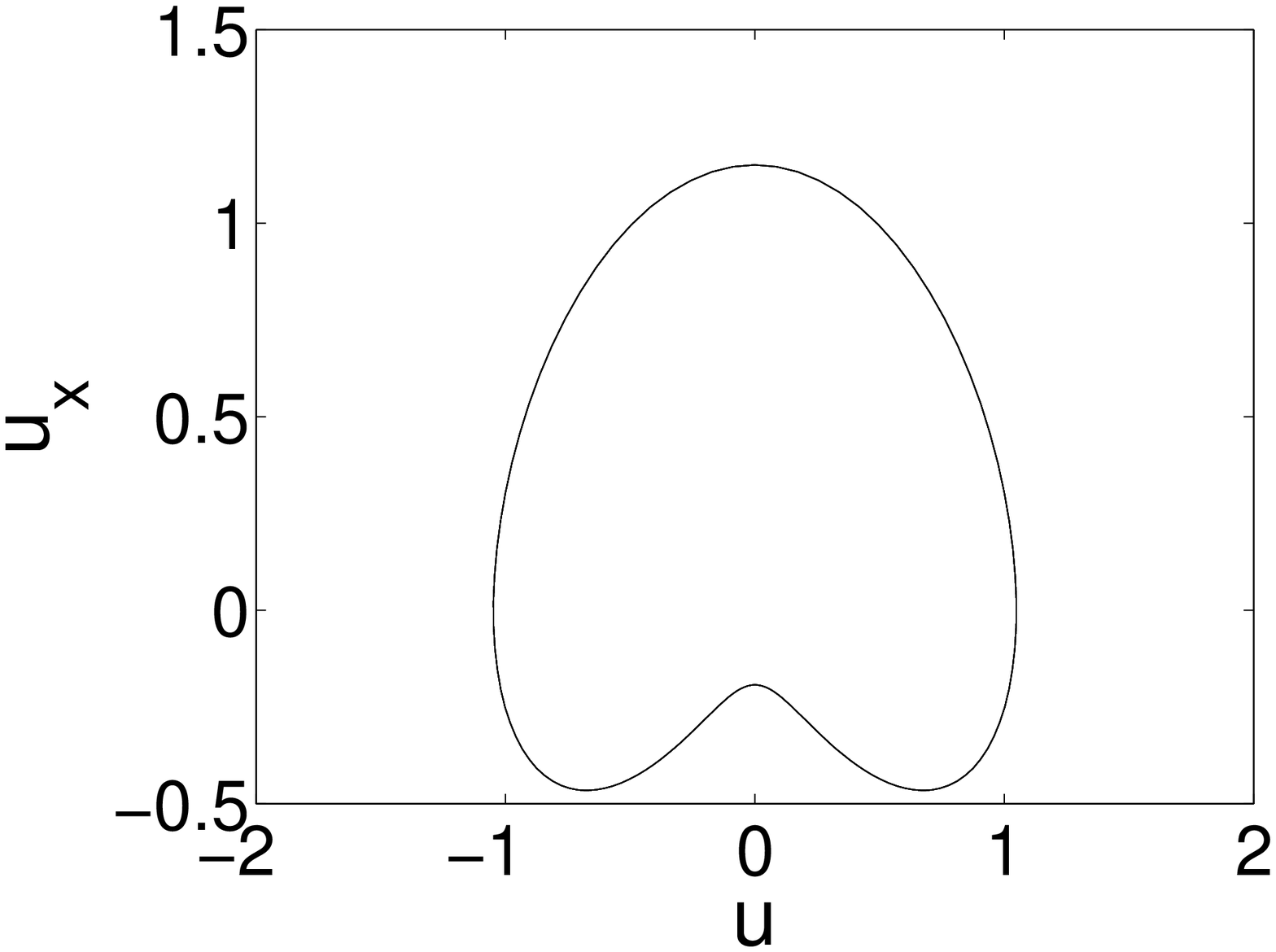}
\hspace{-0.22\textwidth} (a) \hspace{0.22\textwidth}
\includegraphics[width=0.21\textwidth]{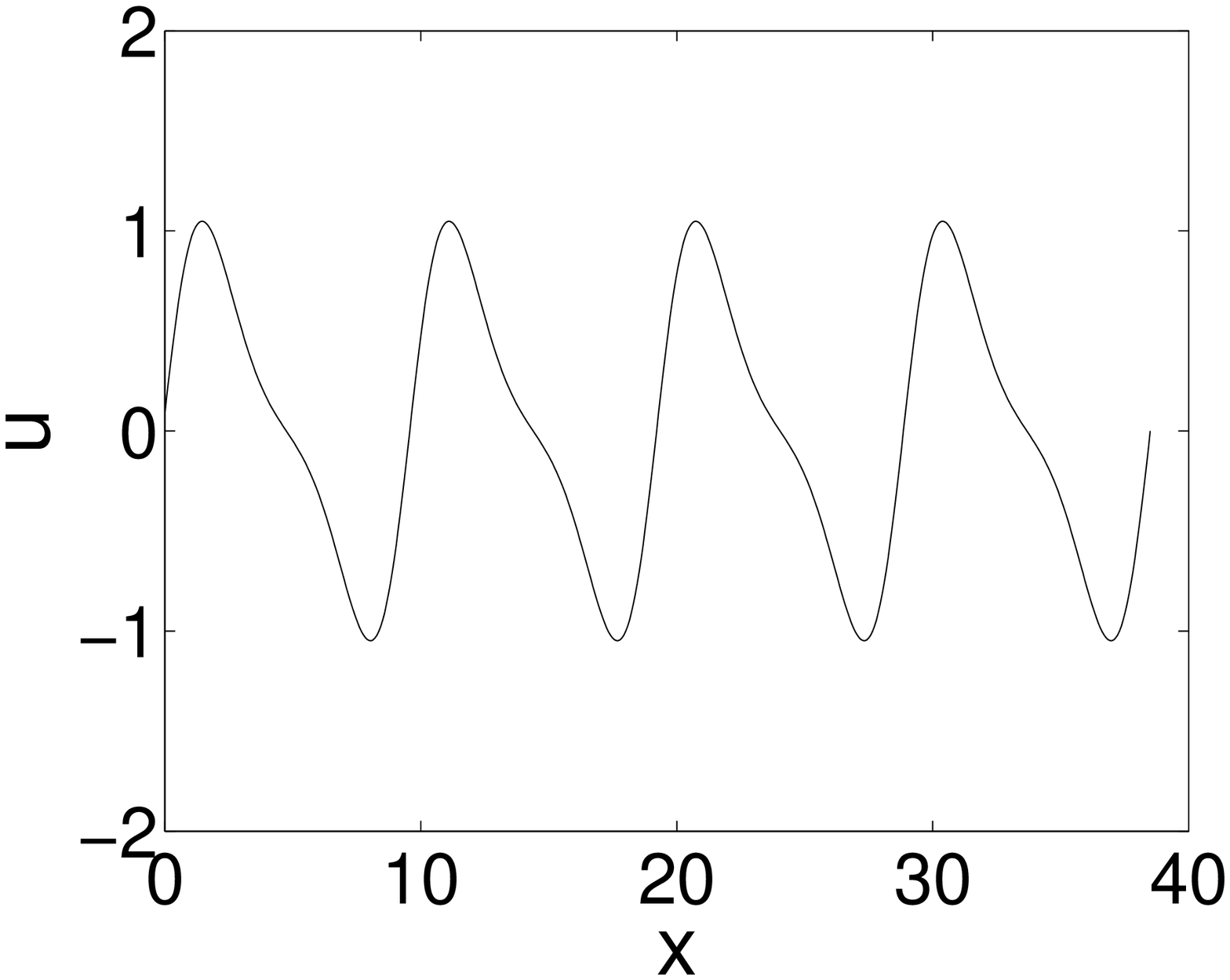}
\hspace{-0.22\textwidth} (b)
    \\
\hspace{-0.22\textwidth}
\includegraphics[width=0.21\textwidth]{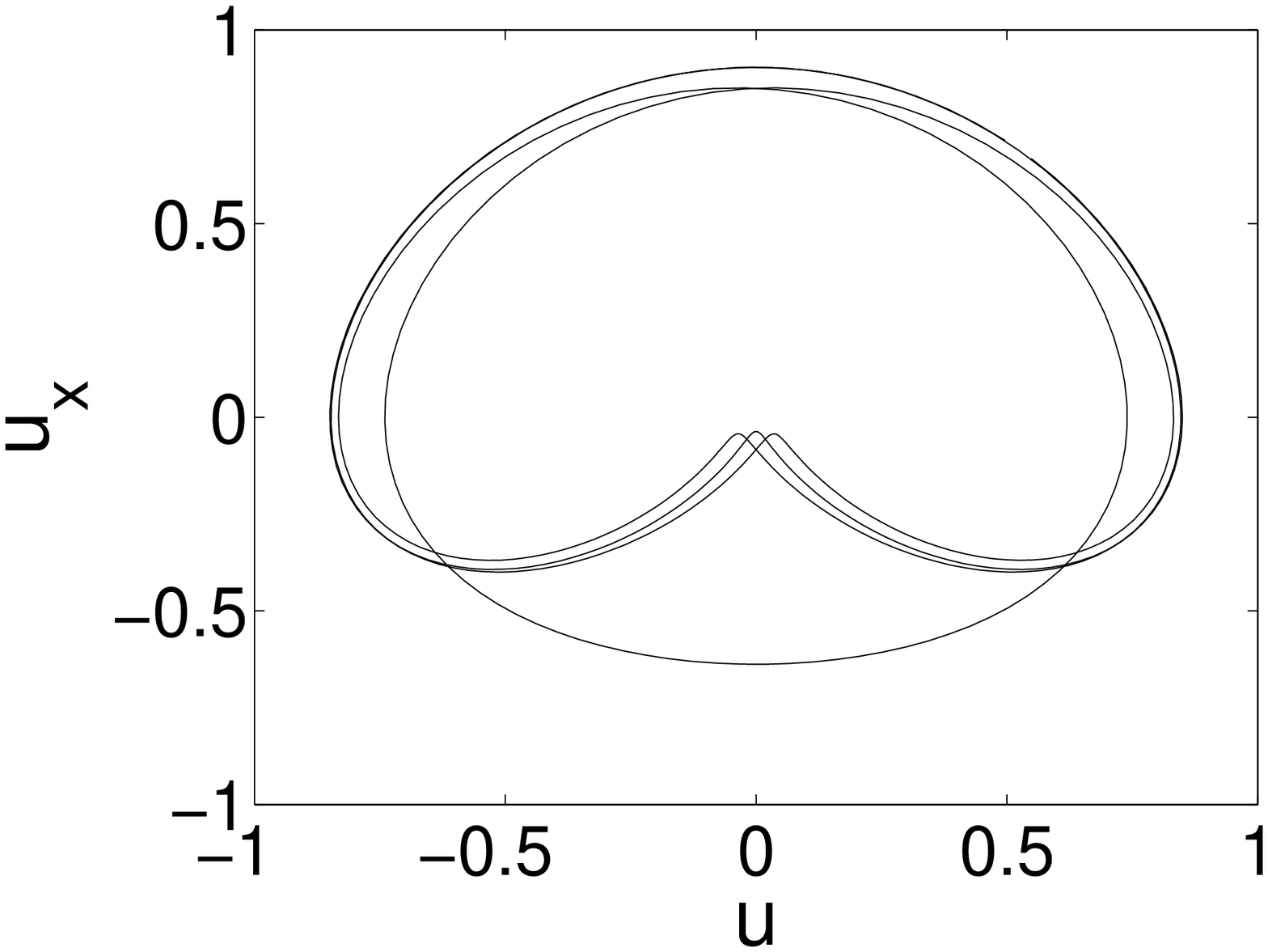}
\hspace{-0.22\textwidth} (c) \hspace{0.22\textwidth}
\includegraphics[width=0.21\textwidth]{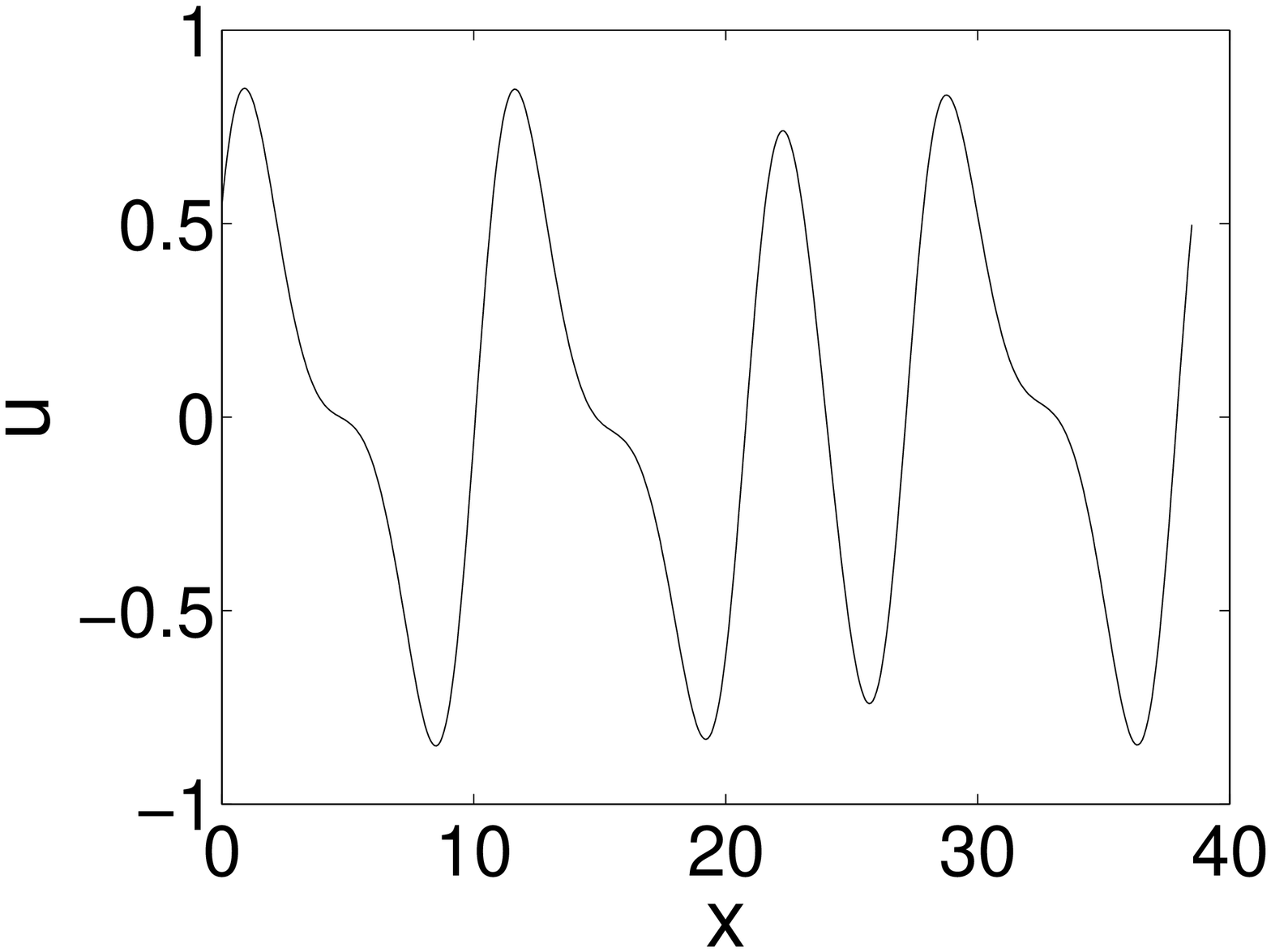}
\hspace{-0.22\textwidth} (d)
    \\
\hspace{-0.22\textwidth}
\includegraphics[width=0.21\textwidth]{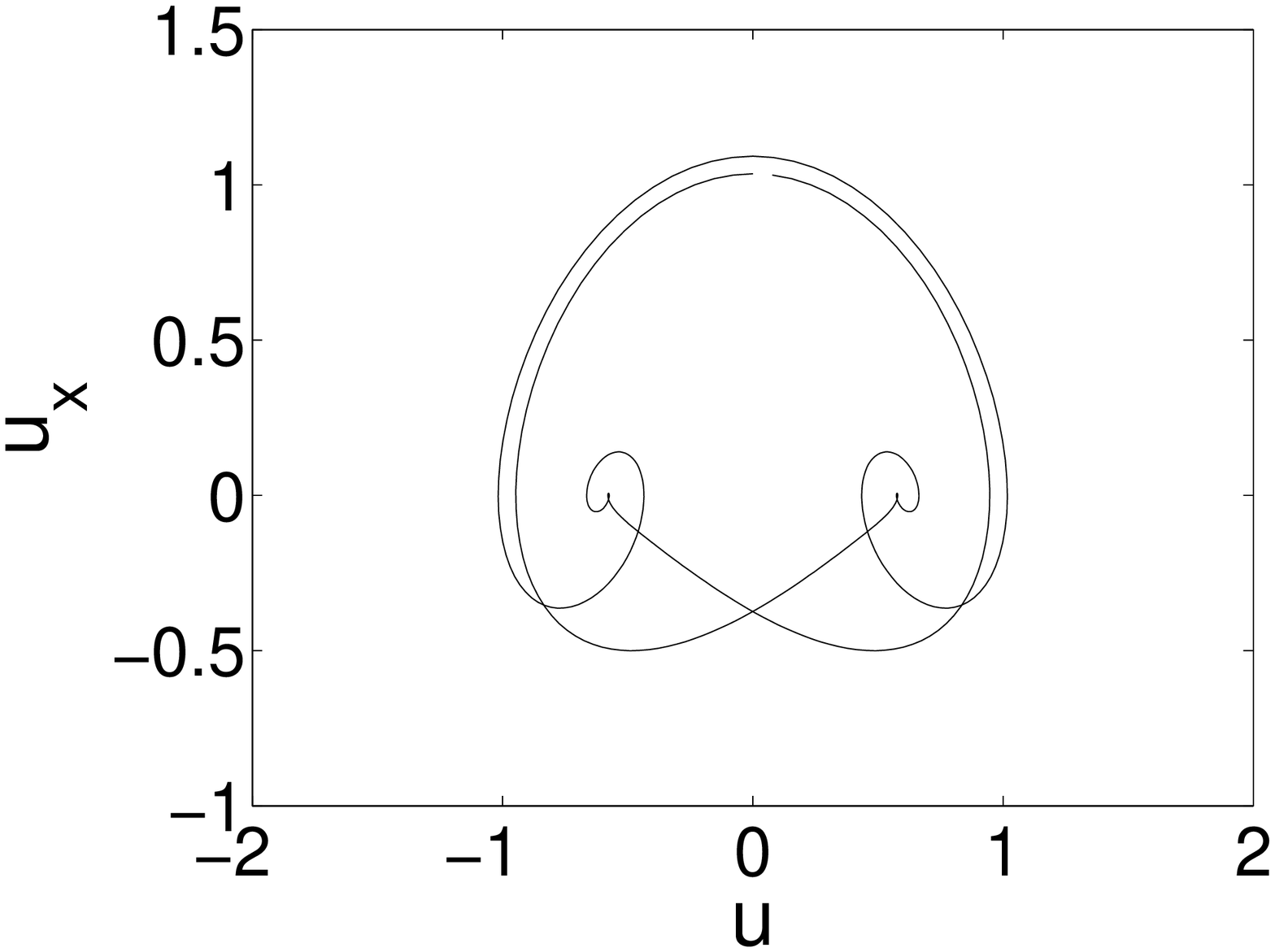}
\hspace{-0.22\textwidth} (e) \hspace{0.22\textwidth}
\includegraphics[width=0.21\textwidth]{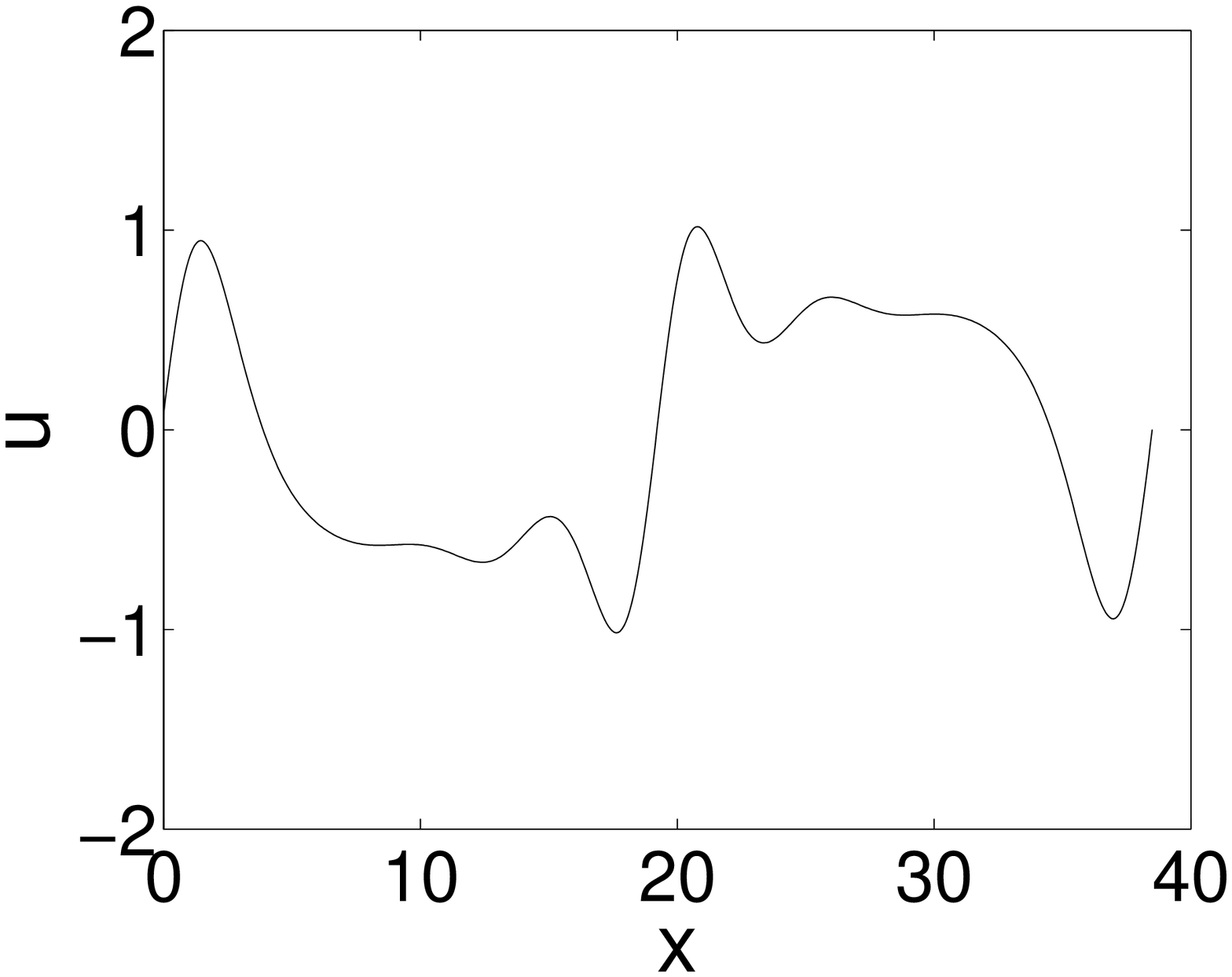}
\hspace{-0.22\textwidth} (f)
    \\
\hspace{-0.22\textwidth}
\includegraphics[width=0.21\textwidth]{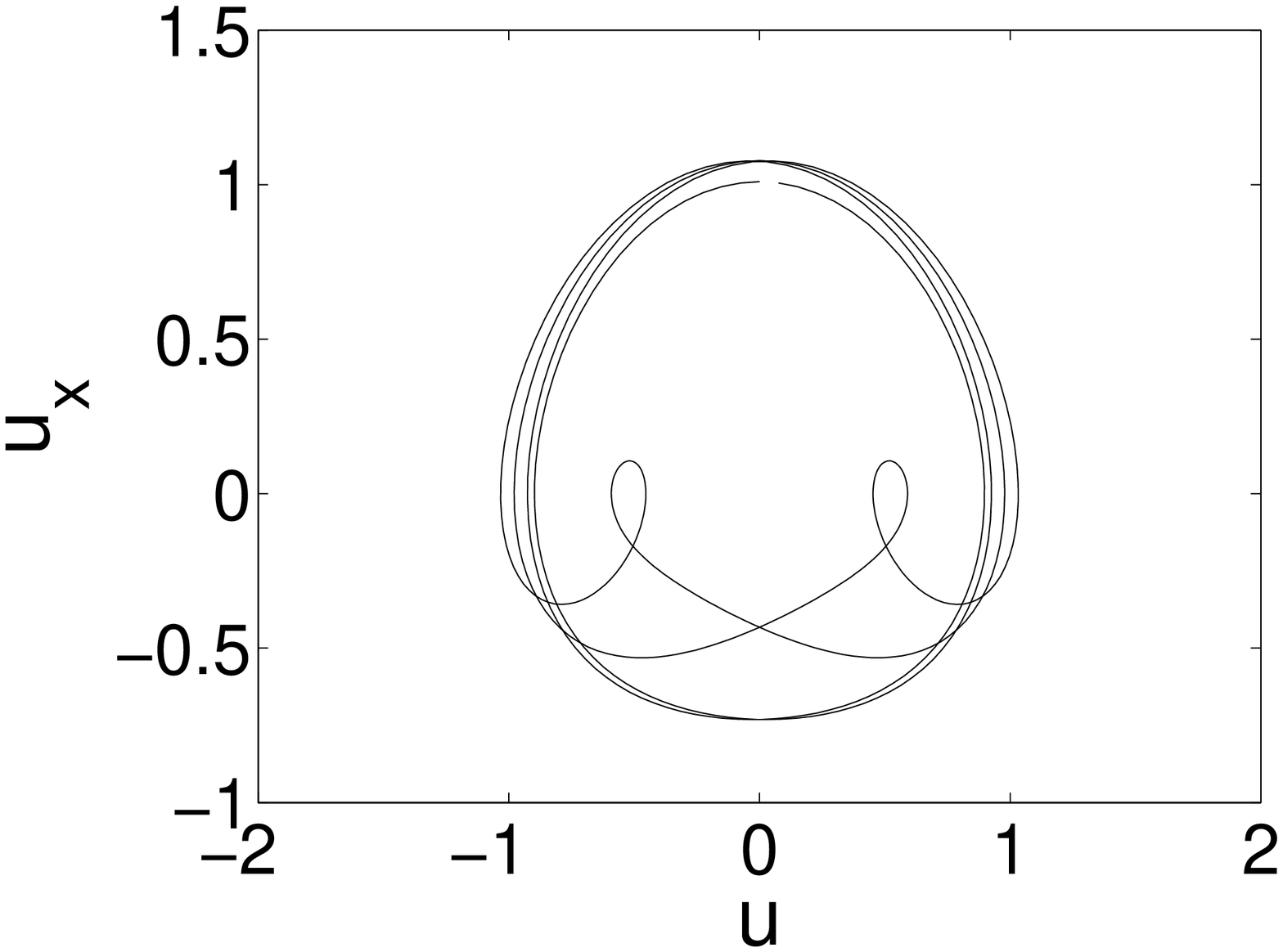}
\hspace{-0.22\textwidth} (g) \hspace{0.22\textwidth}
\includegraphics[width=0.21\textwidth]{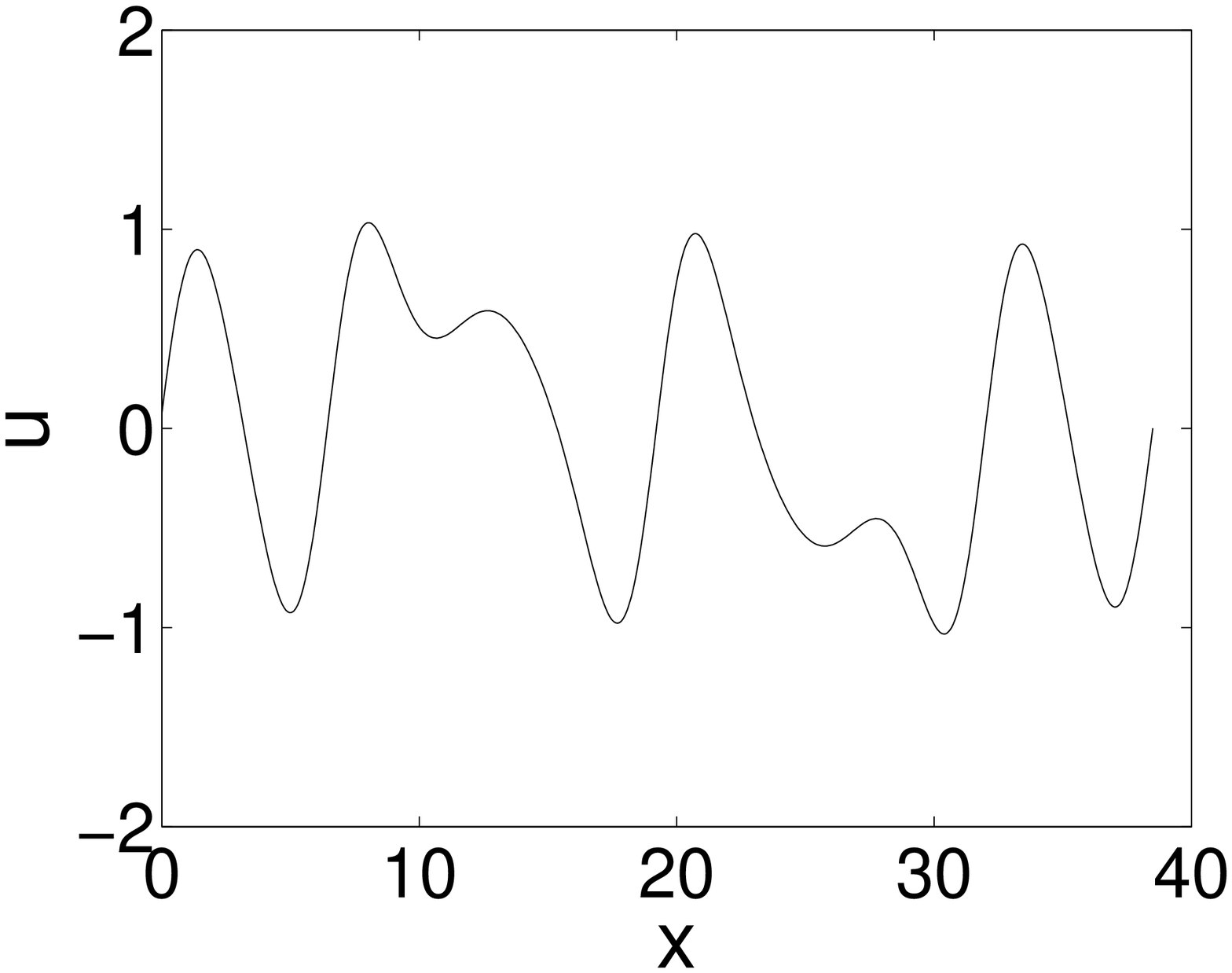}
\hspace{-0.22\textwidth} (h)
    \\
\hspace{-0.22\textwidth}
\includegraphics[width=0.21\textwidth]{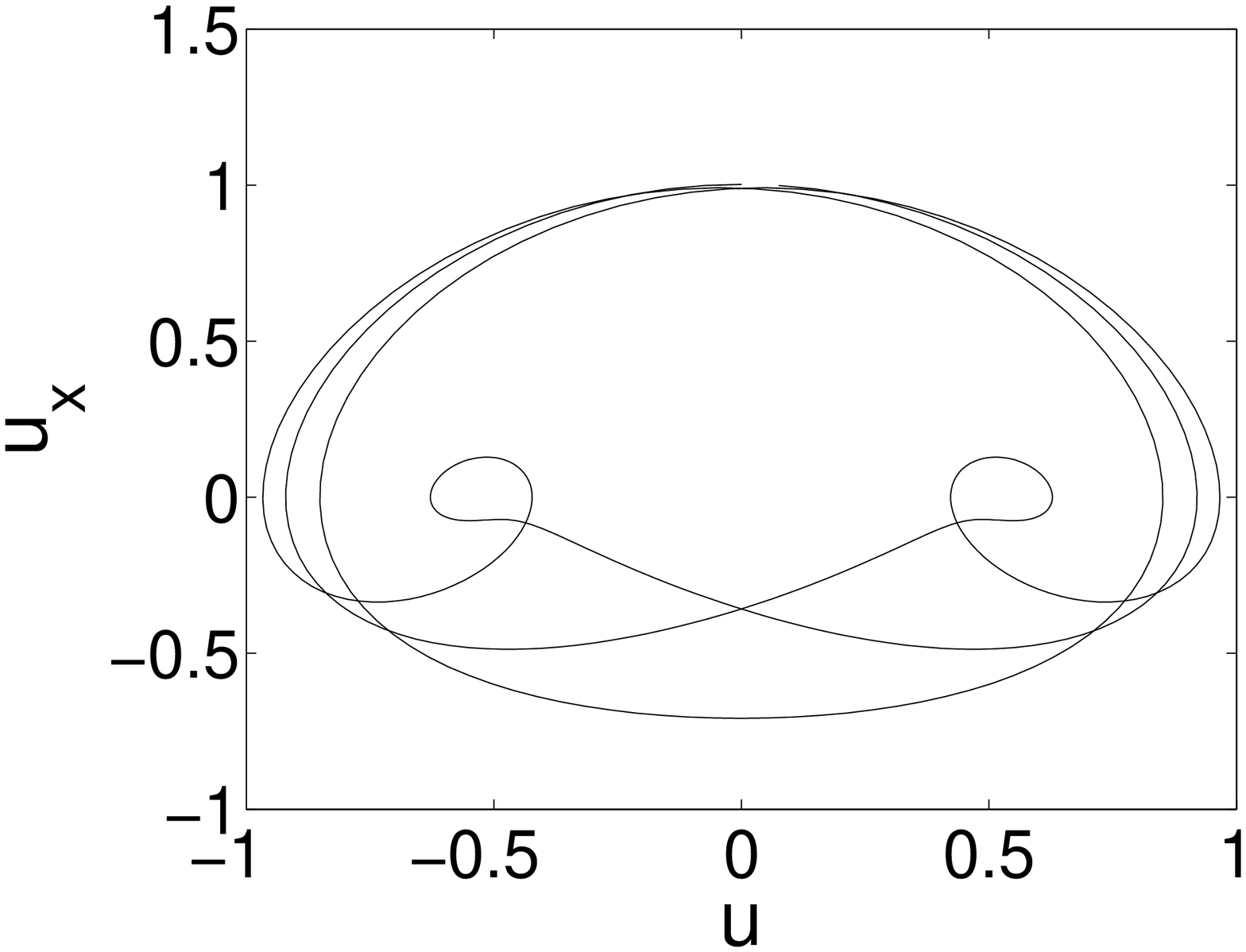}
\hspace{-0.22\textwidth} (i) \hspace{0.22\textwidth}
\includegraphics[width=0.21\textwidth]{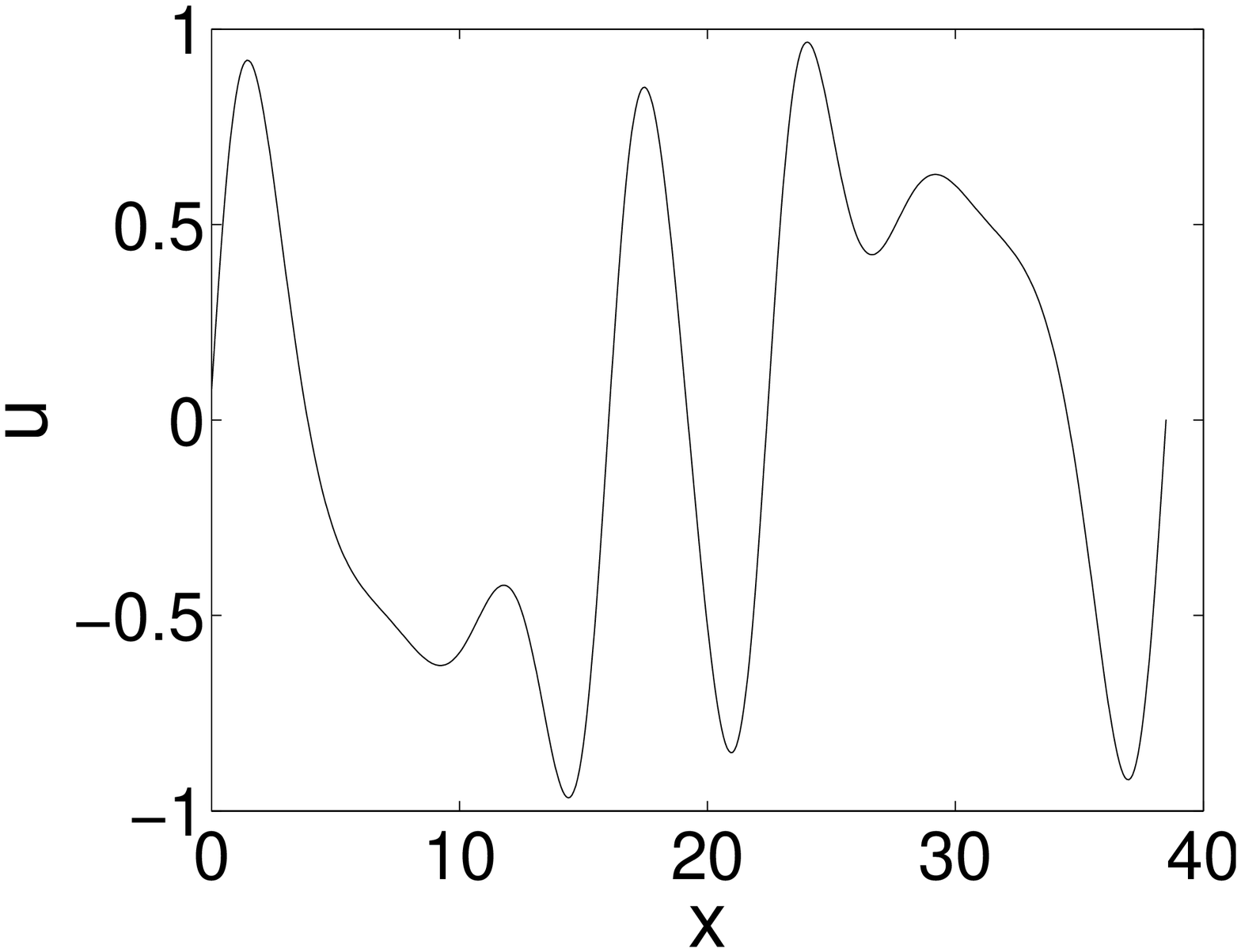}
\hspace{-0.22\textwidth} (j)
\caption[]{
{\Eqva} in Michelson
$[u,u_x]$ representation\rf{Mks86}, and as $u(x)$ spatially
periodic profiles:
(a), (b)
    $C_1$.
(c), (d)
    $C_2$.
(e), (f)
    $R_1$.
(g), (h)
    $R_2$.
(i), (j)
    $T$.
$L=38.5$, antisymmetric subspace.
      }
\label{f:antfix2}
\end{figure}

We pick any point on a typical orbit of \refeq{expan}.
It corresponds to a  loop in
the 3\dmn\ \statesp\  of \refeq{eq:3dks} and so can be used to initialize
the search for an $u(x)$ profile periodic on $[0,L]$.
For $L=38.5\,, \nu=1$
we found in this way several dozen \eqva\ of \refeq{expan}.

%
\begin{table}
\begin{center}
{\small
\lineup
\item[]
\begin{tabular}{@{}ccclll}
\br
$~~~~~$ & & $~~~~~E~~~~~$
                & $\eigRe[1] \pm \,i\,\eigIm[1]$
                & $\eigRe[2] \pm \,i\,\eigIm[2]$
                & $\eigRe[3] \pm \,i\,\eigIm[3]$
\\
\mr
${C_1}$ &     & 0.43646 &0.044     &-0.255 &-0.347  \\[-0.8ex]
        &   & &  $\pm \,i\,$0.261   & $\pm \,i\,$0.431 & $\pm \,i\,$0.463 \\
${C_2}$ &   &  0.25784  &0.33053  & 0.097                  &-0.101 \\[-0.8ex]
        &   &           &         &       $\pm \,i\,$0.243 &  $\pm \,i\,$0.233 \\
\mr
${R_1}$ &  & 0.36602  &  0.011 & -0.215 &-0.358  \\[-0.8ex]
        &  &  & $\pm \,i\,$0.796   & $\pm \,i\,$0.549 & $\pm \,i\,$0.262 \\
${R_2}$ &  & 0.34442 &  0.33223  & -0.001                   & -0.281 \\[-0.8ex]
        &  &         &           &        $\pm \,i\,$0.703  & $\pm \,i\,$0.399  \\
\mr
${T}$  &  & 0.40194   & 0.25480  & -0.07                  &-0.264  \\[-0.8ex]
       &  &           &          &       $\pm \,i\,$0.645 &        \\
\br
\end{tabular}
} 
\end{center}
\caption[]{
    Dynamically important \eqva\ in the antisymmetric
    subspace, periodic b.c.:
    value of the integration constant $E$,
    as defined in \refeq{eq:stdks},
    and the first few least unstable stability
    exponents.
}
\label{t:stationary}
\end{table}

Next we re-initialize the
search by taking the
average of an orbit segment of \refeq{expan},
 with the hope that the typical orbit will
pass through the neighborhood of
important {\eqva}  often. In this way, the number of
dynamically important {\eqva}  is greatly reduced: we find
$10$ solutions, $5$ of which belong to the antisymmetric
subspace, see \reffig{f:antfix2}.
The corresponding $E$ values and leading linear stability eigenvalues
are listed in \reftab{t:stationary}.
No other \eqva\ seem to be dynamically
important for this system size.

In  \reftab{t:stationary}
``$C$'' refers to the center in \reffig{f:antorbspt}\,(b),
``$R$'' to the right. The left {\eqva}  ``$L$'' are omitted, as they are
symmetry partners of ``$R$.''
The ``${C_1}$'' \eqv\ is self-dual with respect to the reflection
\refeq{FModInvSymm}, and the rest come in pairs.

The topology of {\eqva}  is organized
relative to the stationary points $c_+,c_-$ near
the unstable spiral-in manifold of  $c_+$, the
stable spiral-out manifold of  $c_-$.
In \reffig{f:antfix2}\,(a,c), the \eqv\ profiles circle the
pair $c_+,c_-$ as a whole,
and $u$ has $4$ peaks on $[0,L]$, see \reffig{f:antfix2} (b) and (d).
In \reffig{f:antfix2}\,(e,g,i) the
profiles encircle both  \eqva\ as well as each separately.
The circulation around the two {\eqva}   (``of {\eqva}'')
may be used to classify the solutions.
Each circulation gives a peak in the $u$ profile on
$[0,L]$, with big circulations corresponding to large
oscillations and small ones to the secondary oscillations.

\subsection{\Eqva\ and the dynamics}
\label{sect:equidyn}

Empirically the \SIS\ appears to consist of three regions:
the left part ($S_L$), the center
part ($S_C$) and the right part ($S_R$).
By the inversion symmetry \refeq{FModInvSymm},
$S_L$ and $S_R$ are mirror images of each other,
so only one of them needs to be considered, say $S_R$.
Within each region the dynamics takes place on
a chaotic repeller, with orbit occasionally escaping
a region and landing in the next region. Such
rapid transitions show up as ``defects'' in the spatio-temporal
evolution in \reffig{f:antorbspt}\,(a).

Judged by recurrences in the long-time dynamics,
the weakly unstable ${C_1}$ and ${R_1}$ appear to
be the most important {\eqva},
with typical return times $T_i = 2 \pi / \nu_i$:
$T_{C_1}= 24.0$,
$T_{R_1}= 7.89$.
In contrast, the dynamics appears to steer clear of
${C_2}$, \eqv\ unstable in three eigen-directions.
The $T$ \eqv\ appears to
mediate transitions between the side and the center regions.
After checking all projections of
a typical long orbit onto all $[a_i\,,a_j]$
planes, we found that
the dynamics frequents only the neighborhoods of the {\eqva}  listed in
\reftab{t:stationary}.

\begin{figure}[tbp] 
    \centering
\hspace{-0.22\textwidth}
\includegraphics[width=0.21\textwidth]{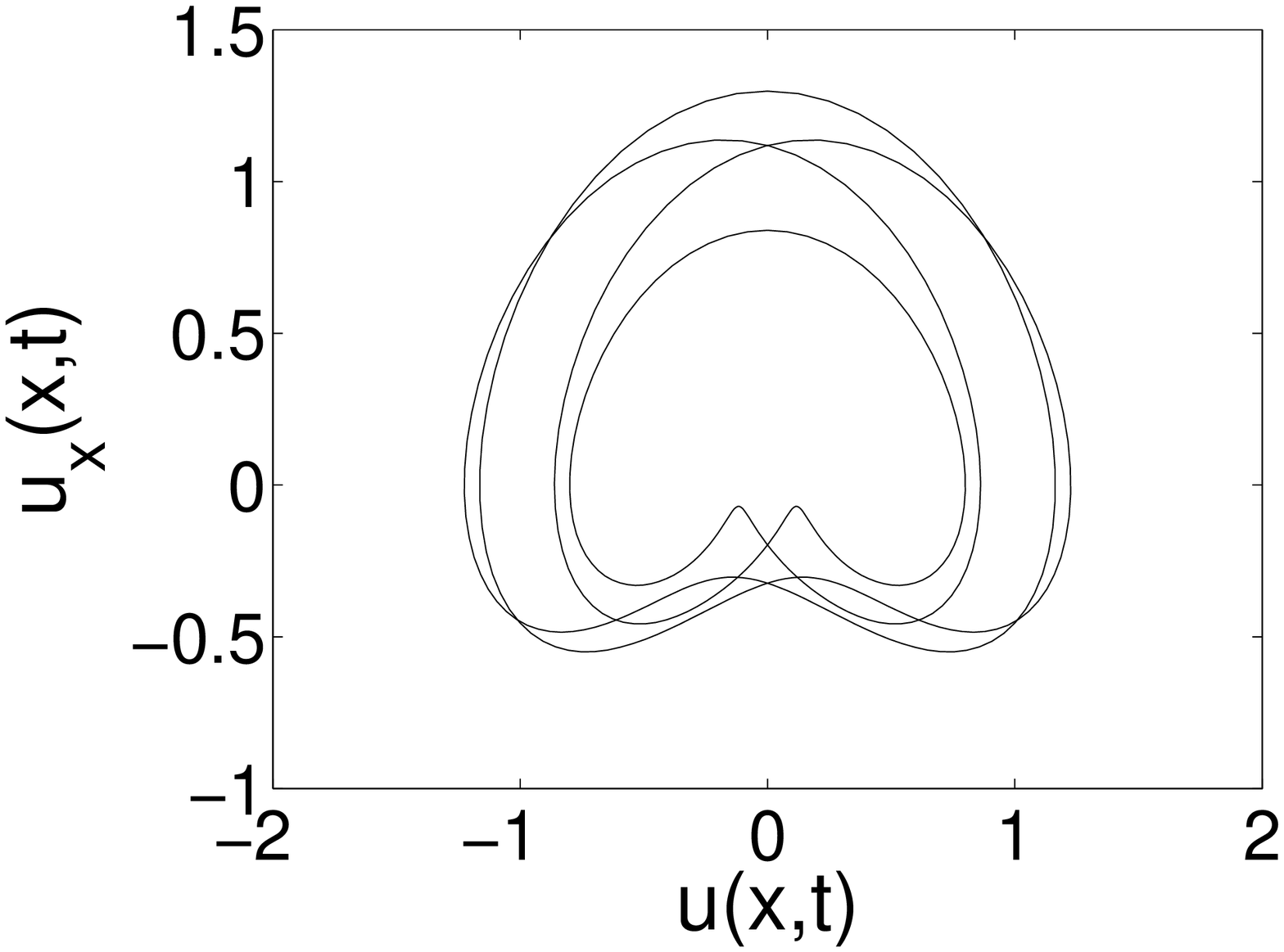}
\hspace{-0.22\textwidth} (a) \hspace{0.22\textwidth}
\includegraphics[width=0.21\textwidth]{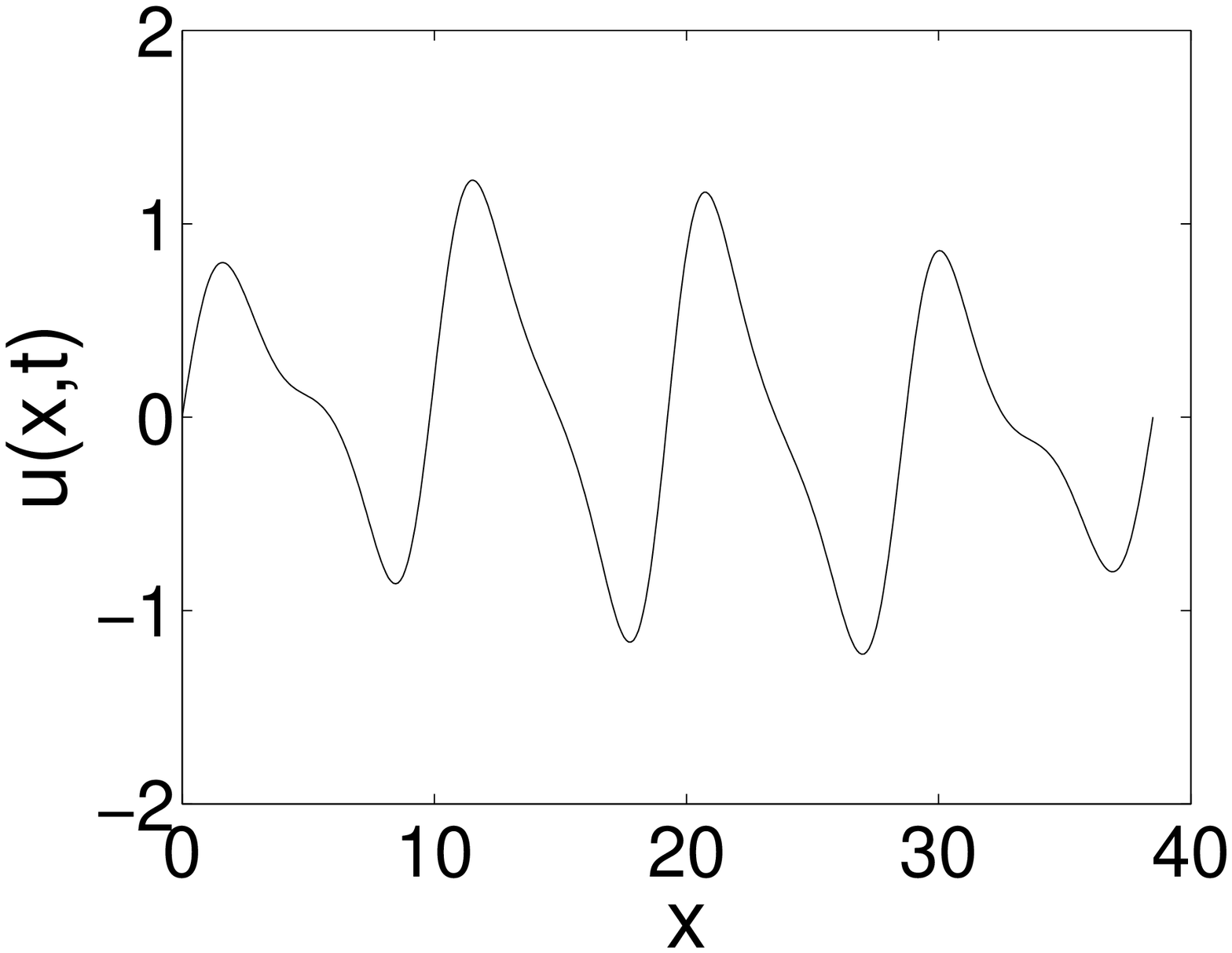}
\hspace{-0.22\textwidth} (b)
    \\
\hspace{-0.22\textwidth}
\includegraphics[width=0.21\textwidth]{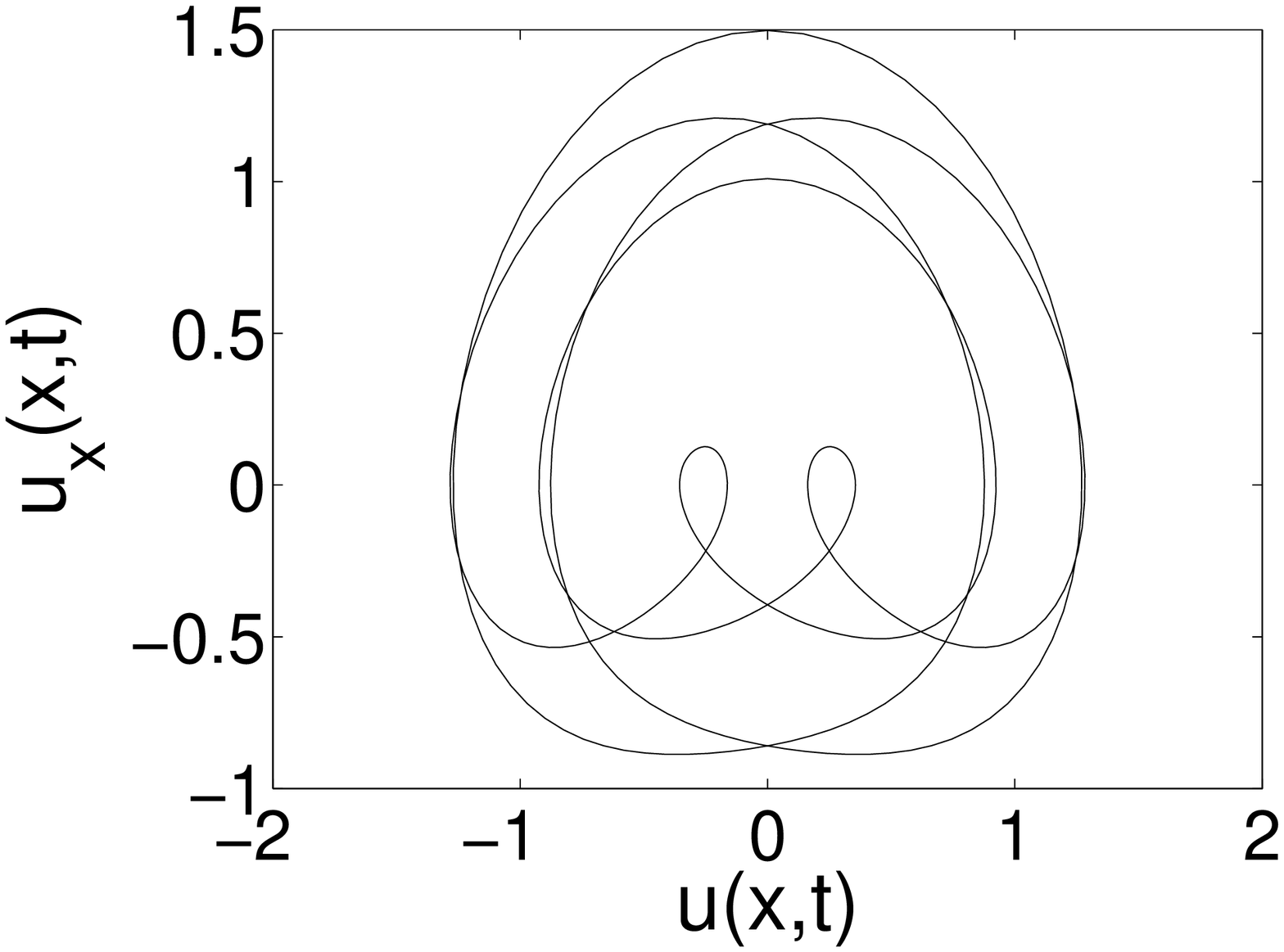}
\hspace{-0.22\textwidth} (c) \hspace{0.22\textwidth}
\includegraphics[width=0.21\textwidth]{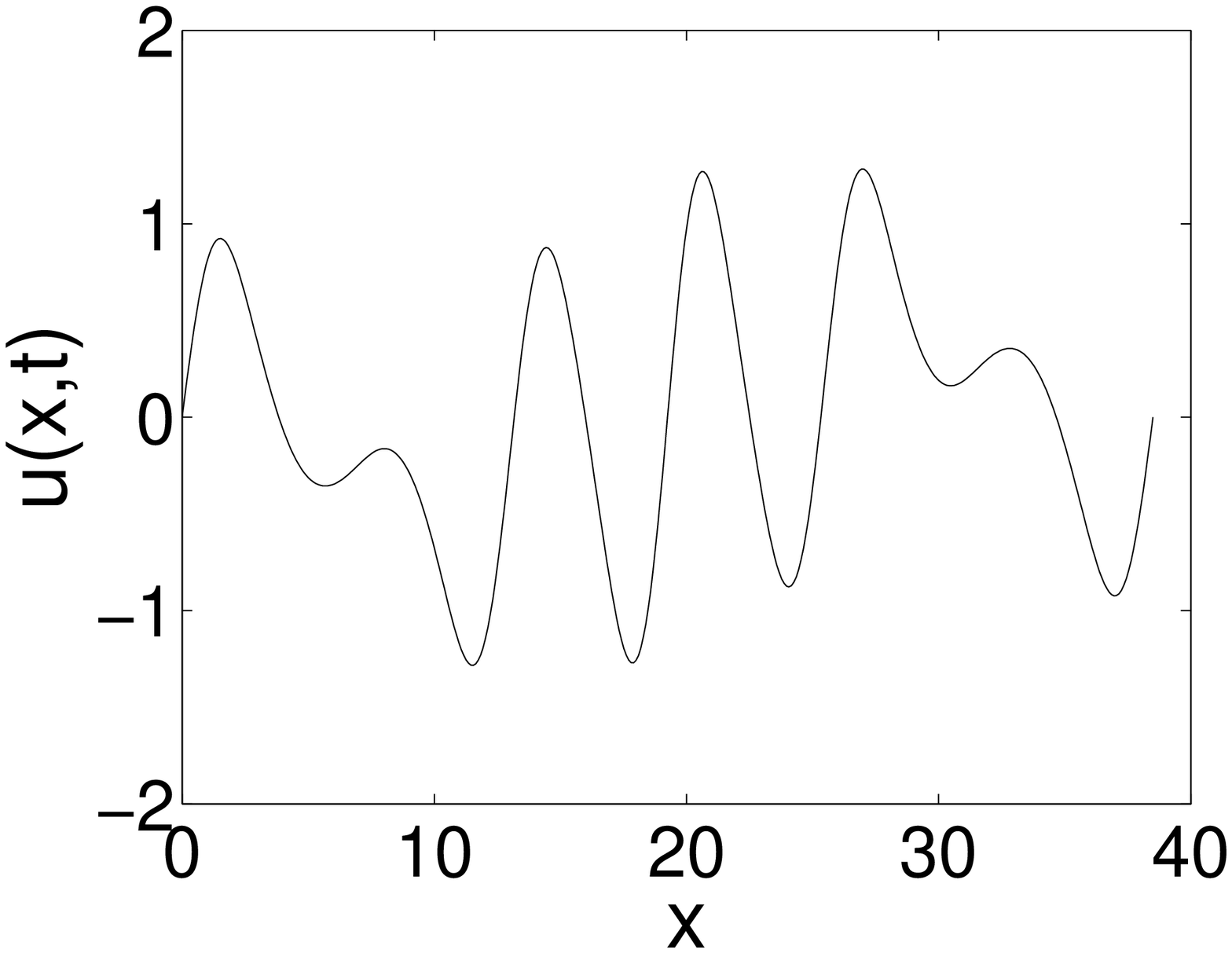}
\hspace{-0.22\textwidth} (d)
    \\
\hspace{-0.22\textwidth}
\includegraphics[width=0.21\textwidth]{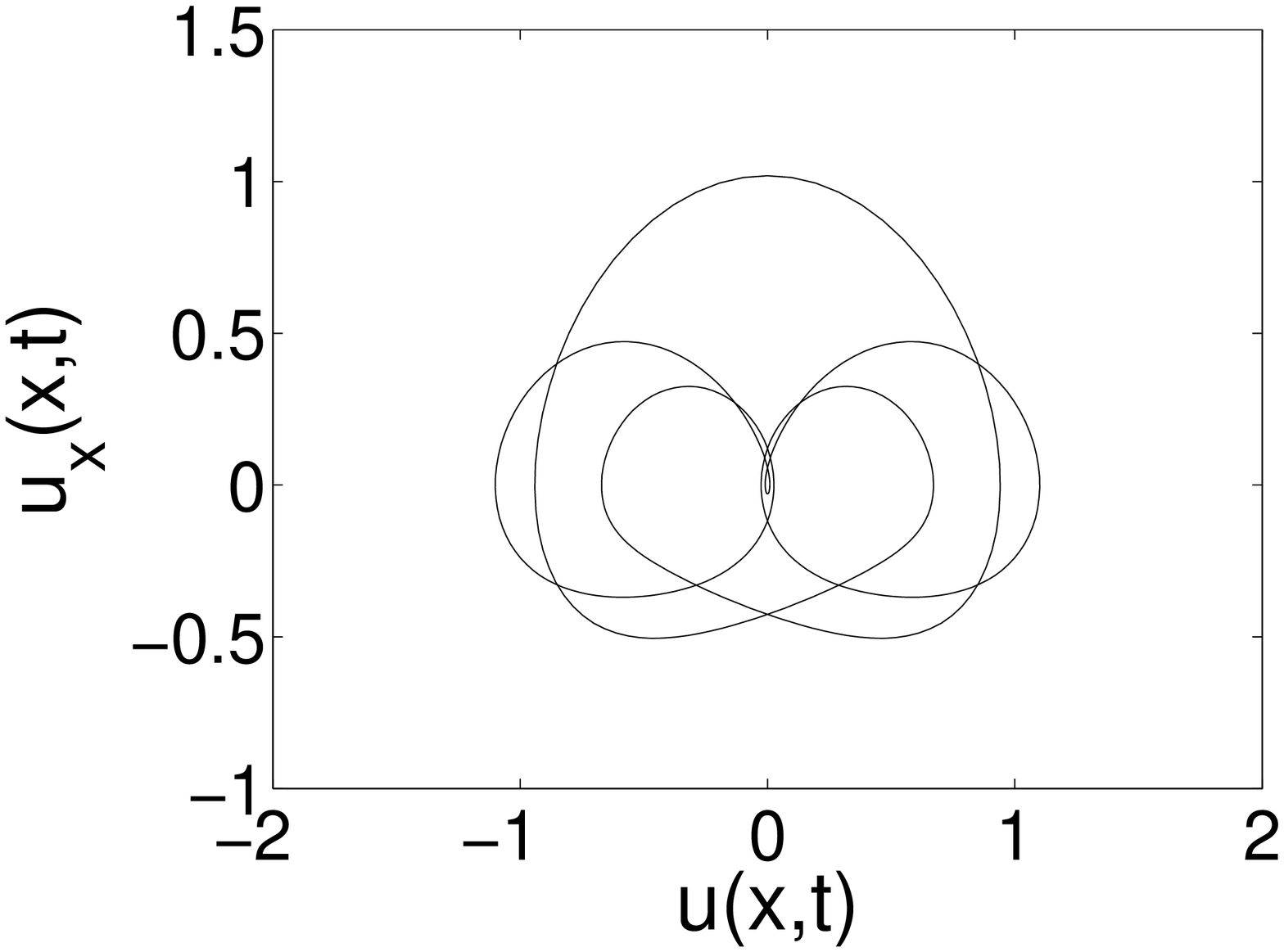}
\hspace{-0.22\textwidth} (e) \hspace{0.22\textwidth}
\includegraphics[width=0.21\textwidth]{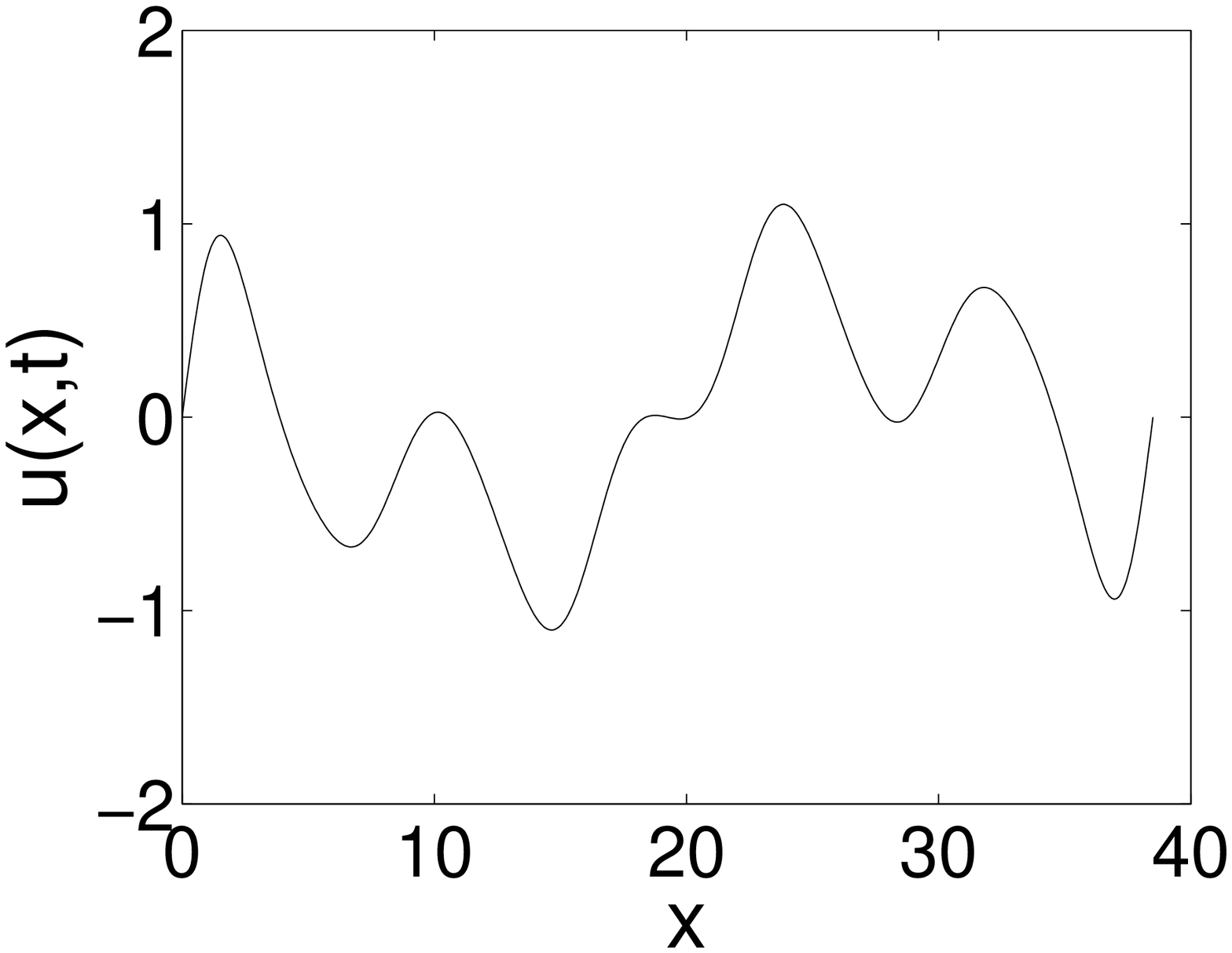}
\hspace{-0.22\textwidth} (f)
\caption[]{
The spatial profile of $u(x,t)$ at a instant $t$ when
(a), (b)
a typical trajectory passes $S_C$;
(c), (d)
or $S_R$;
(e), (f)
or \eqv\ $T$,
the transition region between central region $S_C$
and the side region $S_R$.
Compare with \reffig{f:antfix2}.
      }
\label{f:antorbt1}
\end{figure}

Segments of a typical orbit in the \SIS\  bear
close resemblance to the {\eqva} listed in \reffig{f:antfix2}.
The $(u\,,u_x)$ representations and
the spatial profiles of three typical instants in the
long-time evolution of \reffig{f:antorbt1} resemble the
{\eqva}  shown in \reffig{f:antorbspt}: \reffig{f:antorbt1}\,(c)
is similar to \reffig{f:antfix2}\,(c) and \reffig{f:antorbt1}\,(a)
is similar to \reffig{f:antfix2}\,(a).
The state
in \reffig{f:antorbt1}\,(e), which lies along the transition
from $S_R$ to $S_C$, has no clear \eqv\ counterpart.
So, at this stage,
\eqva\  partition the \statesp\ at the coarsest level into alphabet $\{S_L\,,
S_C\,,S_R\}$.

\section{Recurrent patterns}
\label{sect:kspatt}

In this section we investigate the recurrent patterns in the KSe
dynamics
by constructing the unstable manifolds of dynamically important
 invariant sets.
We define intrinsic curvilinear coordinates
along unstable manifolds, then deduce
return maps and construct approximate symbolic dynamics
which greatly facilitates the search
for {\UPO}s. This program is
quite successful for the ``center,'' $S_C$ part, but less so
for the ``side,'' $S_R$ part.

All numerical work presented here is  for
system size
    $L=38.5 \,, \nu=1$, chosen large enough to exhibit
spatially nontrivial competition between various chaotic patterns.
We find that the $16$ Fourier mode truncation is sufficient for our
purposes\rf{kev01ks,HLBcoh98};
the {\eqva} and \po s
change by a few percent if the number of modes is doubled.
The translational invariance of \refeq{kseq}
in the full {\statesp}
implies that \reqva\ and \po s could play
important role\rf{lop05rel}. The
restriction to the antisymmetric subspace excludes this type of solutions.

\subsection{Poincar\'{e} sections}

\begin{figure}[tbp] 
    \centering
\hspace{-0.22\textwidth}
\includegraphics[width=0.21\textwidth]{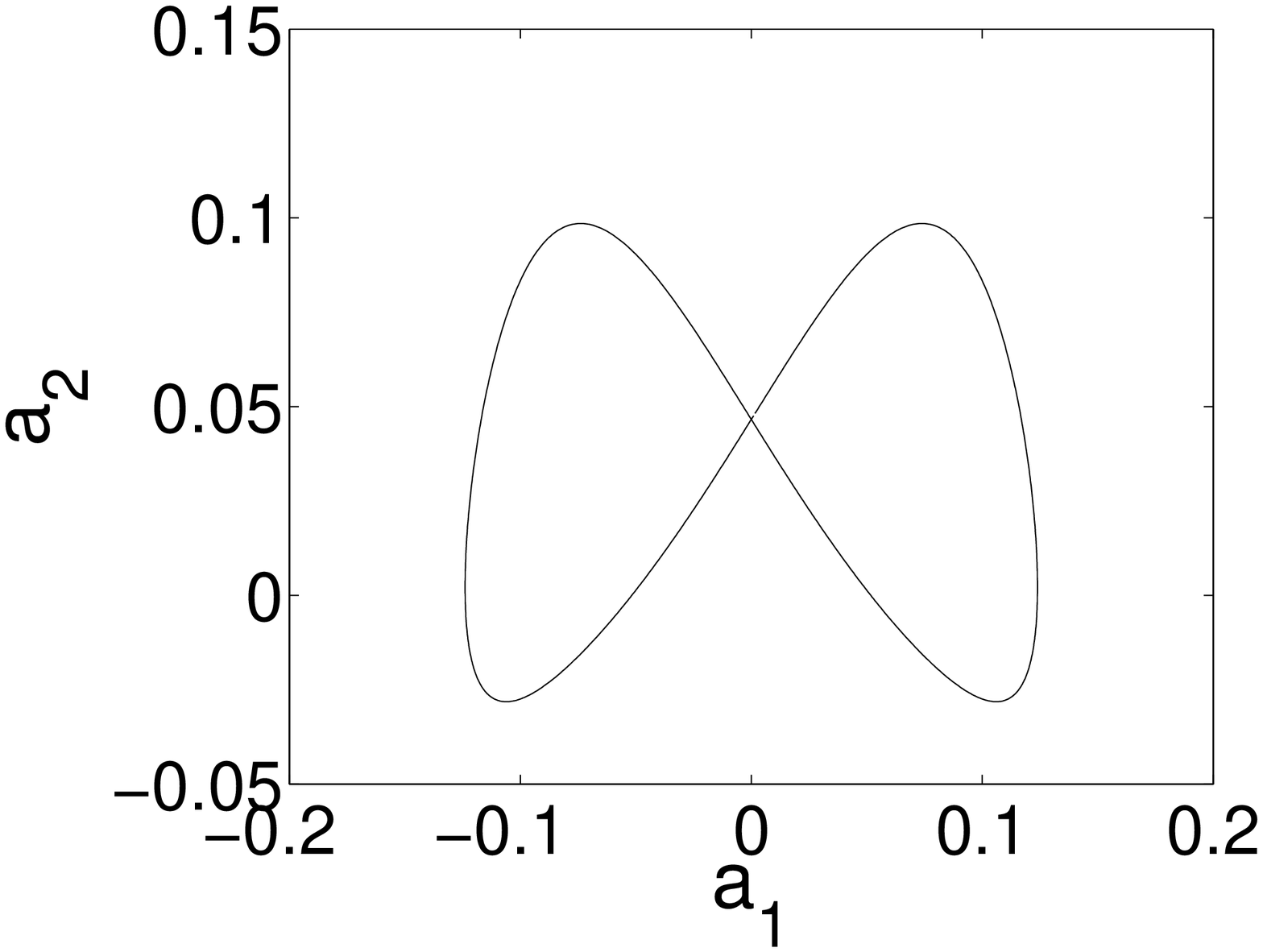}
\hspace{-0.22\textwidth} (a) \hspace{0.22\textwidth}
\includegraphics[width=0.21\textwidth]{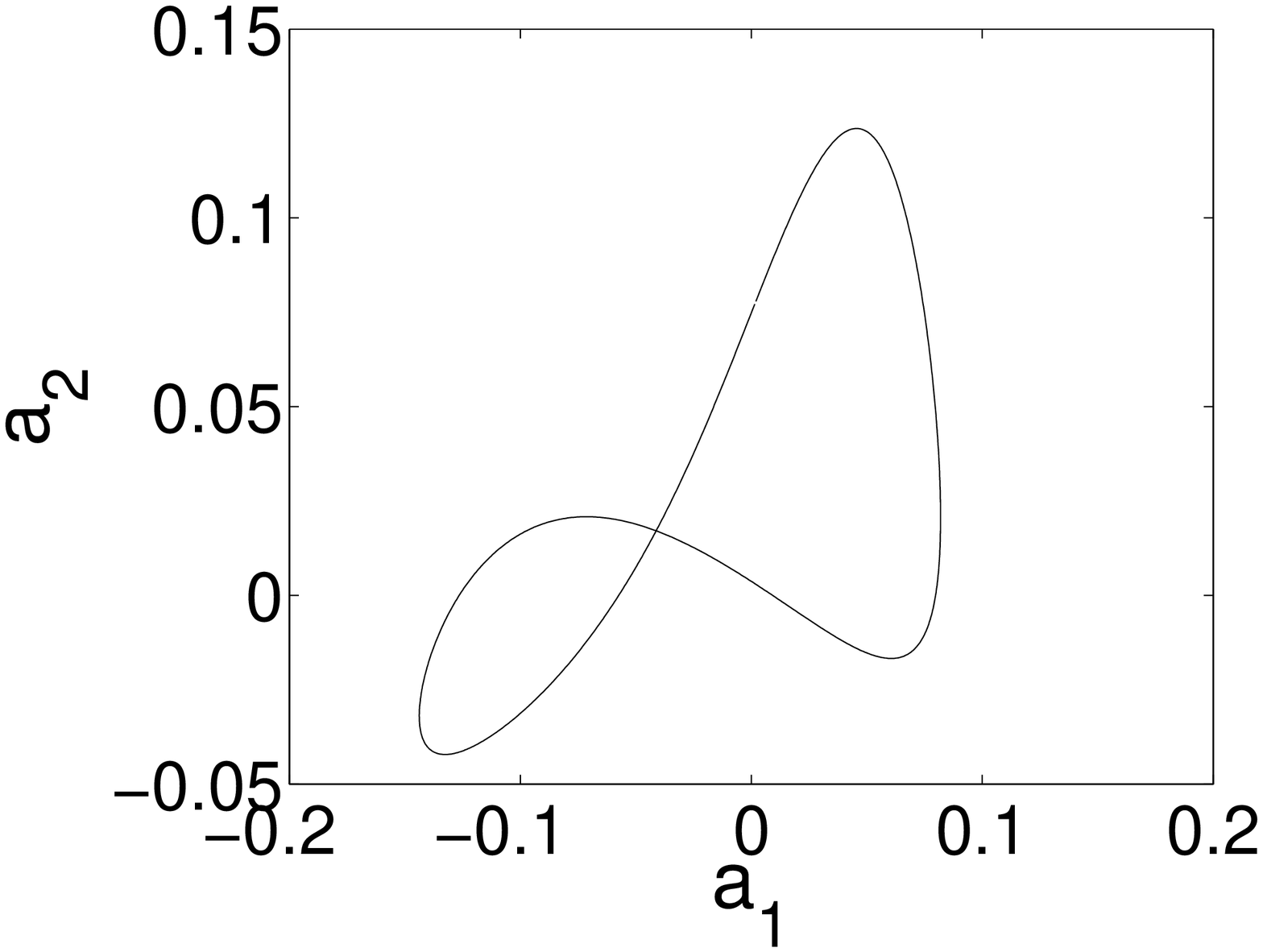}
\hspace{-0.22\textwidth} (b)
    \\
\hspace{-0.22\textwidth}
\includegraphics[width=0.21\textwidth]{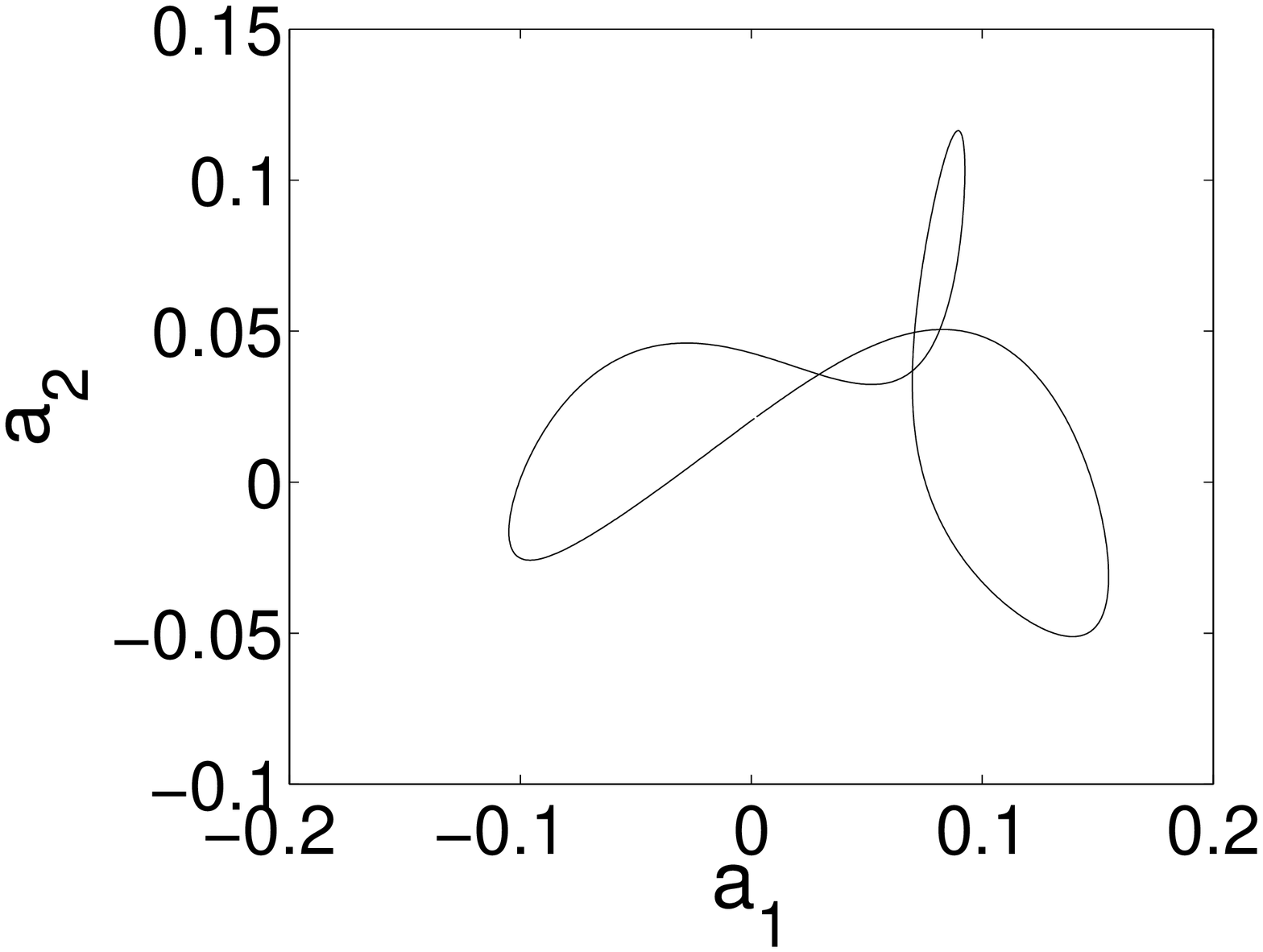}
\hspace{-0.22\textwidth} (c) \hspace{0.22\textwidth}
\includegraphics[width=0.21\textwidth]{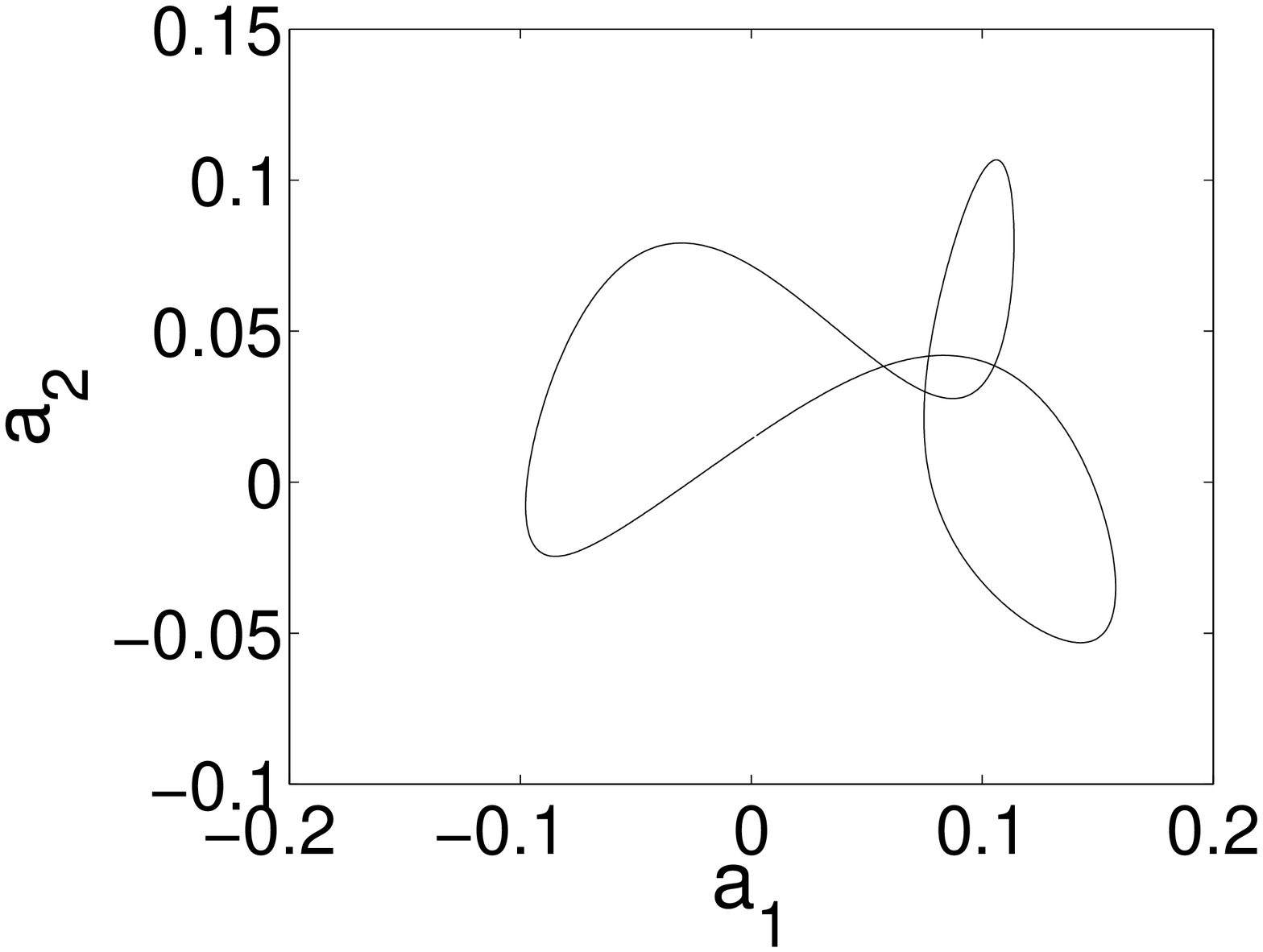}
\hspace{-0.22\textwidth} (d)
\caption[]{
$[a_1,a_2]$ Fourier modes projection of several
shortest \po s in $S_C$. The periods are
(a) $T=25.6095$;
(b) $T=25.6356$;
(c) $T=36.7235$;
(d) $T=37.4083$.
      }
\label{f:ant1p2}
\end{figure}

\begin{figure}[tbp] 
    \centering
\hspace{-0.22\textwidth}
\includegraphics[width=0.21\textwidth]{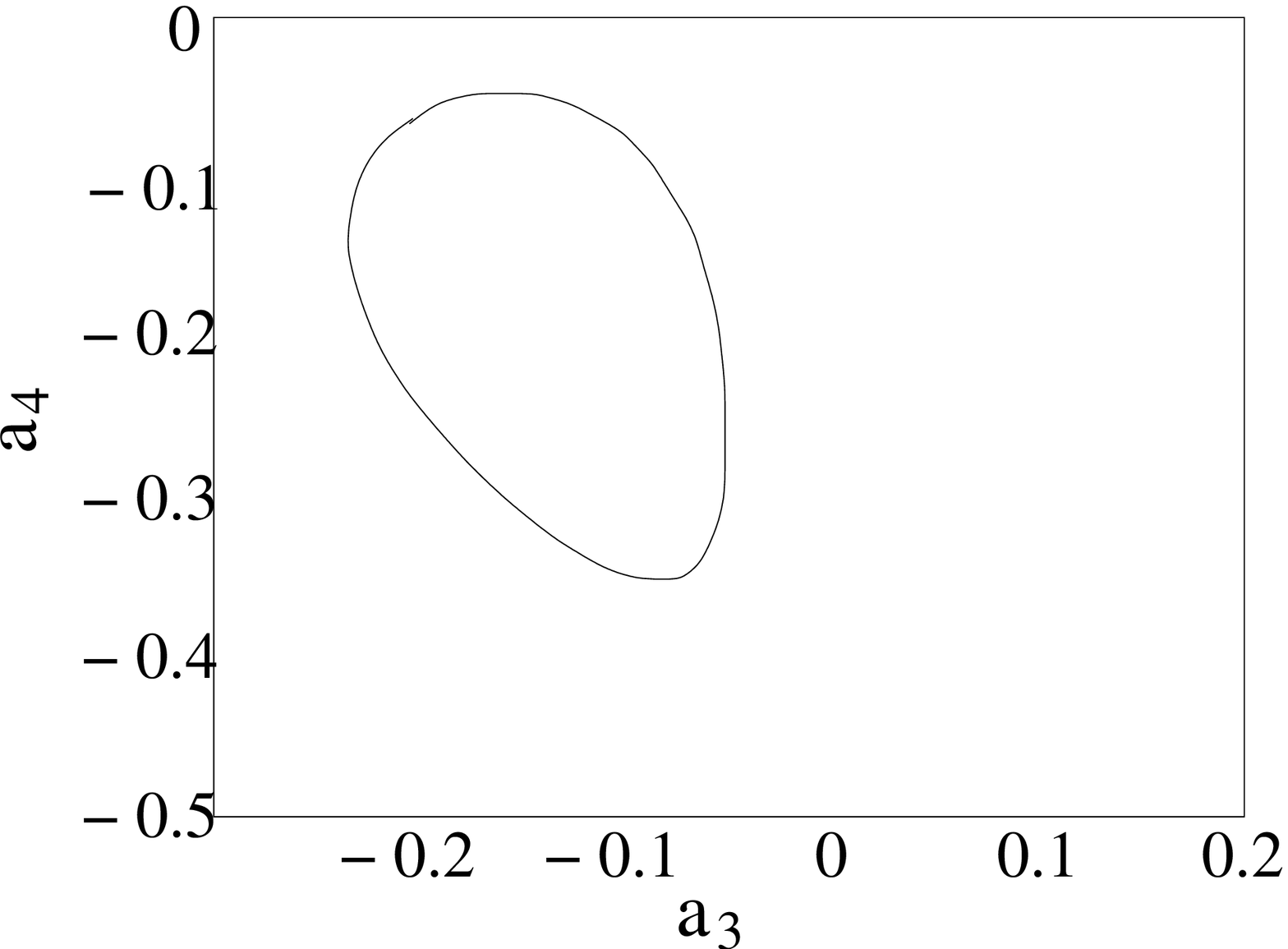}
\hspace{-0.22\textwidth} (a) \hspace{0.22\textwidth}
\includegraphics[width=0.21\textwidth]{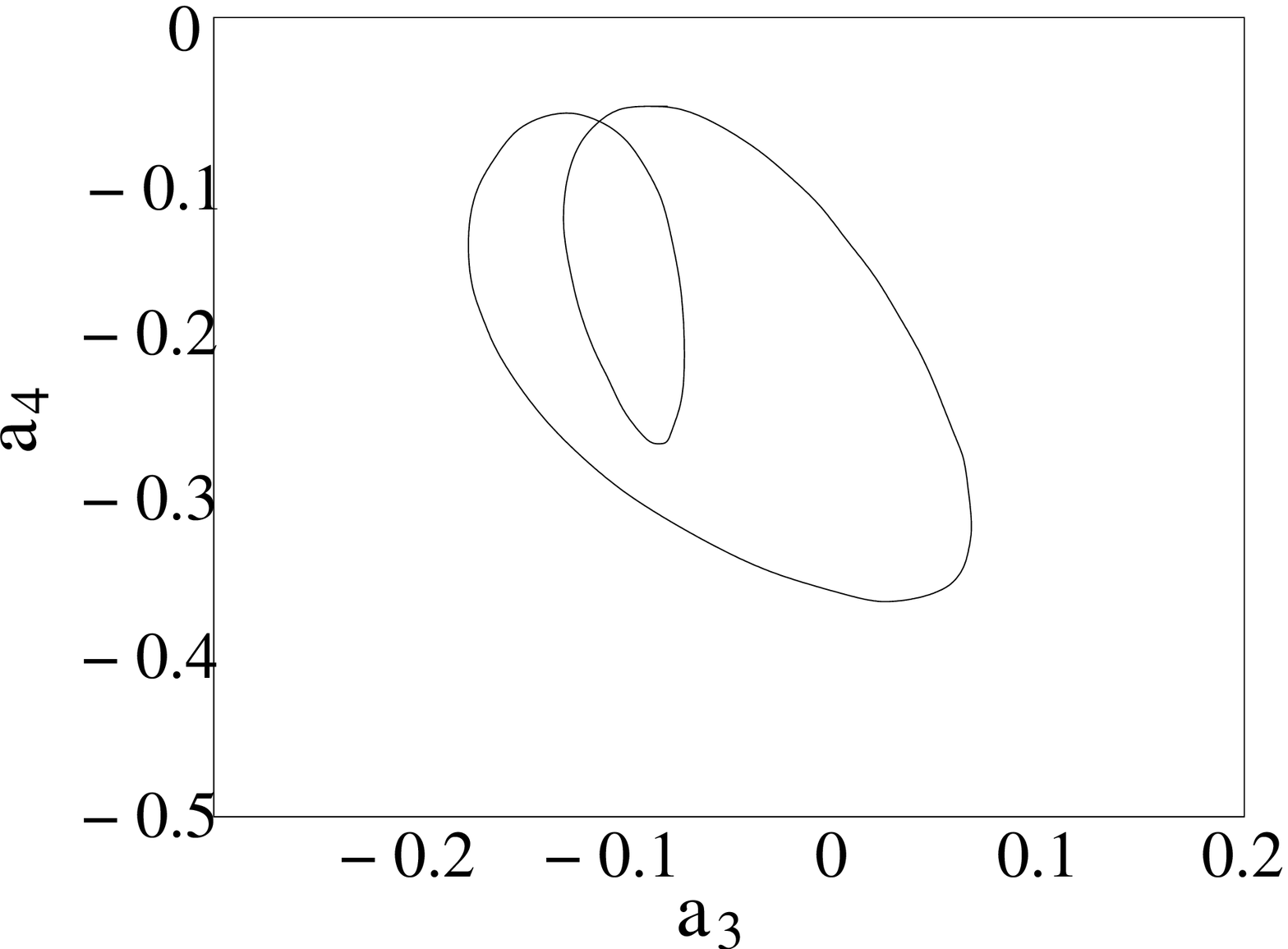}
\hspace{-0.22\textwidth} (b)
    \\
\hspace{-0.22\textwidth}
\includegraphics[width=0.21\textwidth]{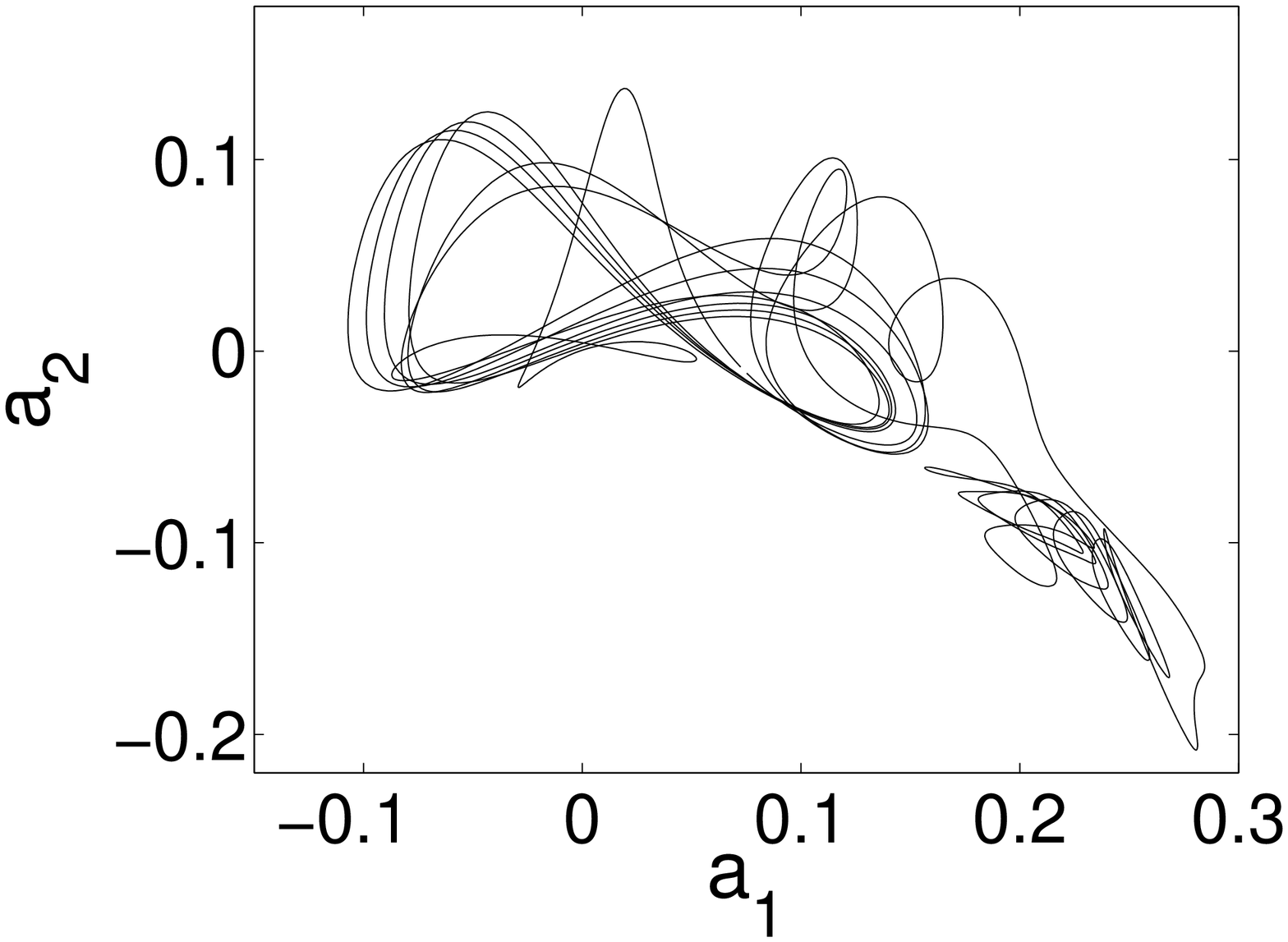}
\hspace{-0.22\textwidth} (c) \hspace{0.22\textwidth}
\includegraphics[width=0.21\textwidth]{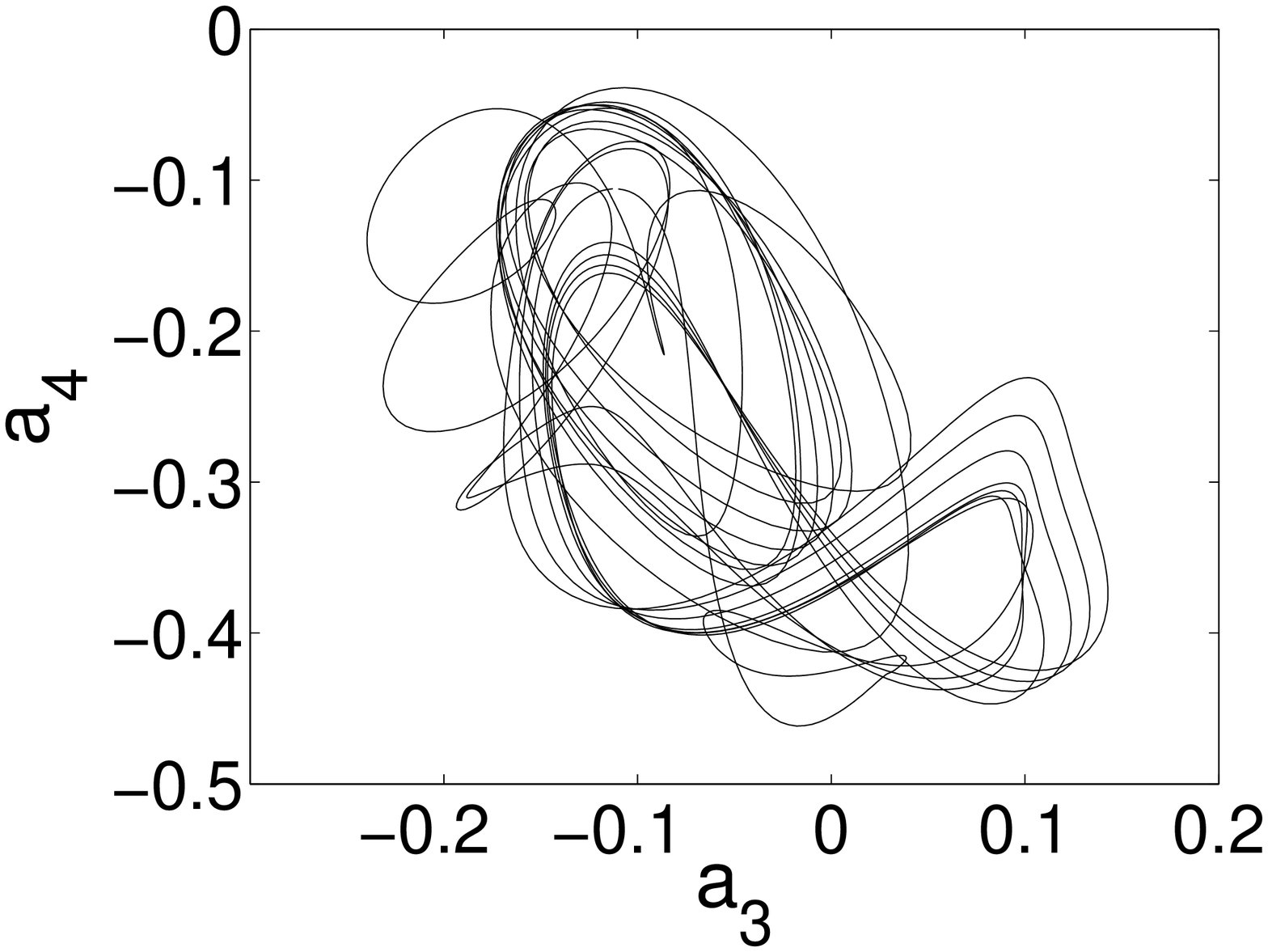}
\hspace{-0.22\textwidth} (d)
\caption[]{
(a)
    The shortest \po\ $p_{R2}$ in $S_R$, $T=12.0798$,
 $[a_3,a_4]$ Fourier modes projection.
(b)
The second shortest \po\ $p_{R1}$ in $S_R$, $T=20.0228$,
$[a_3,a_4]$ projection.
A long \po\ connecting $S_C$ in $S_R$, $T=355.34$:
(c) $[a_1,a_2]$  projection.
(d) $[a_3,a_4]$ projection.
      }
\label{f:antlong}
\end{figure}

We initialize our \po\ searches
by checking for the nearly recurrent orbit segments.
Motivated by \reffig{f:antorbspt}\,(b), we choose the hyperplane $a_1=0$ as our
Poincar\'{e} section for searches within $S_C$.
The shortest {\UPO}s residing in
$S_C$ are displayed in \reffig{f:ant1p2}.
Each of them has a partner related by the reflection
symmetry \refeq{FModInvSymm}, except for the symmetric one,
\reffig{f:ant1p2}\,(a), which is self-dual under the reflection.
The Poincar\'{e} section
for $S_R$ is chosen as $a_4=-0.122$.
The shortest {\UPO}s residing
in $S_R$ are depicted in
\reffig{f:antlong}\,(a,b). Each of them has a symmetry partner in $S_L$.
The one in \reffig{f:antlong}\,(a) is seen to mediate the transition from
$S_R$ to $S_C$, and the one in \reffig{f:antlong}\,(b) will be used to build a
symbolic dynamics in $S_R$.
All \po s that we have found have only one unstable eigendirection,
enabling us
to construct a 1\dmn\ map to model the \statesp\ dynamics on some
chosen Poincar\'{e} section.

\refFig{f:antlong}\,(c,d) shows
a long,  period $T=355.34$ \po\ that communicates between $S_C$
and $S_R$. We initialize the search for
this orbit by choosing a segment of long-time orbit
that communicates between $S_C$ and $S_R$.  Several of such
long communicating periodic orbits are found with similar or longer
period.

The unstable manifolds of all \po s found so far
rarely have more than two unstable eigendirections.
This is consistent with the estimated linear growth of the
Hausdorff dimension of the \SIS\ with the
system size\rf{tajima02}. In general,
there exist many {\UPO}s with many unstable
directions, 
but they do not appear to participate in the asymptotic dynamics.
Although more than $50$ {\UPO}s were found in our preliminary search, we have
no criterion that would preclude the possibility of
more important ones yet to be detected.

In the preliminary search, we chose the Poincar\'{e} section by examining
\reffig{f:antorbspt}. A good
Poincar\'{e} section is essential to the success of all subsequent steps.
It should cut all the orbits in the \SIS\ transversely and
should be numerically convenient;
all our Poincar\'{e} sections are hyper-planes.
After a Poincar\'{e} section is chosen, we
determine the symbolic dynamics by examining
the return map defined on the Poincar\'{e}
section. In our case, the
unstable manifold of the shortest orbit is nearly 1-dimensional and the
dynamics can be approximated by maps defined on 1-d line segments.

\begin{figure} 
\centering
\hspace{-0.22\textwidth}
\includegraphics[width=0.23\textwidth]{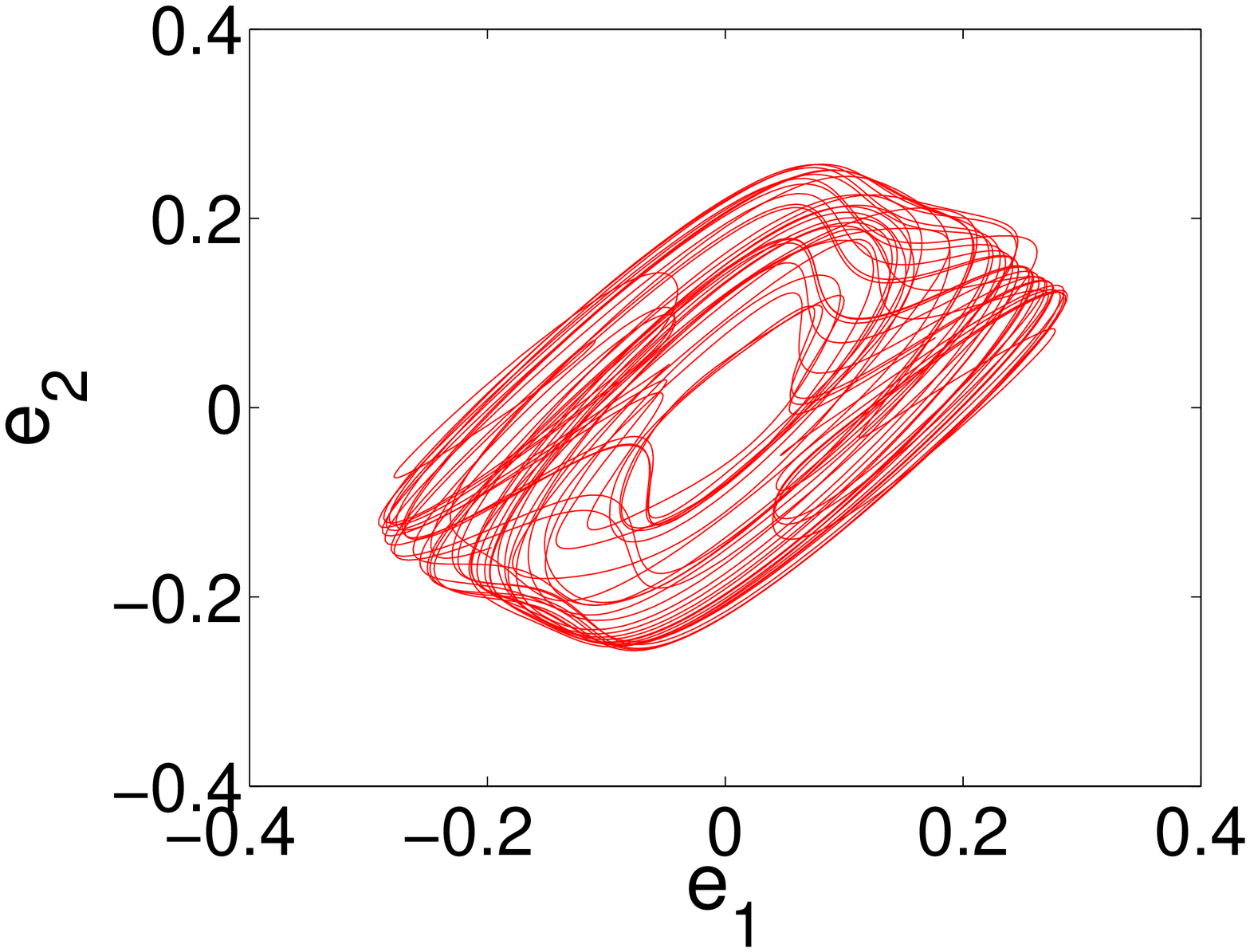}
\hspace{-0.24\textwidth} (a) \hspace{0.22\textwidth}
\includegraphics[width=0.23\textwidth]{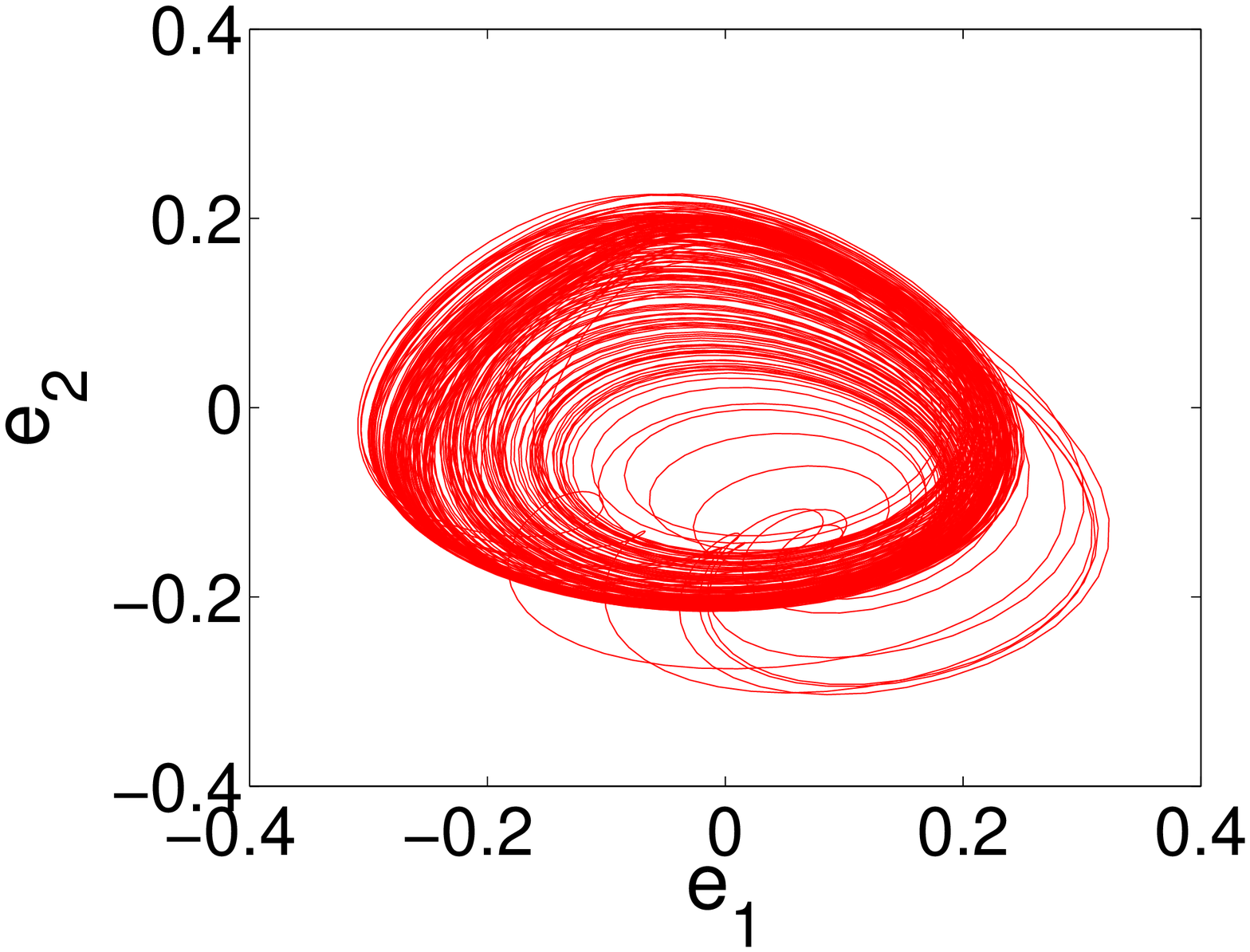}
\hspace{-0.24\textwidth} (b)
\caption{
The projection of a typical \statesp\ trajectory segment onto the
linearized stability eigenplanes of
{\eqva}  $C_1$ and $R_1$.
(a) The projection of part of the center orbit to the eigenplane
    $(e_1,e_2)$
of $C_1$;
(b) The projection of part of the right-sided orbit to the eigenplane
    $(e_1,e_2)$
of $R_1$.
    }
\label{f:eigpjt}
\end{figure}

According to \reftab{t:stationary}, the least unstable {\eqva}  $C_1$
and $R_1$ have 2-$d$ spiral-out unstable manifolds.
In the neighborhood of $C_1$ and $R_1$
the unstable manifolds are well approximated by the corresponding
eigenplanes determined by their complex eigenvalue pairs. The
projections to the eigenplanes of nearby long-time orbit
segments in \reffig{f:antorbspt}\,(b) are displayed in \reffig{f:eigpjt}.
\refFig{f:eigpjt}\,(a) shows a projection onto the eigenplane $e_1-e_2$ of
the center equilibrium $C_1$ (origin in the figure);
the orbit is circling and
avoiding $C_1$. At the left and the right bounds, there are extra orbit
rotations which have
a rotation plane roughly perpendicular to the $e_1-e_2$ eigenplane. It turns
out that this rotation is related to the transition {\eqv}  $T$. Thus,
the dynamics of the central part is controlled by $C_1$ and $T$ (and $T^*$,
the symmetric image of $T$) with the orbit spending most of the time
around $C_1$. An obvious choice of the Poincar\'{e} section is
$\mathcal{P}_C:x_1=0$, where $x_1=(x-x(C_1)) \cdot e_1$ is the
projection of the \statesp\ point $x$ relative
to the equilibrium $C_1$ along the $e_1$ direction.
\refFig{f:eigpjt}\,(b) shows the projection onto the eigenplane $e_1-e_2$ of
the right-side {\eqv}  $R_1$ (at the origin $(0,0)$ in the figure). We
see that most of the time the orbit
circles around $R_1$. There are many small
oscillations near the
bottom, in this projection partially masked by the dense
band. In the following,
$\mathcal{P}_R:x_1=0$ (different from the
center one) is
taken to be the Poincar\'{e} section. This is not a perfect choice as the
occasional tangent passage of the orbit to $\mathcal{P}_R$ can
cause trouble.

\subsection{Curvilinear coordinates, center repeller}

\begin{figure}[tbp] 
    \centering
\hspace{-0.22\textwidth}
\includegraphics[width=0.21\textwidth]{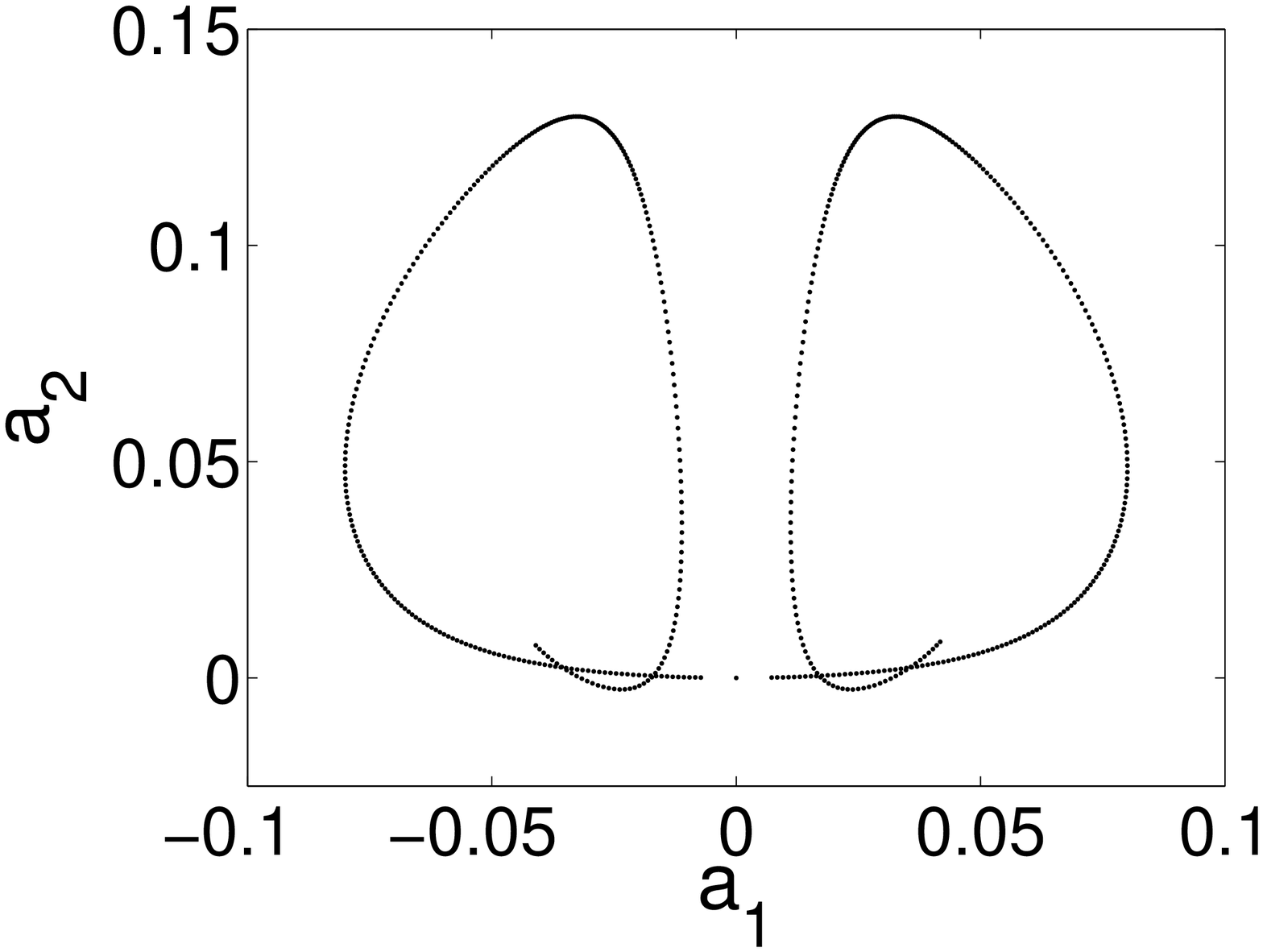}
\hspace{-0.22\textwidth} (a) \hspace{0.22\textwidth}
\includegraphics[width=0.21\textwidth]{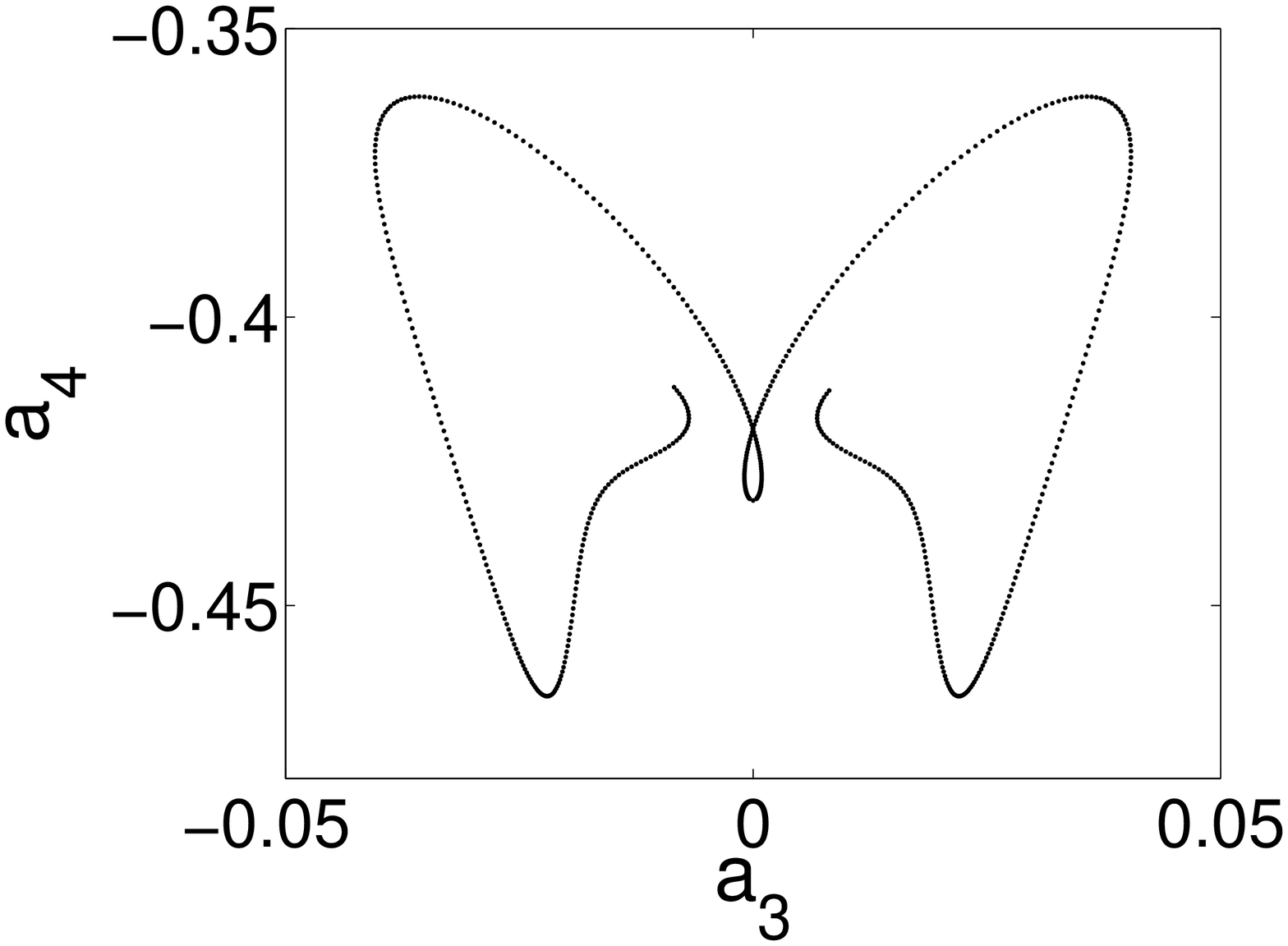}
\hspace{-0.22\textwidth} (b)
    \\
\hspace{-0.22\textwidth}
\includegraphics[width=0.21\textwidth]{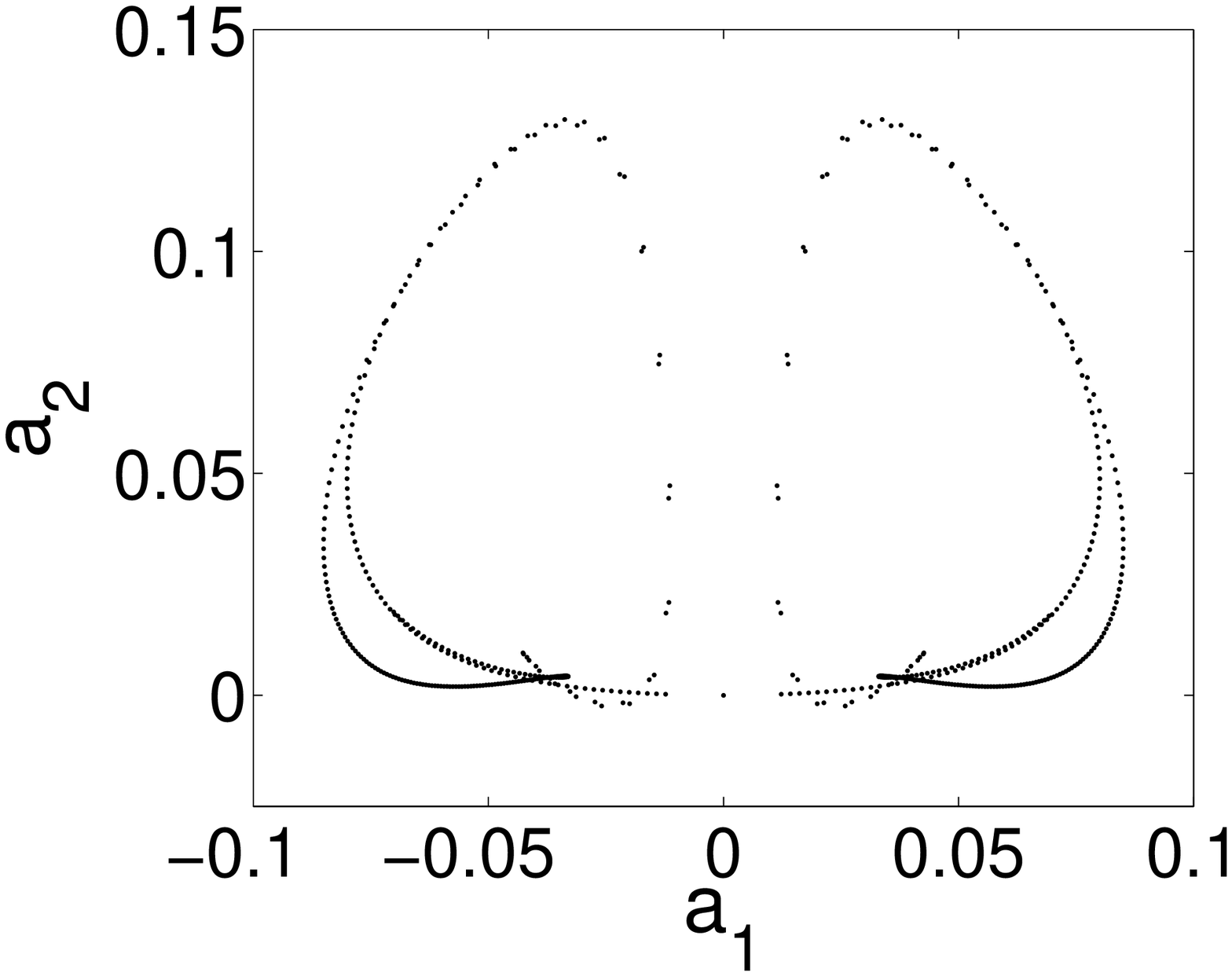}
\hspace{-0.22\textwidth} (c) \hspace{0.22\textwidth}
\includegraphics[width=0.21\textwidth]{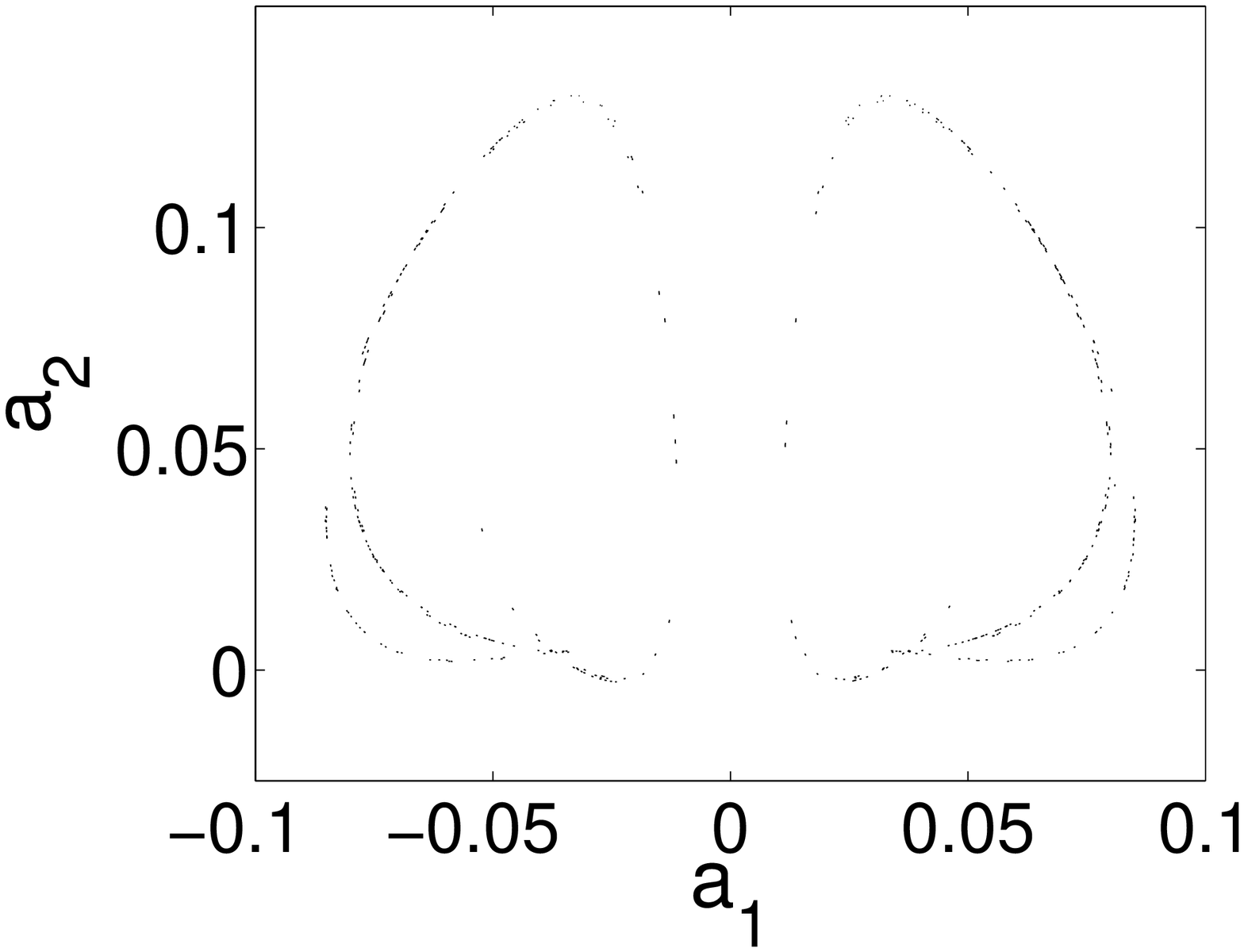}
\hspace{-0.22\textwidth} (d)
    \\
\hspace{-0.22\textwidth}
\includegraphics[width=0.21\textwidth]{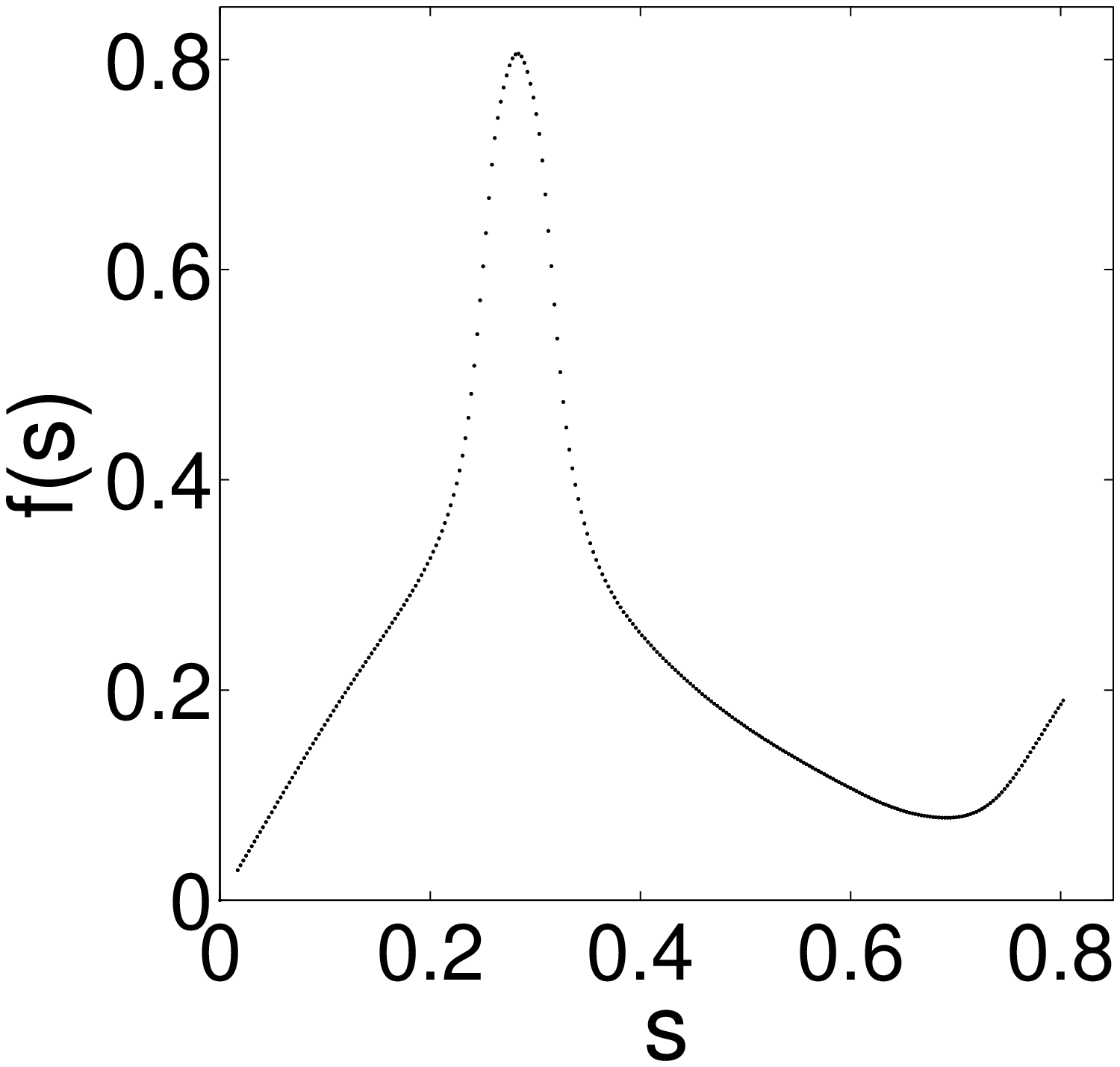}
\hspace{-0.22\textwidth} (e) \hspace{0.22\textwidth}
\includegraphics[width=0.21\textwidth]{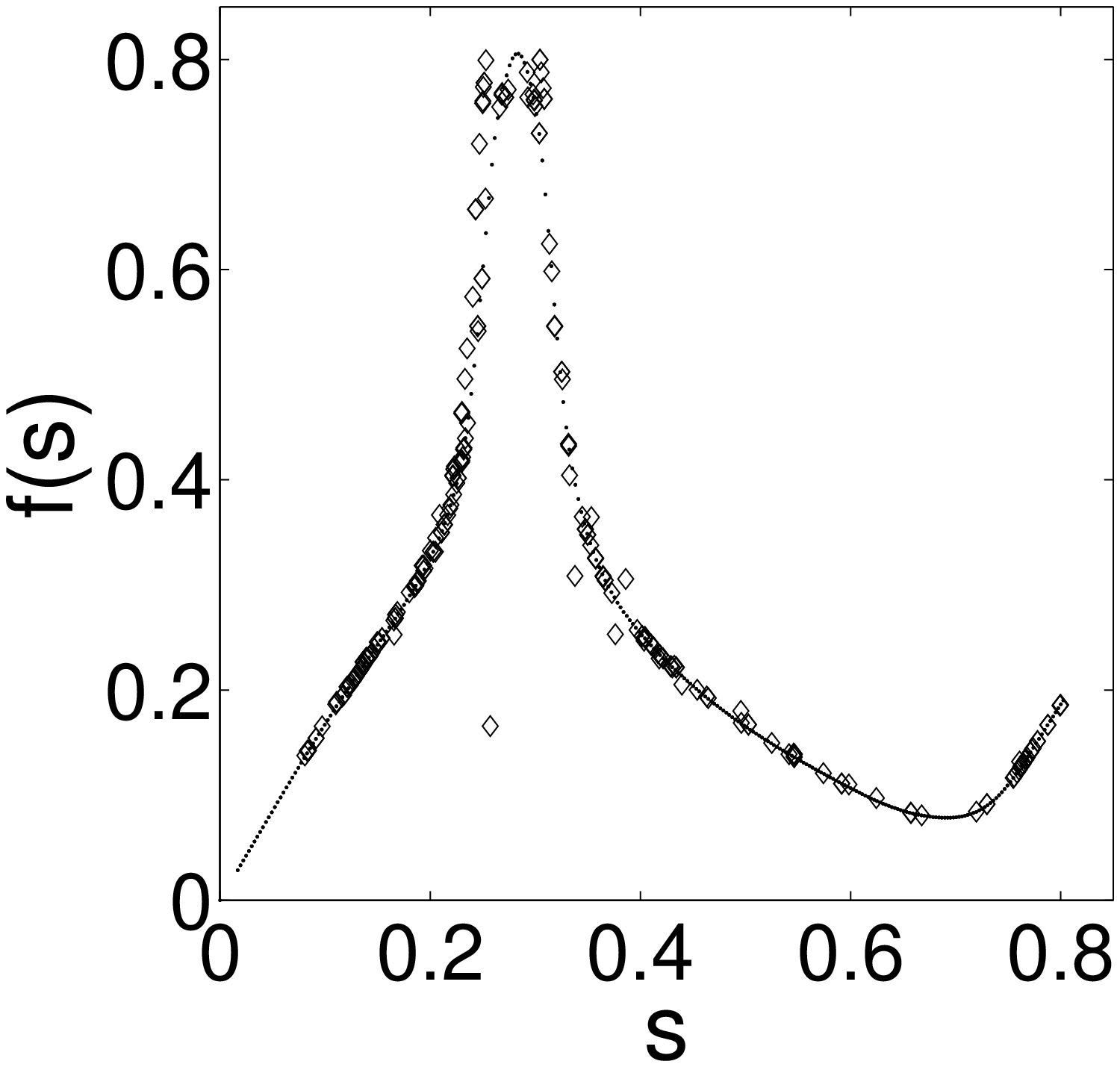}
\hspace{-0.22\textwidth} (f)
\caption[]{
The Fourier modes projection of the unstable manifold
segment $L_C$ of $C_1$
on the Poincar\'{e} section $\mathcal{P}_C$:
(a) $[a_1,a_2]$ projection of the unstable manifold on $\mathcal{P}_C$.
(b) $[a_3,a_4]$ projection ($L_C$ is represented by $564$ points).
(c) Projection of the first iteration of $L_C$.
(d) Projection of all
computed {\UPO}s of the full domain
topological length up to  $n=8$.
The Poincar\'{e} return map on $\mathcal{P}_C$,
the intrinsic coordinate:
(e) The bimodal return map in the fundamental domain,
with the symbolic dynamics given by three symbols $\{0\,,1\,,2\}$.
(f)
The periodic points in the fundamental domain,
overlayed over the return map (e).
      }
\label{f:antmn1}
\end{figure}

By construction, the section $\mathcal{P}_C$ contains $C_1$ and the
eigenvector $e_2$ and is
perpendicular to the $e_1-e_2$ expanding eigenplane of
$C_1$. We compute the intersection of the Poincar\'{e} section
$\mathcal{P}_C$ with the 2-dimensional unstable manifold of $C_1$ by
iterating\rf{carsim}
a set of initial points on the $\pm e_2$
eigenvector, infinitesimally close to $C_1$.
At nearest ``turn-back points'' the unstable manifold
bends back sharply and then nearly retraces itself,
see
\reffig{f:antmn1}\,(a,b).
Denote by  $L_C$ the ``base segment'' of the
unstable manifold between the two nearest turn-back points
that bracket $C_1$.
\refFig{f:antmn1}\,(c) shows the first iterate of $L_C$:
finer structures do develop, but on the whole they lie
close to $L_C$ and should be well described
by the intrinsic curvilinear coordinate that we now define\rf{ks}.

Assign to each $d$-dimensional point $x \in L_C$
a coordinate $s=s(x)$ whose value is the Euclidean
arc length to $C_1$ measured along the 1-dimensional
$\mathcal{P}_C$ section of the $C_1$ unstable manifold.
Next, for a nearby point $x_0 \notin L_C$ determine
the point $x_1 \in L_C$
which minimizes the Euclidean distance $(x_{0}-x_{1})^2$, and
assign arc length coordinate value $s_{0}=s(x_{1})$
to $x_0$.
In this way, an approximate 1-dimensional intrinsic coordinate system
is built along the unstable manifold. This parametrization is useful
if the \SIS\ is sufficiently thin that its perpendicular extent
can be neglected, with every point
on the \SIS\ assigned the nearest point on the base segment $L_C$.

Armed with this intrinsic curvilinear coordinate
parametrization, we are now in a position to
construct a 1-dimensional model of the dynamics on the \SIS.
If $x_n$ is the $n$th Poincar\'e section of a trajectory in
neighborhood of $C_1$, and $s_n$ is
the corresponding curvilinear coordinate,
then $s_{n+1} = f(s_n)$ models the full \statesp\ dynamics
$x_n \to x_{n+1}$. We define $f(s_n)$ as a smooth, continuous
1-dimensional map $f : L_C \to L_C$
by taking $x_n \in L_C$, and assigning to $x_{n+1}$ the nearest
base segment point
$s_{n+1}=s(x_{n+1})$.

This Poincar\'e return map is multi-modal and,
due to the discrete symmetry \refeq{FModInvSymm},
antisymmetric under the $s \to -s$ reflection.
This discrete symmetry
\refeq{FModInvSymm} can be profitably used to
simplify the symbolic dynamics\rf{DasBuch}.
Defining the $s \geq 0$ segment as the
fundamental domain, the return map for the fundamental
domain is partitioned by two points $\{0.2825,0.6905\}$
into a 3-letter alphabet $\{0\,,1\,,2\}$, see
\reffig{f:antmn1}\,(e).
The chaos generating mechanism -
stretching and folding - is clearly illustrated by this return map.
The dynamics can now
be approximated by a subshift of finite type in the space
of symbol sequences built from
these three letters\rf{GL-Gil07b,DasBuch}.

The distribution of the points on
the return map is highly non-uniform.
This is consistent with the Poincar\'e section \reffig{f:antmn1}\,(c)
for which the vertical segments in the middle
contain fewer representative points than other segments.
Our studies indicate that the dynamics
in this region of \statesp\ is controlled by the
``transition'' {\eqv}  $T$. $T$ has a much stronger
repelling rate ($\mu_T\sim 0.2548$) than the $C_1$ {\eqv},
($\mu_{C_1}\sim 0.04422$, see \reftab{t:stationary}).
A line segment is strongly stretched by a close
passage to $T$, and its representative points
get very sparse upon one Poincar\'{e} iteration. This indicates
that the dynamics in the center
region of \statesp\ is coarsely organized by the
two {\eqva},  $C_1$ and $T$.

\begin{table*}
\caption[]{
Cycles up to topological length
4 for the ``center'' repeller in the fundamental domain;
cycles up to topological
length 8 for the ``side'' repeller.
Listed are the topological
lengths, the itineraries $p$, the largest Lyapunov exponents
    \(
\eigExp[p]=\ln |\ExpaEig_{p,1}|/\period[p]
\,,
    \)
periods \period[p], the first three (four) stability eigenvalues.
Our approximate symbolic dynamics fails to resolve cycles
$0001$ and $0001^*$.
    }
\begin{center}{
\begin{tabular}{lllllll}
\multicolumn{5}{l}{~~~``Center'' periodic orbits} \\
~~~ & $p$ & ~~~\eigExp[p] & $~~~\period[p]$
          & ~~~$\ExpaEig_1$ & ~~~$\ExpaEig_3$ & ~~~$\ExpaEig_4$ \\
\hline
1 &1     & 0.0946  &12.8047 &-3.3581  & 0.20299             & 0.008861 \\
2 &01    & 0.07363 &25.6356 &-6.6028  & 0.004697            &-0.0003854  \\
3 &001   & 0.05814 &38.7241 &-9.5     & 0.0001186           &3.687817783$\cdot10^{-5}$ \\
  &011   & 0.0698  &38.4520  &14.6402  & 5.0398$\cdot10^{-5}$& 0.00029158 \\
  &002   & 0.04558 &37.8160  &8.6967   &1.01538$\cdot10^{-4}$& 5.1672$\cdot10^{-5}$\\
4 &0001  & 0.04647 &52.5998 &-11.5219 & 1.3912$\cdot10^{-5}$&-3.4661$\cdot10^{-7}$\\
 &$0001^*$ & 0.05511 &57.7721 &-24.1326 & 4.1012$\cdot10^{-5}$&-6.6332$\cdot10^{-5}$\\
  &0011  & 0.05736 &51.8915 &19.6194  &-7.5089$\cdot10^{-5}$
                                      & 2.3526+$i\,$2.4938$\cdot10^{-8}$ \\
  &0111  &0.07977  &51.2393 & -59.5683& 0.0001284           &-1.0469$\cdot10^{-8}$ \\
  &0002  &0.047    &57.3329 &  14.802 & 7.8781$\cdot10^{-6}$&-5.0674$\cdot10^{-7}$ \\
  &0012  &0.05103  &58.0857 & -19.3765&-6.9811$\cdot10^{-5}$&-3.5525$\cdot10^{-8}$ \\
\hline
\\[1.5ex]
\multicolumn{5}{l}{~~~``Side'' periodic orbits} \\
~~~ & $p$ & ~~~\eigExp[p] & $~~~\period[p]$
          & ~~~$\ExpaEig_1$ & ~~~$\ExpaEig_3$ & ~~~$\ExpaEig_4$ \\
\hline
1 &0     &0.02747  &20.0228 &-1.7333  &-0.433  &0.001158+$i\,$0.0002633 \\
2 &01    &-0.006163&40.0565 &-0.5569+$i\,$0.5479 & 1.4892$\cdot10^{-6}$
                                               &1.2363$\cdot10^{-6}$  \\
6 &000001   &0.01113  & 120.1658 &-3.8073  &-0.05628  &-4.03$\cdot10^{-10}$  \\
7 &0000101  &0.02363  & 140.196  &-27.4733 &-0.006032 &-6$\cdot10^{-12}$ \\
8 &00001001 &0.03375  & 160.2561 & 223.2796& 0.0005662&-1.202$\cdot10^{-9}$ \\
\hline
\end{tabular}
}
\end{center}
\label{t:orbitsf}
\end{table*}

The return map $f(s)$ leads to approximate finite
Markov diagrams and the associated symbolic dynamics,
enabling us to search for {\UPO}s in a systematic fashion. Up
to length $4$, the
pruning rule implied by \reffig{f:antmn1}\,(e) precludes symbol sequences
$\{21,22,202,2010\}$.
We find all {\UPO}s
of period $n$ within $S_C$ by first determining the
zeros of the 1-dimensional map $f^n(s)$
({\em i.e.}, the $n$-periodic points of $f(s)$),
and then initiating the \descent\ by the corresponding
points in the full \statesp.  Whenever the 1\dmn\ model map
is a good representation of the full \statesp\ dynamics, this approach
yields a \po\ for every admissible itinerary, of arbitrary period.

\begin{figure}[tbp] 
    \centering
\hspace{-0.22\textwidth}
\includegraphics[width=0.21\textwidth]{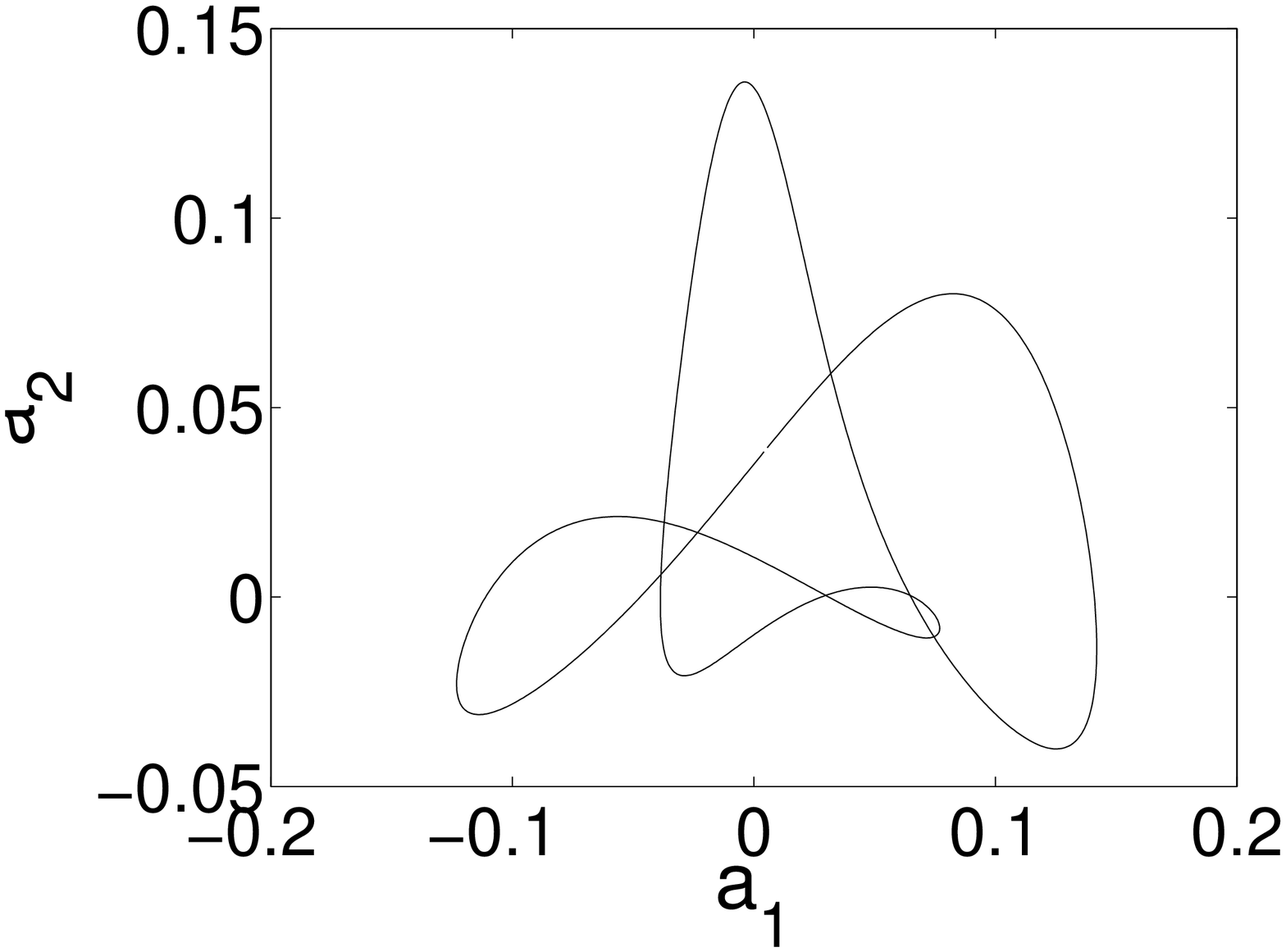}
\hspace{-0.22\textwidth} (a) \hspace{0.22\textwidth}
\includegraphics[width=0.21\textwidth]{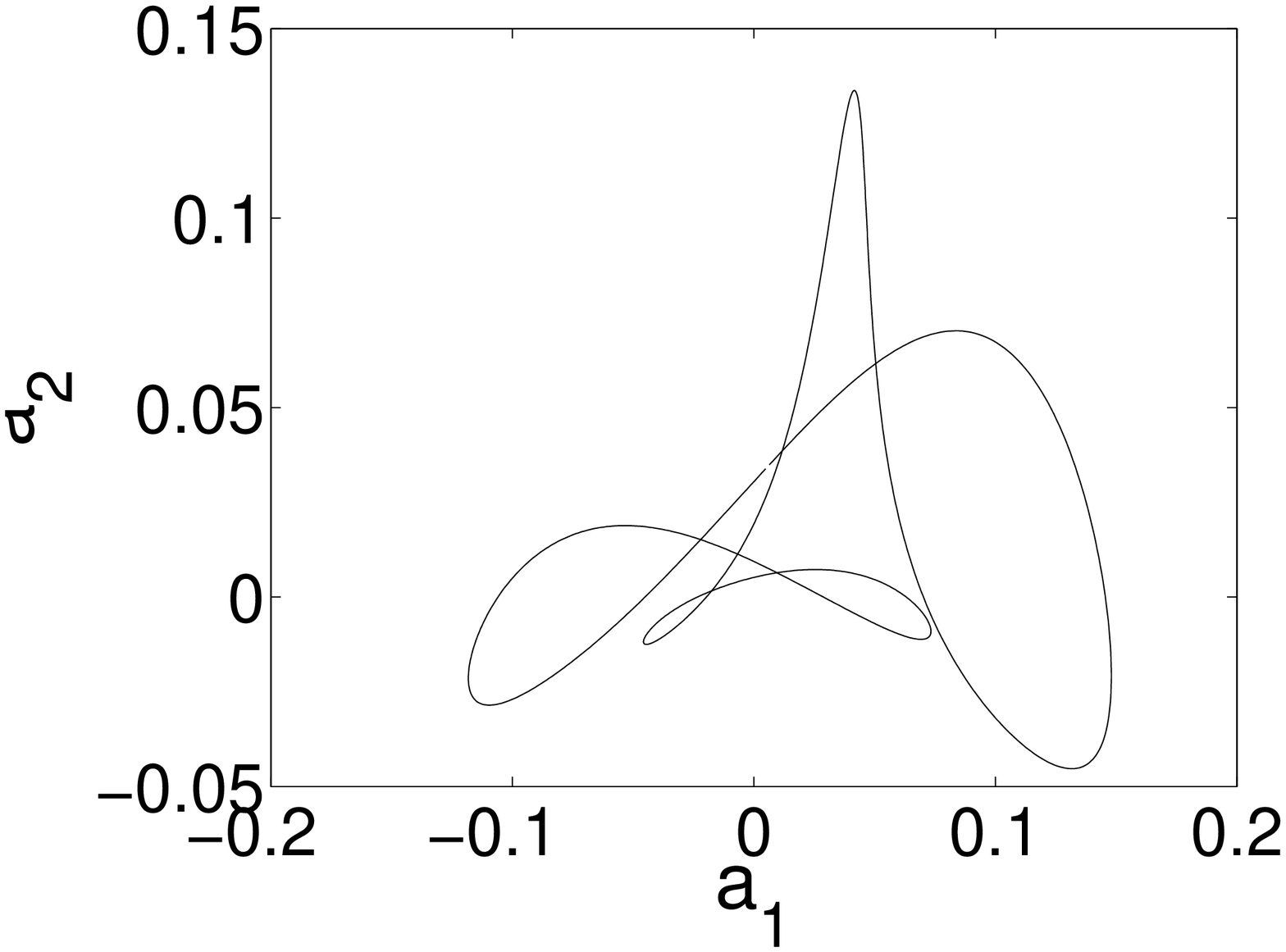}
\hspace{-0.22\textwidth} (b)
    \\
\hspace{-0.22\textwidth}
\includegraphics[width=0.21\textwidth]{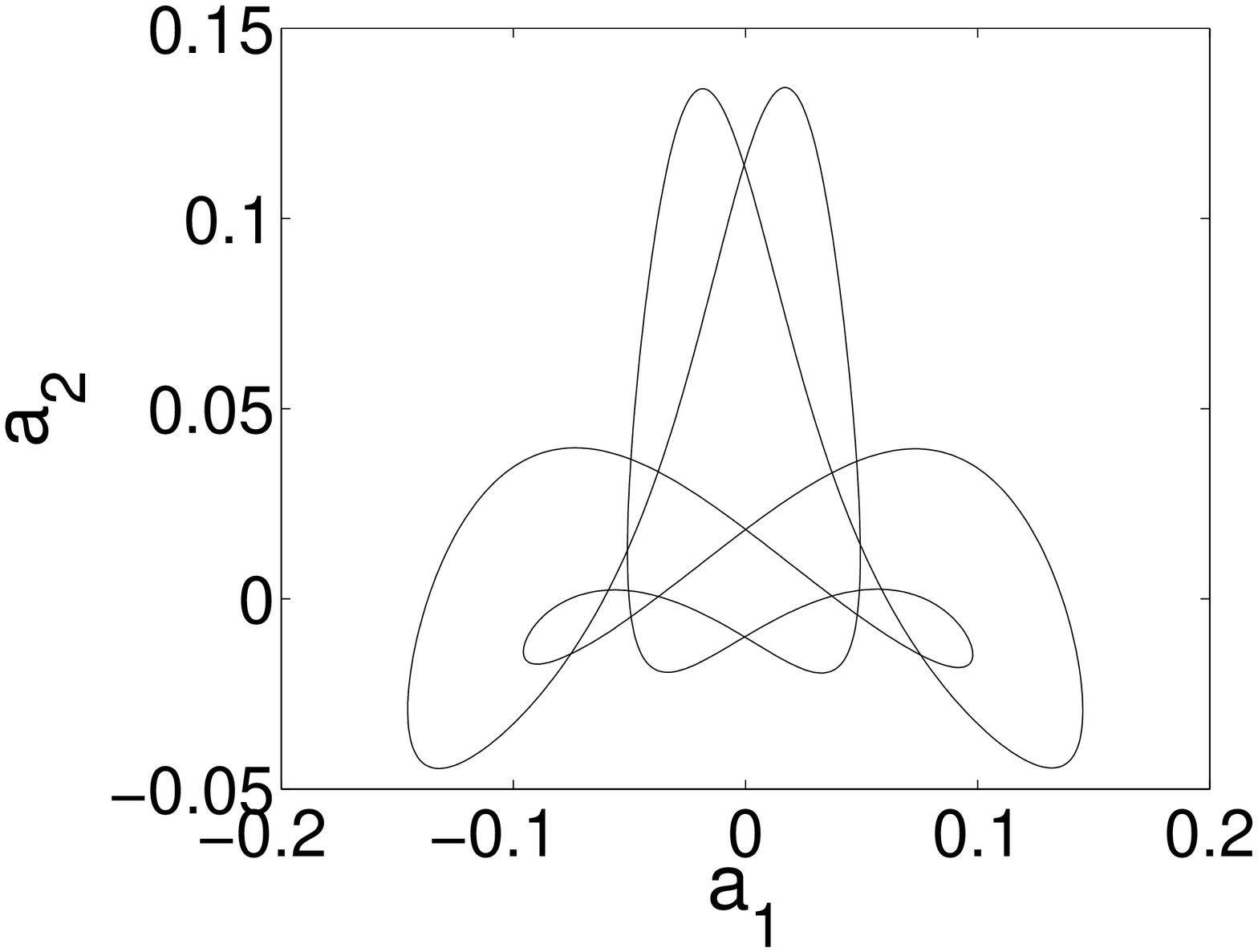}
\hspace{-0.22\textwidth} (c) \hspace{0.22\textwidth}
\includegraphics[width=0.21\textwidth]{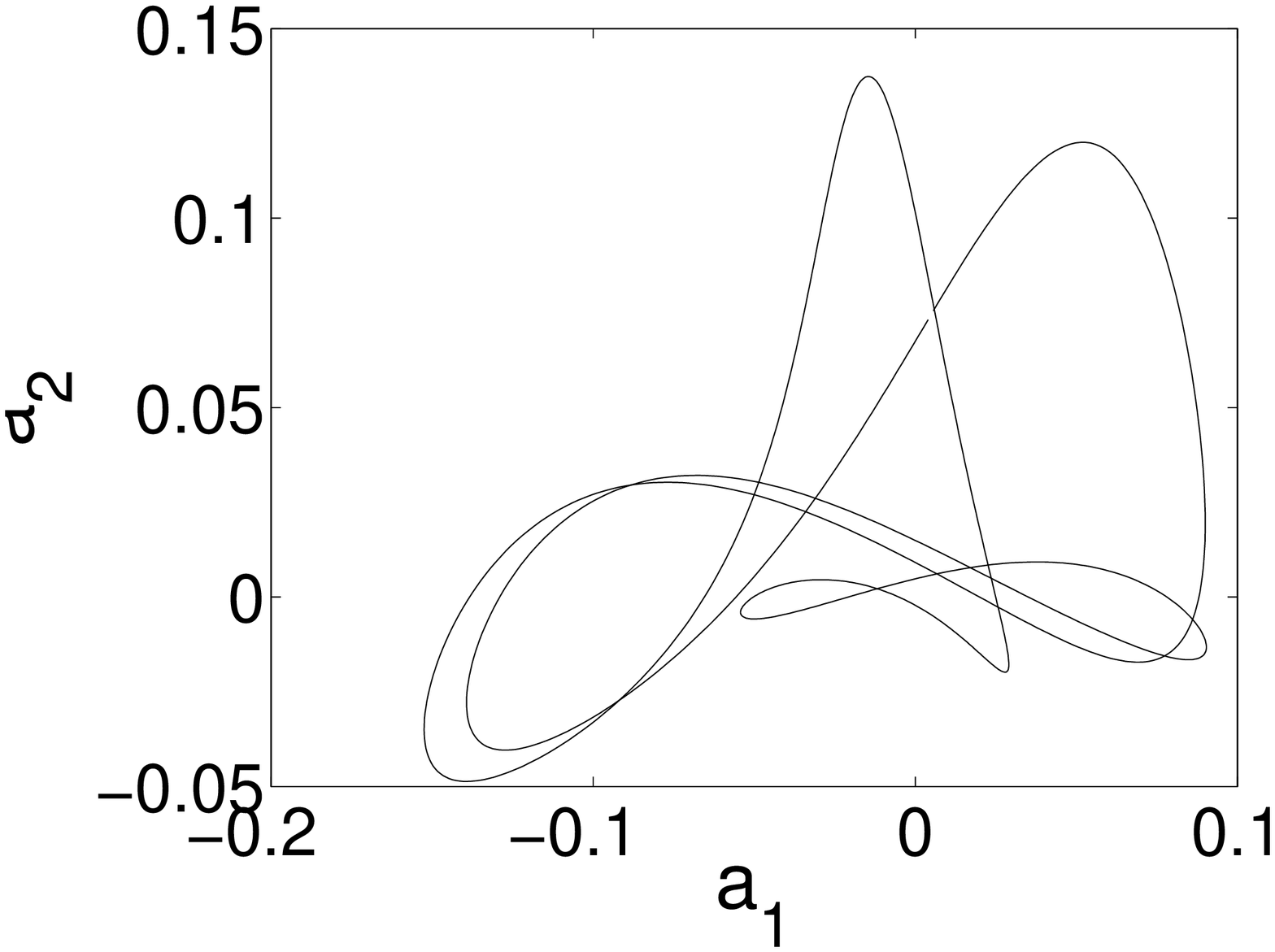}
\hspace{-0.22\textwidth} (d)
\caption[]{
The $[a_1,a_2]$ Fourier modes projection of two cycles in
$S_C$ listed in \reftab{t:orbitsf}:
(a) cycle $0011$,
(b) cycle $0012$.
Two short typical cycles not listed in the table:
(c) a symmetric {\UPO} in $S_C$ with the period $T=77.4483$,
(d) an asymmetric {\UPO}
in $S_C$ with the period $T=81.3345$.
      }
\label{f:antcycl8}
\end{figure}

For the ``center'' region $S_C$  the return map of
\reffig{f:antmn1}\,(e,f) appears to
be a good description of full \statesp\ dynamics. We found all admissible
orbits up to length $8$,
and list the orbits of fundamental domain periods up to $4$ in
\reftab{t:orbitsf}.
The cycles with $n$ odd are symmetric under the reflection
\refeq{FModInvSymm}, while the cycles with $n$ even
are either self-dual or have a symmetry partner.
Few examples are plotted in \reffig{f:antcycl8}.

All {\UPO}s we found in $S_C$ have one unstable eigendirection,
with other eigendirections highly
contracting, which justifies partially the
1\dmn\ description of
the dynamics. \refFig{f:antmn1}\,(d) shows
the $\mathcal{P}_C$  periodic points
of periods up $8$; the set
agrees well with the unstable manifold section
of \reffig{f:antmn1}\,(c).
 \refFig{f:antmn1}\,(f) displays the
return map in the fundamental domain reconstructed
from these periodic points. It matches well the map
\reffig{f:antmn1}\,(e), with the exception of an ``outlier'' cycle,
denoted $0001^*$ in \reftab{t:orbitsf}, which
shares the symbol sequence with cycle $\{0001\}$. The same ``outlier''
is visible in the center of
the lower-left quadrant of \reffig{f:antmn1}\,(d).
This point lies on another branch of the attractor, defined by a
different turn-back point. Cycles depicted in
\reffig{f:ant1p2}\,(a,c,d) and \refFig{f:antcycl8}\,(c,d)
also belong to other branches, and are  not listed in
\reftab{t:orbitsf}.
There are infinitely many turning points and
unstable manifold folds;
our description accounts for only one important part of $S_C$.

To summarize: the approximate 1-dimensional dynamics based on the closest
turn-back pair interval bracketing the $C_1$ \eqv\ unstable manifold
suffices to establish existence of a Smale horseshoe in this neighborhood.
A more accurate description would require inclusion of
further turn-backs\rf{hansen,hansen1d}.

\subsection{Side repeller}

\begin{figure}[tbp] 
    \centering
\hspace{-0.22\textwidth}
\includegraphics[width=0.21\textwidth]{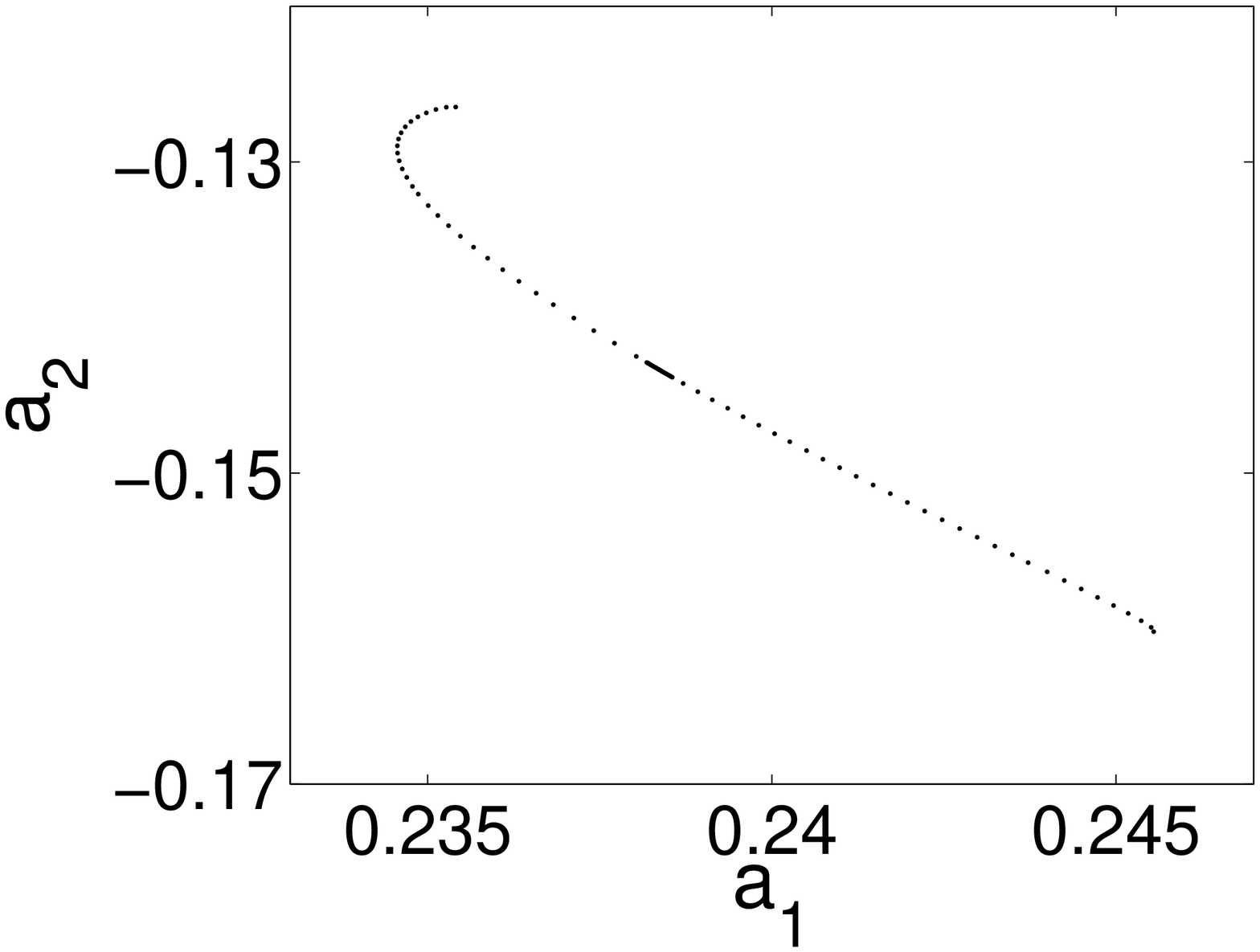}
\hspace{-0.22\textwidth} (a) \hspace{0.22\textwidth}
\includegraphics[width=0.21\textwidth]{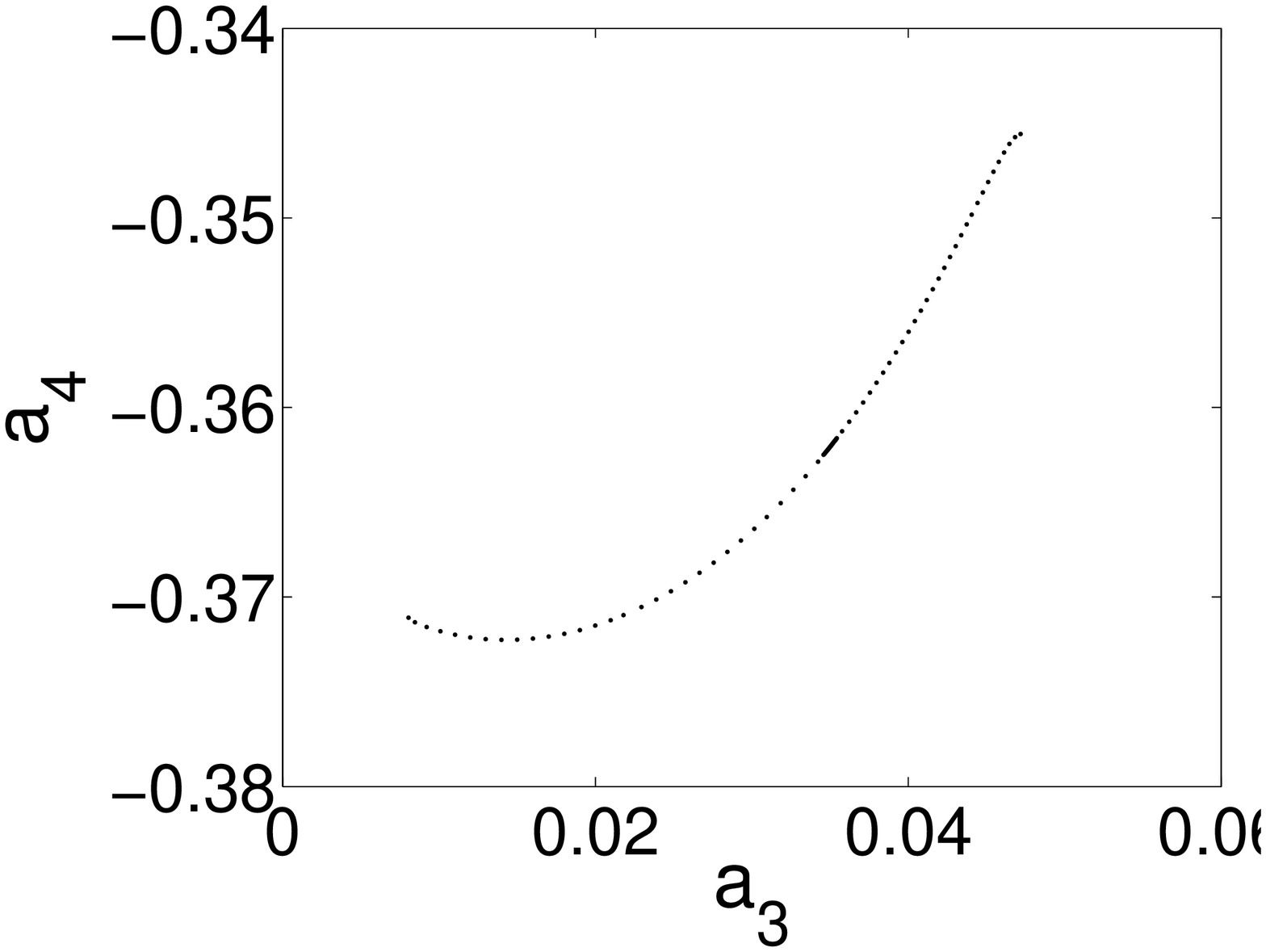}
\hspace{-0.22\textwidth} (b)
    \\
\hspace{-0.22\textwidth}
\includegraphics[width=0.21\textwidth]{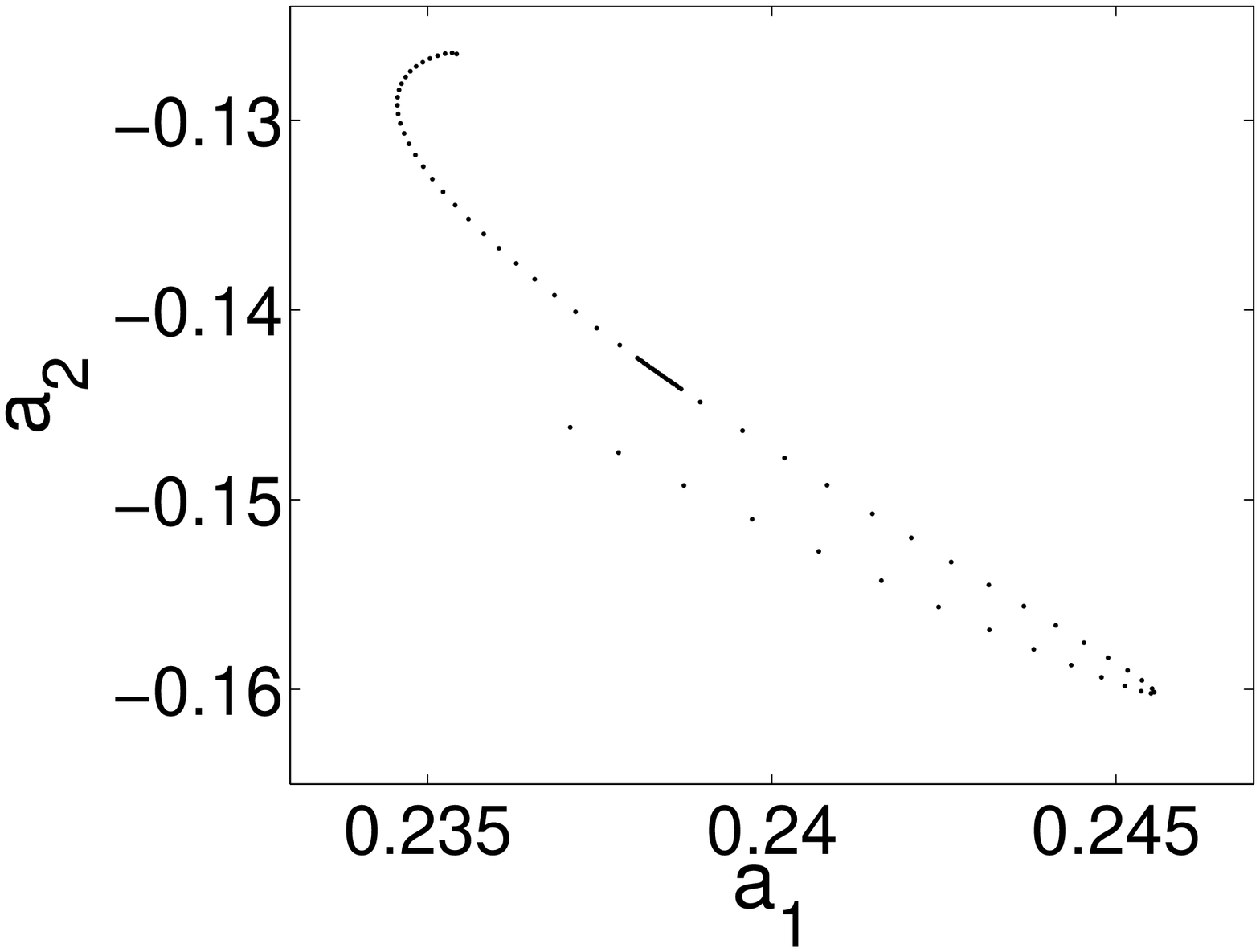}
\hspace{-0.22\textwidth} (c) \hspace{0.22\textwidth}
\includegraphics[width=0.21\textwidth]{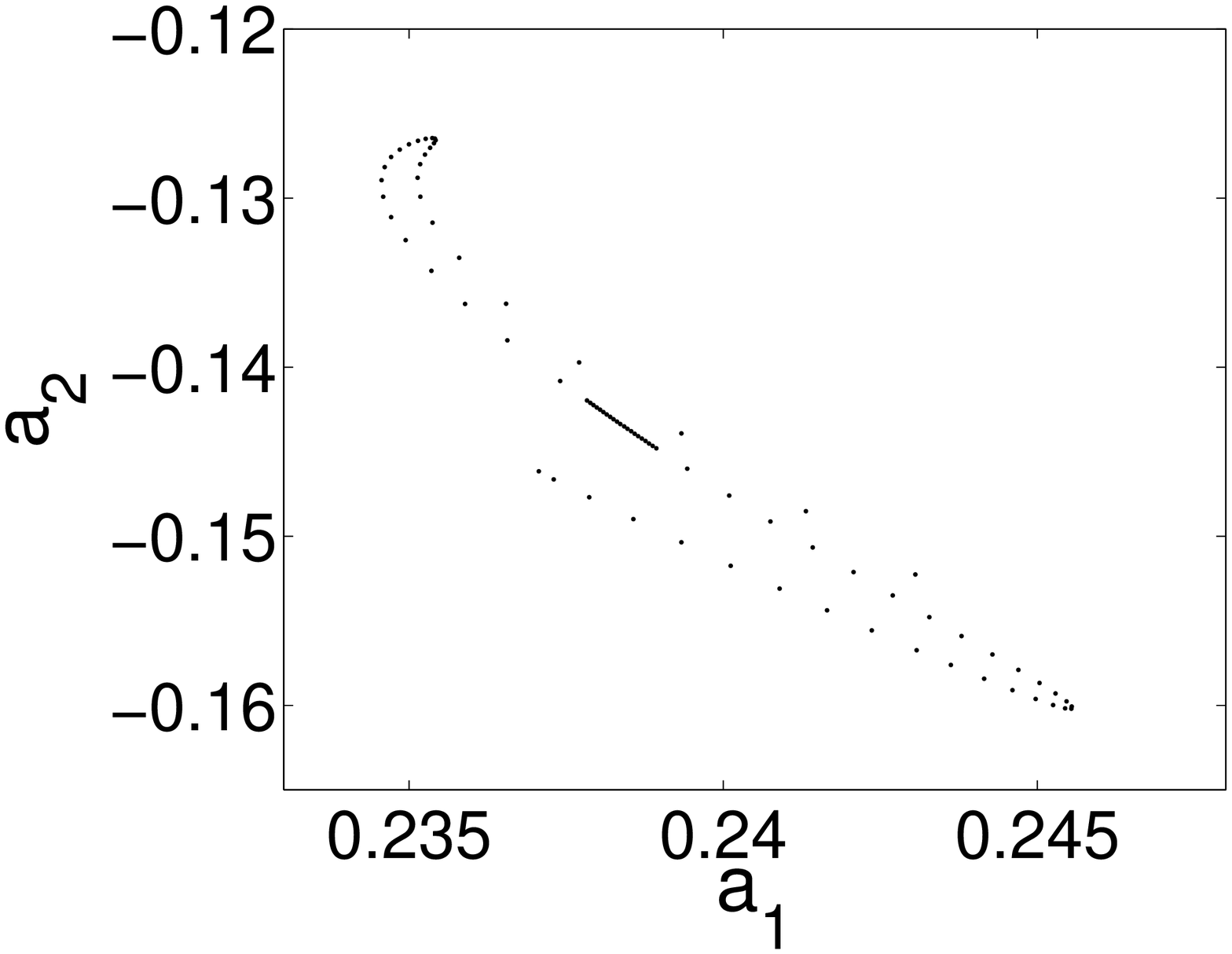} 
\hspace{-0.22\textwidth} (d)
    \\
\hspace{-0.22\textwidth}
\includegraphics[width=0.21\textwidth]{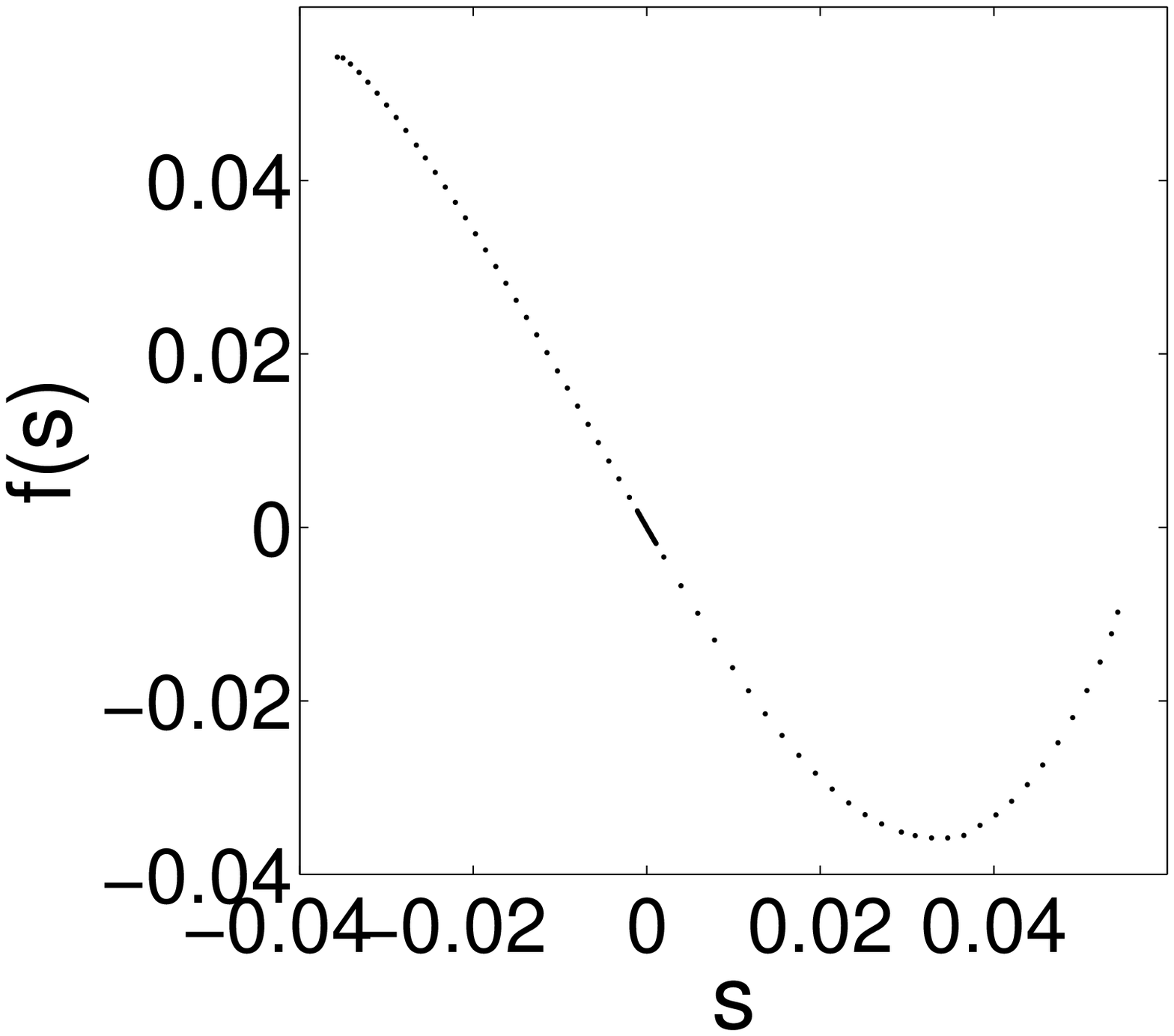}
\hspace{-0.22\textwidth} (e) \hspace{0.22\textwidth}
\includegraphics[width=0.21\textwidth]{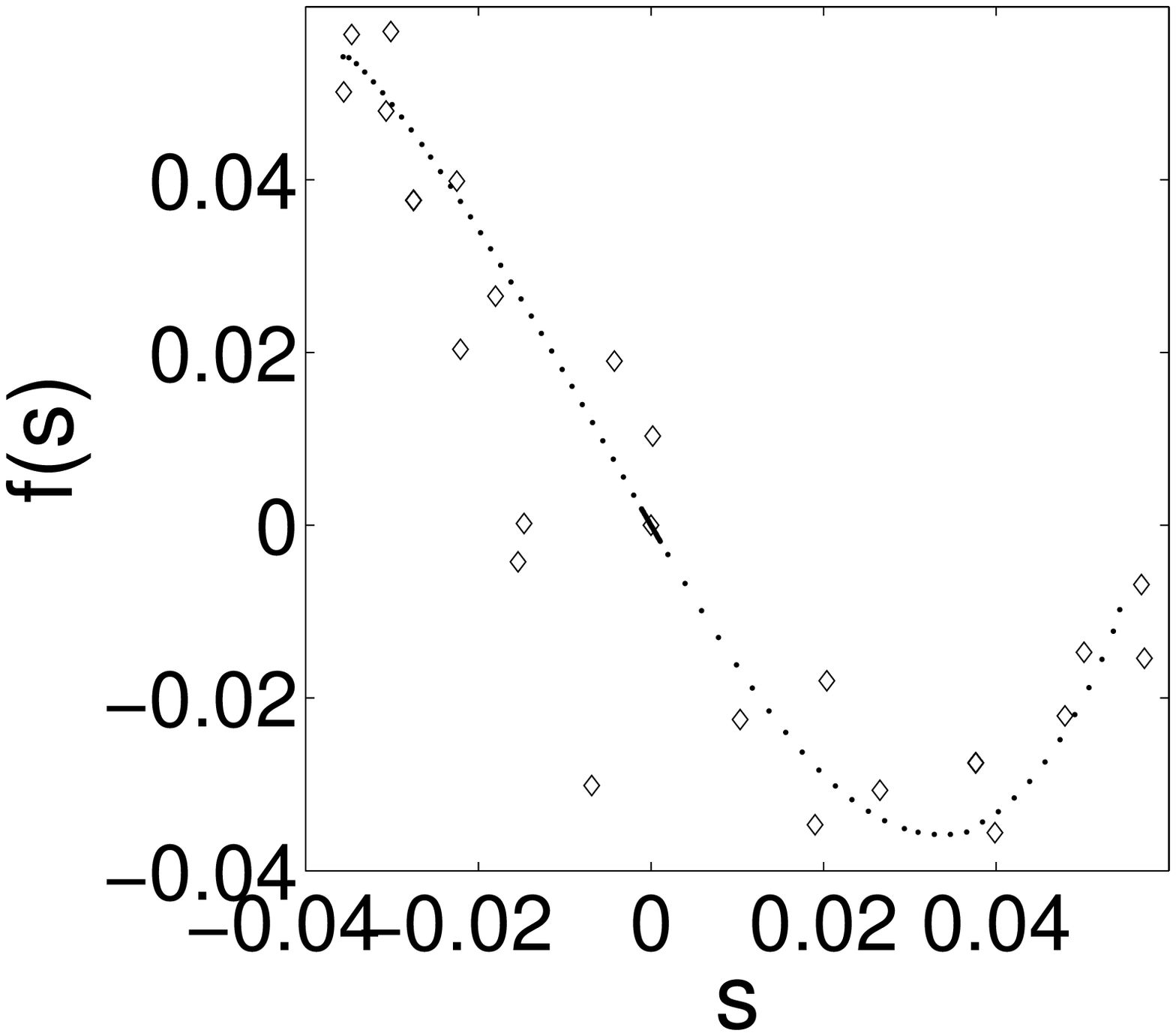}
\hspace{-0.22\textwidth} (f)
\caption[]{
The Fourier modes projections
of the unstable manifold $L_R$ of 0 cycle
on the Poincar\'{e} section $\mathcal{P}_R$:
(a) $[a_1,a_2]$ projection of the unstable manifold on $\mathcal{P}_R$.
(b) $[a_3,a_4]$ projection of the unstable manifold on $\mathcal{P}_R$.
Projection of (c) the first iteration of $L_R$,
(d) the second iteration of $L_R$.
The Poincar\'{e} return map on $\mathcal{P}_R$,
intrinsic coordinate:
(e) The return map is unimodal, with alphabet
$\{0\,,1\}$.
(f) The return map reconstructed from the periodic points (diamonds).
For comparison,
the unstable manifold return map is indicated by the dots.
      }
\label{f:antmn2}
\end{figure}

The dynamics within the ``side'' repeller $S_R$ appears more
complicated than the one discussed above for the
``center'' $C_R$. In this case
our local Poincar\'e return map captures only a small (though important)
subset of the $S_R$ {\nws}. For the $S_R$
case, we utilize the unstable manifold of the $n=1$ cycle shown in
\reffig{f:antlong}\,(b) and listed in \reftab{t:orbitsf} as
cycle ``0.''
The unstable manifold has two intersections
with the Poincar\'{e}
section $\mathcal{P}_R$; we take the upper one (along $e_2$) as the
origin from which to measure arc length $s$.
1\dmn\ unstable eigenvector of
the 0 cycle  fundamental matrix is then used to
construct the 1-dimensional, arc-length
parametrized base segment of the
unstable manifold $L_R$, as in the
preceding section and \refref{carsim}. We represent it
numerically by $82$ points.
\refFig{f:antmn2}\,(a,b) shows the $[a_1,a_2]$ and $[a_3,a_4]$
Fourier modes projection of $L_R$.

\refFig{f:antmn2}\,(c) shows the first
iterate of $L_R$, with a clear
indication of a turn-back.
\refFig{f:antmn2}\,(d) indicates a second turn-back arising from the second
iteration of $L_R$. Higher
iterations show still finer turn-back structures. The approximate
return map displayed in \reffig{f:antmn2}\,(e) is unimodal, with
the two symbols $\{0\,,1\}$ representing
the two monotone laps, partitioned by the critical point at $s=0.034$.
The corresponding pruning rules
up to length $8$ precludes the existence of sequences $\{11,1001,10000\}$,
so the set of admissible
subsequences up to length $8$ is $\{0,01,0001,000101,00010101\}$.

The search based on this symbolic dynamics yields cycles up to length $n=8$
listed in \reftab{t:orbitsf}. Note that the $01$ cycle is attractive, and thus
not part of the repeller $S_R$.
This approximate symbolic dynamics
predicts that all cycles of lengths $n=6,7,8$ are pruned;
nevertheless such cycles do exist. Furthermore,
the admissible sequences such as $0001$ have
no corresponding \UPO s.
\refFig{f:antmn2}\,(f)
shows the unstable manifold return map superimposed on
the one constructed from the periodic
points (diamonds). The return map here captures
only the gross features of the
overall dynamics of the {\UPO}s.
All this indicates that in this case our
simple 1-dimensional model
dynamics is not as a good description as that for $S_C$.

\begin{figure}[tbp] 
\centering
\hspace{-0.22\textwidth}
\includegraphics[width=0.21\textwidth]{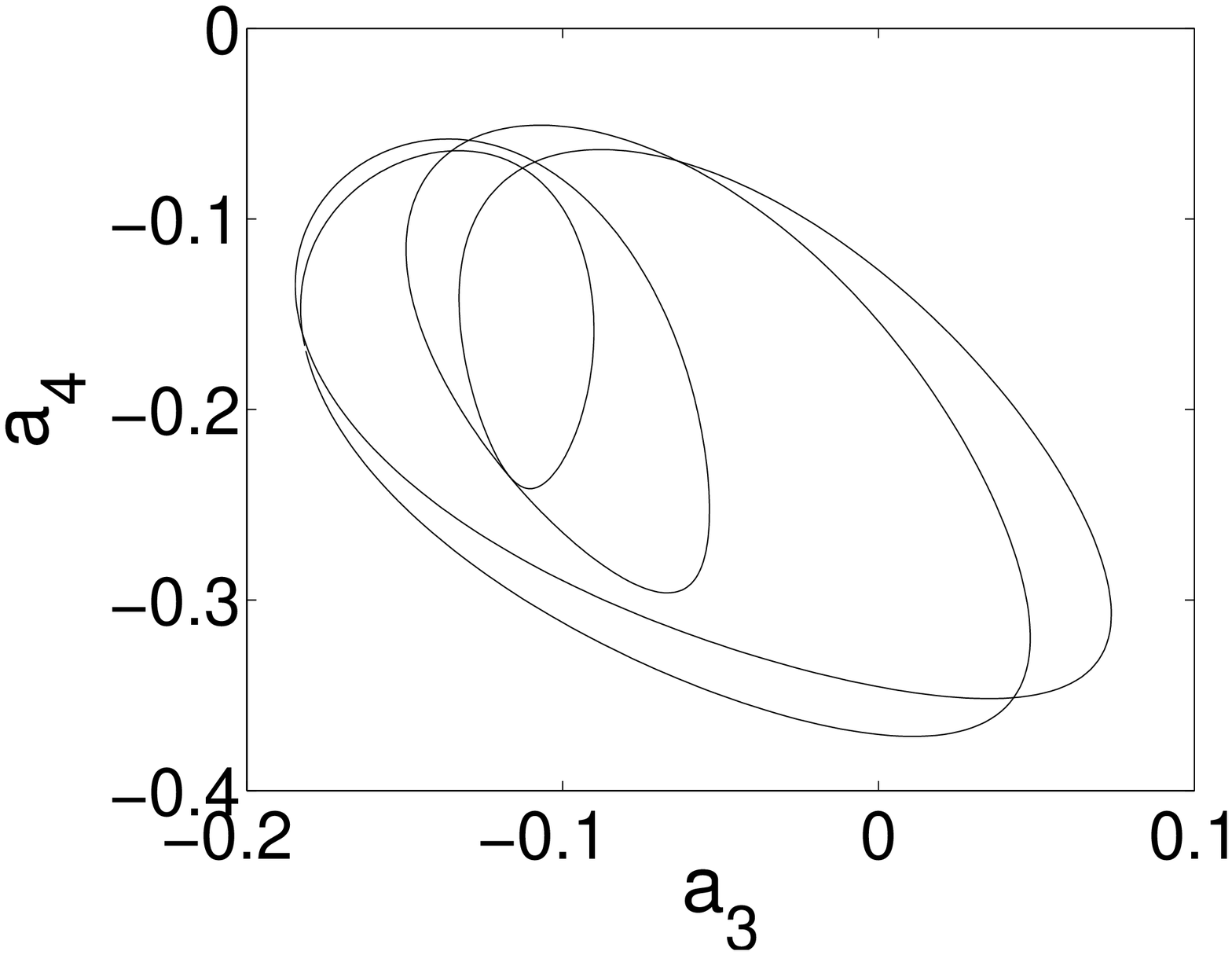}%
\hspace{-0.22\textwidth} (a) \hspace{0.22\textwidth}%
\includegraphics[width=0.21\textwidth]{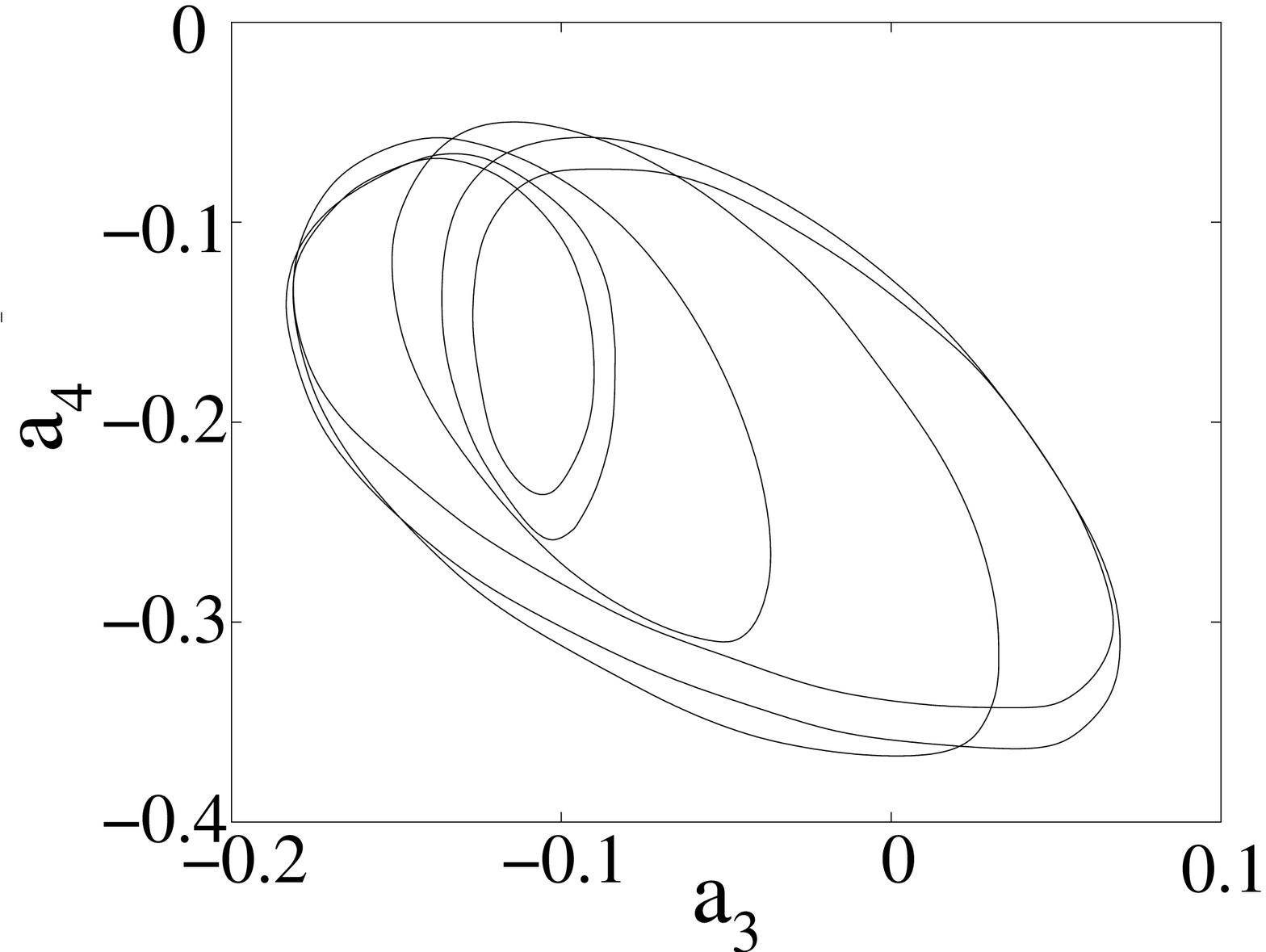}%
\hspace{-0.22\textwidth} (b)
\caption[]{
(a) $[a_3,a_4]$ Fourier modes projection of the
 attracting \po\ $01$ within $S_R$.
(b) A nearby {\UPO} within $S_R$.
      }
\label{f:antcyc1}
\end{figure}

To summarize: the \descent\ enables us to determine
all symbolic dynamics admissible {\UPO}s on $S_C$ up to
topological length $n=8$.
The Poincar\'e return map guided  \descent\ searches for
$S_R$ were not as successful, yielding only the small set of  {\UPO}s
listed in \reftab{t:orbitsf}.

However, even partial knowledge of symbolic dynamics
implied by the approximate return map of
\reffig{f:antmn2}\,(e) leads
to a very interesting discovery of the
{\em attracting} \po\ of period $T=40.0565$,  \reffig{f:antcyc1}\,(a).
A nearby {\UPO} of period $T=60.1112$ of similar
appearance is shown
in \reffig{f:antcyc1}\,(b).
In studies of turbulence in fluids only known stable
solutions are the laminar \eqva, but,
from the dynamical systems point of view, for higher-dimensional
flows any number
of coexisting attractors can exist.
The above nontrivial stable state would
never be observed in a long-time, random initial conditions
numerical simulation of the flow, as
its immediate basin of attraction is an exceedingly small island embedded
within the ``sustained turbulence'' region of \statesp.

\section{Summary}
\label{sect:sum}

The recurrent patterns program
was first implemented in detail\rf{ks} on the 1-d
Kuramoto-Sivashinsky system at the onset of chaotic dynamics.
For these specific parameter values many recurrent patterns
were determined numerically, and the periodic-orbit theory
predictions tested.
In this paper we venture into a large {\KS} system, just large
enough to exhibit ``turbulent'' dynamics of topologically richer
structure, arising through
competition of several unstable
coherent structures.
Both papers explore dynamics confined to the
antisymmetric subspace, space for which \po s characterize
``turbulent'' dynamics.
\refRef{SCD07} studies {\KS} in the full periodic domain,
where relative periodic orbits due to the continuous translational
symmetry play a key role, and \refref{GHCW07}
applies the lessons learned to a full 3$D$ Navier-Stokes flow.
In this context Kawahara and Kida\rf{KawKida01} have
demonstrated that the recurrent patterns can be determined
in turbulent hydrodynamic flows
by explicitly computing  several important unstable spatio-temporally periodic
solutions in the 3-dimensional plane  Couette turbulence.

We have applied here the ``recurrent pattern program''
to the Kuramoto-Sivashinsky system in a periodic domain,
antisymmetric subspace,
in a larger domain size than explored previously\rf{ks}.
The
\statesp\  {\nws}  for the system of this particular size
appears to consist of three
repelling Smale horseshoes and orbits communicating between them.
Each subregion is characterized by
qualitatively different spatial $u$-profiles in the 1\dmn\ physical space.
The ``recurrent patterns,'' identified in this  investigation by
nearby \eqva\ and \po s, capture well the
\statesp\ geometry and dynamics of the system. Both the
\eqva\ and \po s are efficiently determined by
the {\descent} method. The
\eqva\ so determined, together with their unstable
manifolds, provide the global frame for the {\nws}.
We utilize these unstable manifolds to build 1\dmn\ curvilinear
coordinates along which the
infinite-dimensional PDE dynamics is well approximated by 1-dimensional
return maps and the associated symbolic dynamics. In principle, these
simple models of dynamics enable us to systematically classify and
search for
recurrent patterns of arbitrary periods.
For the particular examples studied, the approach works well for the
``central'' repeller but not so well for the ``side'' repeller.

Above advances are a proof of principle, first steps in the
direction of implementing the recurrent patterns program.
But there is a large conceptual
gap to bridge between what has been achieved, and what needs to be done:
Even the flame flutter has been probed only in its weakest turbulence
regime, and it is an open question to what extent Hopf's vision remains
viable as such spatio-temporal systems grow larger and more turbulent.

\bibliography{./nonlind}

\end{document}

%% file: defsKS.tex
\newcommand{\KS}{Kuramoto-Siva\-shin\-sky}
\newcommand{\KSe}{Kuramoto-Siva\-shin\-sky equa\-ti\-on}
\newcommand{\tildeL}{\ensuremath{\tilde{L}}}

\newcommand{\UPO}{unstable periodic orbit}
\newcommand{\SIS}{non--wandering set}
\newcommand{\po}{periodic orbit}

\newcommand{\eqv}{equilibrium}

\newcommand{\eqva}{equilibria}
\newcommand{\Eqva}{Equilibria}

\newcommand{\reqva}{relative equilibria}

\newcommand{\statesp}{state space}

\newcommand{\recurrStr}{recurrent coherent structure}

  \newcommand{\Preliminary}[1]{}

  \newcommand{\YL}[1]{}
  \newcommand{\PC}[1]{}

\newcommand{\rf}     [1] {~\cite{#1}}
\newcommand{\refref} [1] {Ref.~\cite{#1}}
\newcommand{\refRef} [1] {Ref.~\cite{#1}}
\newcommand{\refrefs}[1] {Refs.~\cite{#1}}

\newcommand{\refeq}  [1] {(\ref{#1})}

\newcommand{\reffig} [1] {Fig.~\ref{#1}}

\newcommand{\refFig} [1] {Fig.~\ref{#1}}

\newcommand{\reftab} [1] {Tab.~\ref{#1}}

\newcommand{\refsect}[1] {Sec.~\ref{#1}}

\newcommand{\beq}{\begin{equation}}

\newcommand{\eeq}{\end{equation}}
\newcommand{\ee}[1] {\label{#1} \end{equation}}
\newcommand{\bea}{\begin{eqnarray}}

\newcommand{\eea}{\end{eqnarray}}
\newcommand{\barr}{\begin{array}}
\newcommand{\earr}{\end{array}}

\newcommand{\BER}[1]{{\mbox{\footnotesize BER}}} 

\newcommand{\ExpaEig}{\ensuremath{\Lambda}}

\newcommand\period[1][ ]{\ensuremath{T_{#1}}}   

\newcommand{\eigExp}[1][ ]{\ensuremath{\lambda_{#1}}}   
\newcommand{\eigRe}[1][ ]{\ensuremath{\mu_{#1}}}        
\newcommand{\eigIm}[1][ ]{\ensuremath{\nu_{#1}}}        

\newcommand{\descent}{Newton descent}

\newcommand{\costFct}{cost function}    
\newcommand{\Loop}{L}
\newcommand{\pVeloc}{v}         
\newcommand{\lSpace}{\tilde{x}}     
\newcommand{\lVeloc}{\tilde{v}}     

\newcommand{\dmn}{$\!-\!d$}             

\newcommand{\nws}{non--wandering set}

\newcommand {\id}{{\ \hbox{{\rm 1}\kern-.6em\hbox{\rm 1}}}}

